%% file: masterfile.tex
\documentstyle[querform2,twoside,edocstyle,epsf]{article}
\let\a=\alpha    \let\b=\beta     \let\g=\gamma    \let\d=\delta
\let\e=\epsilon
            \let\q=\theta       
\let\k=\kappa
\let\l=\lambda  \let\m=\mu      \let\n=\nu            \let\x=\xi
\let\p=\pi
\let\r=\rho          \let\t=\tau            
\let\f=\phi
\let\c=\chi        \let\y=\psi       \let\vq=\vartheta       \let\vf=\varphi
\let\w=\omega
\let\X=\Xi                  \let\S=\Sigma      
\let\F=\Phi         \let\G=\Gamma
\let\Y=\Psi              \let\D=\Delta     \let\L=\Lambda
      \let\del=\partial     
\newcommand{\ra}{\rightarrow}

\newcommand{\cint}[1]{\oint_C \frac{d#1}{2\pi}}
\newcommand{\ccint}[2]{\oint_{#1} \frac{d#2}{2\pi}}
\newcommand{\Comment}[1]{\par\bigskip{\footnotesize #1}\par\bigskip}
\newcommand{\wh}[1]{\widehat{#1}}
\newcommand{\Mat}{{\fett M}}
\newcommand{\ZN}{{\cal Z}_N}
\newcommand{\tr}{\mbox{Tr}\,}
\newcommand{\Det}{\mbox{Det}\,}
\newcommand{\Real}{{\goth Re}}
\newcommand{\Imagine}{{\goth Im}}
\newcommand{\Math}[1]{\mbox{$#1$}}
\newcommand{\dV}[1]{\frac{\d}{\d V(#1)}}
\newcommand{\VL}[1]{{\schnorkel L}_{#1}}
\newcommand{\Graph}[1]{\parbox{1.5cm}{\epsfxsize=1.5cm\epsfbox{#1}}}
\newcommand{\FGraph}[1]{\parbox{1.5cm}{\epsfxsize=1.1cm\epsfbox{#1}}}
\newcommand{\SF}{{\eckig \F}}
\newcommand{\MYb}{{\bar{\fett \Y}}}
\newcommand{\MY}{{\fett \Y}}
\newcommand{\Mvf}{{\fett \F}}
\newcommand{\MF}{{\fett F}}
\newcommand{\qb}{\bar{\q}}
\newcommand{\Bdangle}[1]{\Bigl \langle \!\! \Bigl \langle \, #1
                   \, \Bigr \rangle \hspace{-1.2mm} \Bigr \rangle}
\newcommand{\dangle}[1]{\langle \!\langle \, #1
                   \, \rangle \!\rangle}

\newcommand{\Wick}[2]{\MY_#1\, \ldots\, \MYb_#2
                        \put(-3,-2){\line(-1,0){11}}
                        \put(-3,-2){\line(0,1){1}}
                        \put(-14,-2){\line(0,1){1}}}
\newcommand{\Wicke}[1]{\MY_#1\, \ldots\, 
                        \put(-3,-2){\line(-1,0){6}}
                        \put(-3,-2){\line(0,1){1}}
                        \put(-9,-2){\line(0,1){1}}}
\newcommand{\eWick}[2]{ \ldots\, \MYb_#1
                        \put(-3,-2){\line(-1,0){6}}
                        \put(-3,-2){\line(0,1){1}}
                        \put(-9,-2){\line(0,1){1}}}
\newcommand{\binom}[2]{\left( \matrix{#1\cr#2}\right)}
\newcommand{\SL}[1]{{\schnorkel L}_{#1}}
\newcommand{\SG}[1]{{\cal G}_{#1}}
\newcommand{\Z}{{\bubble Z}}
\newcommand{\Jh}[1]{\widehat{J}_{#1}}
\newcommand{\bh}[1]{\widehat{\y}_{#1}}
\newcommand{\SEM}{{\cal Z}_S}
\newcommand{\AB}{A\, B \put(-1.5,-1){\line(0,1){.5}}
                                         \put(-1.5,-1){\line(-1,0){2.5}}
                                         \put(-4 ,-1){\line(0,1){.5}}}
\newcommand{\Pfaff}{\mbox{Pfaff}\,}
      \let\pa=\partial      \let\bm=\bibitem
\newcommand{\dY}[1]{\frac{\d}{\d \Y (#1)}}
\newcommand{\Blangle}{\Bigl \langle}
\newcommand{\Brangle}{\Bigr \rangle}
\newcommand{\Bangle}[1]{\Bigl \langle \, #1
                    \,  \Bigr \rangle}
\newcommand{\XL}{( \X_1 - \L_1 )}
\newcommand{\Yop}{\widehat{\eckig \Y} \, \circ\, }
\newcommand{\Yopx}{\widehat{\eckig \Y}_x \, \circ\, }
\newcommand{\Vop}{\widehat{\schnorkel V}^\prime \, \circ \, }
\newcommand{\px}[1]{\frac{1}{(p-x)^{#1}}}
\newcommand{\py}[1]{\frac{1}{(p-y)^{#1}}}
\newcommand{\lra}{\leftrightarrow}
\newcommand{\bra}[1]{\Bigl [ \, #1 \, \Bigr ]}
\newcommand{\mat}[1]{\mbox{$ #1 $}}
\newcommand{\Fbg}{\mat{F^{\mbox{\scriptsize bos}}_g}\  }
\newcommand{\Ffg}{\mat{F^{\mbox{\scriptsize ferm}}_g}\  }
\newcommand{\MX}{{\fett X}}
\newcommand{\ML}{{\fett L}}
\newcommand{\DL}{\del_{\fett L}}
\newcommand{\MQ}{{\fett \L}}
\newcommand{\DQ}{\del_{\fett \L}}
\newcommand{\dl}[1]{\frac{\del}{\del l_{#1}}}
\newcommand{\dq}[1]{\frac{\del}{\del \q_{#1}}}
\newcommand{\dm}[1]{\frac{\del}{\del \m_{#1}}}
\ohnelabels
\ohnezitat
\mitbilder
\begin{document}
\pagestyle{empty}
\TITELSEITE{
 \begin{center} \null \leftline{ITP-UH 27/95, \,\, 
DESY 95-234 \hfill ISSN 0418 - 9833}
\leftline{December 1995}\vskip 2.5cm
 {\LARGE\sc Supersymmetric Generalizations of Matrix Models} \\
  \large \vfill
     Vom Fachbereich Physik der Universit\"at Hannover \\ 
     zur Erlangung des Grades \\
     Doktor der Naturwissenschaften\\
     Dr.\ rer.\ nat. \\ genehmigte Dissertation \\ 
     von \\
     \vfill
     {\bf Jan Christoph Plef\/k\/a, M.Sc. } \\ 
     geboren am 31.\ Januar 1968 in
     Hanau \\ \vskip 1cm
   {\bf 1995}
  \end{center} }{\input abs}{\null \leftline{\phantom{jaja}}
\vfill \centerline{ {\it
Sissi und Timm gewidmet 
}}\vfill}
\cleardoublepage
\Inhaltsverzeichnis
\textanfang
\anfangInhalt
\zitat{Nature is simple if you look at it in the right way. For example
I believe that God created two dimensions. One time and one space dimension.
Nothing is simpler. \\[1mm] [1]}{AS}
\kapitel{Introduction}
\input{introduction}
\zitat{}{}
\Kapitel{The Hermitian Matrix Model}{hermitianmm}
\input{hermitianmm}

\zitat{}{}
\Kapitel{The $c=-2$ Matrix Model}{c2mm}
\input{c2mm}
\spalteInhalt
\zitat{}{}
\Kapitel{The Supereigenvalue Model}{supereigen}
\input{supereigen}
\spalteInhalt
\zitat{}{}
\Kapitel{The External Field Problem}{external}
\input{external}
\zitat{}{}

\endeInhalt
\cleardoublepage
\thispagestyle{empty}
\null
\centerline{\large\sc Lebenslauf}
\par
\bigskip\bigskip
\begin{tabular}{lcl}
1974 -- 1978 &: &    Besuch der Grundschule, ein Jahr hiervon in \\&&
 Los Angeles, Kalifornien. \\ &&\\
1978 -- 1987&: &     Eleonorenschule Gymnasium Darmstadt. \\&&   \\
Mai 1987&: &  Abiturpr\"ufung.\\ &&\\
Juli 1987 -- Sept. 1988&:&  Zivildienst am Alice--Hospital 
                                             in Darmstadt.\\ &&\\
Okt. 1988 -- Aug. 1991 &:&  Studium der  Physik (Diplom) an der 
  Technischen\\&&  Hochschule Darmstadt. \\&&\\
Oktober 1990 &:&  Vordiplom in Physik.\\ &&\\
Sept. 1991 -- Dez. 1992 &:&  Graduate Student in Physik an der
  \\&& Texas A\&\relax M University, College Station, TX, USA.  \\&& \\
Dezember 1992 &:&  Master of Science. \\ && \\
seit Februar 1993 &:&  Doktorand am Institut f\"ur Theoretische 
Physik\\&&  der Universit\"at Hannover. \\ && \\
April -- Dez. 1993 &:& Wissenschaftlicher Mitarbeiter.\\&&\\
seit Januar 1994   &:& Promotionsstipendiat der 
Studienstiftung des \\
&& Deutschen Volkes. 
\end{tabular}
\end{document}

%% file: abs
%
\null
\centerline{\large\sc Supersymmetrische Verallgemeinerungen von}
\centerline{\large\sc Matrix Modellen}
\bigskip
\centerline{\bf Zusammenfassung}
\medskip
Die vorliegende Arbeit besch\"aftigt sich mit der supersymmetrischen
Verallgemeinerung von Matrix-- und Eigenwertmodellen. Nach einer kurzen
Einf\"uhrung in das Hermitesche Ein--Matrix--Modell wenden wir uns dem
\Math{c=-2} Matrix--Modell zu. In seiner Formulierung durch ein matrixwertiges
Superfeld ist dieses Modell invariant unter Supersymmetrietransformationen
auf Matrixebene. Wir zeigen die Existenz einer Nicolai--Abbildung dieses
Modells auf ein freies Hermitesches Matrix--Modell und diskutieren
seine diagrammatische Entwicklung. Korrelationsfunktionen f\"ur
quartische Potentiale und beliebigen Genus werden berechnet, welche
die Stringsuszeptibilit\"at von \Math{c=-2} Liouville--Theorie im Skalenlimes
aufweisen. Wir zeigen auf, wie sich diese Ergebnisse zum Z\"ahlen
supersymmetrischer Graphen verwenden lassen.
\newline
Daraufhin studieren wir das Supereigenwertmodell, der bis heute
einzige erfolgreiche diskrete Zugang zur Quantisierung von 2d 
Supergravitation. Das Modell wird in einer superkonformen Feldtheorie
Formulierung durch Forderung von Super--Virasoro Bedingungen konstruiert.
Die vollst\"andige L\"osung wird mit Hilfe der Momentenmethode
hergeleitet. Diese erm\"oglicht die Berechnung der freien Energie und
aller Multi--Loop--Korrelatoren auf beliebigem Genus und f\"ur allgemeine
Potentiale. Die L\"osung wird im diskreten Fall und im Doppelskalenlimes
pr\"asentiert. Explizite Resultate bis Genus zwei werden angegeben.
\newline
Es folgt eine Diskussion der supersymmetrischen Verallgemeinerung des
externen Feld Problems. Die diskreten 
Super--Miwa--Transformationen des Supereigenwertmodells k\"onnen
im Eigenwert-- und im Matrixfall angegeben werden. Eigenschaften von
externen Supereigenwertmodellen werden diskutiert, die genaue Form des
zum gew\"ohnlichen Modell korrespondierenden externen Supereigenwertmodells 
konnte jedoch noch nicht hergeleitet werden.
\par
\vfill
\leftline{{\bf Referent:} Prof.\ Dr.\ O.\ Lechtenfeld}
\leftline{{\bf Koreferent:} Prof.\ Dr.\ N.\ Dragon}
\leftline{{\bf Tag der Promotion:} 29.11.1995}
\par
\pagebreak\null
\centerline{\large \sc Supersymmetric Generalizations of Matrix Models}
\centerline{\phantom{\large\sc von Matrix Modellen}}
\bigskip
\centerline{\bf Abstract}
\medskip
In this thesis generalizations of matrix and eigenvalue models involving
supersymmetry are discussed. Following a brief review of the Hermitian
one matrix model, the \Math{c=-2} matrix model is considered. Built
from a matrix valued superfield this model displays supersymmetry on
the matrix level. We stress the emergence of a Nicolai--map of this
model to a free Hermitian matrix model and study its diagrammatic
expansion in detail. Correlation functions for quartic potentials on
arbitrary genus are computed, reproducing the string susceptibility of
\Math{c=-2} Liouville theory in the scaling limit. The results may be
used to perform a counting of supersymmetric graphs. 
\newline
We then turn to the supereigenvalue model, today's only successful discrete
approach to 2d quantum supergravity. The model is constructed in a
superconformal field theory
formulation by imposing the super--Virasoro
constraints. The complete solution of the model is given in the moment 
description, allowing the calculation of the free energy and the multi--loop 
correlators on arbitrary genus and for general potentials. The solution
is presented in the discrete case and in the double scaling limit. Explicit
results up to genus two are stated.
\newline
Finally the supersymmetric generalization of the 
external field problem is addressed. We state the discrete
super--Miwa transformations of the supereigenvalue model on the
eigenvalue and matrix level. Properties of external supereigenvalue 
models are discussed, although the model corresponding to the ordinary
supereigenvalue model could not be 
identified so far.
\vfill
\centerline{\bf Acknowledgments}
\medskip
I wish to express my sincere gratitude to Prof.\ Olaf Lechtenfeld for his
guidance and steady support during my time as his student.
\newline
Moreover I would like to thank E.\ Abdalla, P.\ Adamietz, G.\ Akemann,
L.\ Alvarez--Gaum\'e, J.\ Ambj\o rn, J.\ Bischoff, 
N.\ Dragon, R.\ Emparan, S.\ Kharchev, A.\ Mironov
and A.\ Zadra for valuable discussions, comments and remarks related
to the work of this thesis.
\newline
Finally I thank the Studienstiftung des Deutschen Volkes for financial
support.
\newline
And most of all I thank Mieke Bauer for putting up with me during the ups and
downs of this work.

%% file: introduction
%
%
Random matrices have been studied in physics since the work of E.\ Wigner
in the 1950's.
Initially proposed as an effective model for higher excitations in
nuclei, they have found numerous applications in various fields
throughout the years. In high--energy and mathematical physics matrix
models experienced a great renaissance following the discovery of their
relevance for the quantization of two dimensional gravity and 
bosonic string theory in 1985. Only recently the supersymmetric
generalization of matrix models has been addressed. This thesis
reports on the construction, solution and interpretation of such
generalized models, focusing on the work of the author.
\par
Matrix models have the appealing property that they are exactly solvable
in the limit of infinite matrix size $N$. Subleading corrections to arbitrary 
order in \Math{1/N} may be computed by iterative means, while results
for finite $N$ are few. The analysis of the diagrammatic expansion of the
Hermitian matrix model reveals that solving the model is equivalent to
performing the sum over random equilateral triangulations of 
two--dimensional surfaces of fixed genus. 
\par
String theory on the other hand is today's most popular candidate for a 
unification of quantum field theory and gravity. It is built on the 
attractive idea of replacing particles by string--like one--dimensional
objects, which sweep out a two--dimensional ``world--sheet'' as they 
evolve in time. According to Polyakov this theory is to 
be interpreted as two--dimensional quantum gravity, in which the 
world--sheet replaces two--dimensional space--time, and where 
the coordinates of the string form matter fields coupled to the 
two--dimensional gravity theory. Interactions correspond to topology 
changes and are encoded in the genus of 
the world--sheet, which may be viewed as an ``inflated'' Feynman graph. 
Thus the task in
a path--integral quantization is to perform an integral over the two 
dimensional geometries of the world--sheet and sum over their genera. 
\par
Precisely this integration may be performed with the help of the matrix 
model by taking a ``continuum limit'' of the above--mentioned exact
solution, i.e.\ when the triangles of the discretized surfaces become dense. 
However, this rather unconventional method of performing a path integral 
only works for toy models of bosonic strings with world--sheets of
Euclidean signature living in \Math{D=c\leq 1} 
dimensions. Extending these techniques to dimensions greater than one
is a hard problem still unsolved.
Due to the exact solvability of matrix models new 
non--perturbative techniques were introduced to these
low--dimensional string theories. We will review the Hermitian matrix
model and its correspondence to random surfaces in chapter I.
\par
Of potential phenomenological relevance, however, are superstring
theories, which in contrast to the bosonic string have bosons and fermions
in their spectrum. The Neveu--Schwarz--Ramond superstring is
to be interpreted as two dimensional supergravity coupled to a matter
action of superfields. It is thus very desirable to find a supersymmetric 
generalization of the matrix model technique in order to perform the 
corresponding path integral over super--geometries. There are (at least) 
three different ways to proceed.
\par
In a geometrical approach one puts superstrings on a random lattice and
then tries to establish a map to an adequate supersymmetric matrix model.
But due to the known problems with supersymmetry and fermions 
in lattice field theory this seems to be rather hopeless. However, the
alternative superstring theory of Green and Schwarz does not suffer
from these problems, as here the fields on the world--sheet remain bosonic
and supersymmetry is only present on the external space--time level.
A lattice version of this superstring theory has been formulated and
studied. Still, the relation to a solvable supersymmetric matrix
model is an open problem (for more details see refs.\ \cite{GSLatt}).
Even classically the Green--Schwarz string is supersymmetric only
in \Math{D\geq 4} dimensions, a domain that seems to be hardly reachable for 
solvable matrix models.
\par
Alternatively one can consider supersymmetric generalizations of
matrix models and hope that their critical behaviour reveals properties
of two--dimensional supergravity. However, no example is known in the
literature where this is the case. Nevertheless, the analysis of these
models is interesting in its own right from the viewpoint of random
matrices. We shall therefore study the \Math{c=-2} matrix
model in chapter II, a model displaying supersymmetry on the level of 
matrices. In its ``continuum limit'' this model describes the 
propagation of bosonic strings in minus two--dimensional space--time. 
\par
The final and successful approach seeks for a generalization of the 
integrable structure of the Hermitian matrix model. This structure at
the heart of the Hermitian matrix model resides in a set of Virasoro 
constraints.
It turns out that, by imposing a generalization in the form
of super--Virasoro constraints, one obtains the
supereigenvalue model of Alvarez--Gaum\'e, Itoyama, Ma\~nes and
Zadra, which generalizes the eigenvalue formulation of the Hermitian
matrix model. 
Due to the eigenvalue (rather than matrix) character of this model there is no
geometrical interpretation in terms of triangulated super--Riemann
surfaces at hand. The supereigenvalue model is exactly solvable for 
general polynomial potentials in the limit of an infinite number $N$ of 
eigenvalues. Moreover, all subleading corrections in \Math{1/N} are
determined through an iterative process. We will describe this solution
in chapter III in the ``discrete'' and ``continuum'' cases. The continuum
results reveal that the supereigenvalue model indeed describes the
coupling of minimal superconformal theories to two dimensional quantum
supergravity. 
\par
As we shall see, the supereigenvalue model has a large number 
of similarities to the Hermitian matrix model. This motivated the
author to study the external field problem in this context as well.
In chapter IV we review the external Hermitian matrix model and
present some preliminary results of a generalization to the supersymmetric
case.
\par
The presentation of this thesis is intended to be self contained, only
basic knowledge of conformal field theory is assumed. For a
review on this topic see e.g.\ ref.\ \cite{CFT}. There is a large number
of reviews on matrix models and two--dimensional quantum gravity, e.g.\
refs.\ \cite{2dGra}, but only the recent ref.\ \cite{Book} contains the 
supereigenvalue model.
\par
The parts of this thesis containing the solution of the supereigenvalue
model have been published priorly in refs.\ \cite{Ple1,Ple2}.

%% file: hermitianmm
%
%
%
Generally speaking matrix models are quantum
field theories where the field is a $N\times N$ real or complex  
matrix ${\fett M}(x)$. We shall consider the
simple case of the Hermitian one matrix model in $D=0$ dimensions,
which due to its simplicity in exactly solvable in the limit
of infinite matrix size $N$. 
This model has been 
intensively studied in the literature. In the following we give a 
basic introduction to the model, as well as a review on a collection
of more detailed aspects which will be relevant for the generalization to the
supersymmetric case.
\Subkapitel{The Model}{model}
The Hermitian one matrix model is defined by the partition function
\beq
\ZN [ g_k]= e^{N^2\, F[g_k]}=
\int {\cal D} \Mat \, \exp \Bigl [ - N\, \tr V(\Mat )\Bigr ],
\eeq[hermmmodel]
where $\Mat_{ij}$ is a \Math{N\times N} Hermitian matrix. 
\Math{ F[ g_k]} is the free energy. 
The measure for the ``path integral'' of this zero dimensional theory is 
given by
\beq
{\cal D}\Mat = \prod_{i<j}\, d \Real (\Mat_{ij})\, d \Imagine (\Mat_{ij}) \,
\prod_{i}\, d \Mat_{ii},
\eeq[hermeasure]
and we consider the most general polynomial matrix potential with coupling
constants $g_k$
\beq
V(\Mat)= \sum_{k=0}^\infty g_k\, \Mat^k.
\eeq[hermpotential]
Note the $U(N)$ invariance \Math{\Mat \mapsto U^\dagger\, \Mat \, U}
of  the ``action'' \Math{\tr V(\Mat)}. This may be
used to diagonalize the Hermitian matrix \Math{\Mat= U^\dagger\, {\fett D}
\, U}, where \Math{{\fett D}=\mbox{diag}(\l_1,\ldots ,\l_N)} is the
diagonal matrix. By performing the change of variables from $\Mat$
to $\l_i$ and $U$, the integral over the unitary group factors out and
we are left with the integration over the eigenvalues \Math{\l_i}. 
The Jacobian of this transformation is \Math{\prod_{i<j}(\l_i-\l_i)^2}.
\par
\bigskip{\footnotesize
This may be seen by considering
the norm of the infinitesimal variation of \Math{\Mat= U^
\dagger\, {\fett D}\, U}
\beql
| \,\d\Mat\,|^2&=& \sum_{i,j}\, \d\Mat_{ij}\,\d\Mat_{ji}\, =\, \tr (\d\Mat )^2
\zeile
&=& \tr \Bigl ( - U^\dagger\,\d U\, U^\dagger\, {\fett D}\, U + U^\dagger
\,\d {\fett D}\, U + U^\dagger\, {\fett D}\, \d U\, \Bigr )^2\zeile
&=&  \tr (\d{\fett D})^2 - 2i\,\tr [\d{\fett D},{\fett D}]\,\d u + 
         2\,\tr (-\d u\, {\fett D}\,\d u\,{\fett D} + (\d u)^2\,{\fett D}^2\,),
\nonumber\eeql                 
where we have introduced \Math{\d u= i\, \d U\, U^\dagger={\d u}^\dagger}.
The second term in the last expression vanishes as $\d{\fett D}$ and
${\fett D}$ are diagonal. We then find
\beqx
| \,\d\Mat\,|^2 = \sum_i (\d\l_i )^2 + \sum_{i,j} (\l_i -\l_j )^2\, |\d u_{ij}|^2.
\eeqx
Note that the independent variables are the variation of the eigenvalues
\Math{\d\l_i} and \Math{\Real \d u_{ij}}, \Math{\Imagine \d u_{ij}} for
\Math{i<j}. From the last equation we obtain the Jacobian of \Math{\Mat
\mapsto \l_i, U} which is \Math{\sqrt{G}}
where $G$ is the metric tensor, explicitly
\beqx
G= \prod_{i\neq j}\, (\l_i -\l_j)^2 \quad \Rightarrow \quad
\sqrt{G}= \prod_{i<j}\, (\l_i -\l_j )^2 \equiv \D^2 (\l).
\eeqx
}
\par\bigskip
The Hermitian one matrix model \gl{hermmmodel} may then be
written in the eigenvalue representation
\beq
\ZN [ g_k]= c_N\, \int_{-\infty}^{\infty} (\prod_{i=1}^N d\l_i )\,
\prod_{i<j} (\l_i - \l_j)^2 \, \exp \Bigl [ -N\,\sum_{i=1}^N \, V(\l_i 
)\Bigr ],
\eeq[hermeigenvaluemodel]
where \Math{c_N} is the volume of the \Math{U(N)} group. Note the relation
of the Jacobian \Math{\D^2 (\l)} to the van der Monde determinant
\Math{\D (\l) = (-)^{N(N-1)/2}\, \Det (\l^{j-1}_i)}.
\Subsubkapitel{Loop Insertion Operator}
Expectation values in the Hermitian one matrix model are defined in 
the usual way as
\beq
\langle\, {\schnorkel O}(\Mat)\, \rangle = \frac{1}{\ZN}\,
\int {\cal D} \Mat \, {\schnorkel O}(\Mat)\,
\exp \Bigl [ - N\, \tr V(\Mat )\Bigr ].
\eeq[hermaverage]
A similar expression holds in the eigenvalue picture.
 It is very convenient to work with the one--loop
correlator
\beq
W(p)= \frac{1}{N}\,\sum_{k=0}^\infty\, \frac{\langle\,\tr \Mat^k\,\rangle}
{p^{k+1}},
\eeq[herm1loopcorrelator]
which acts as a generating functional for the amplitudes
\Math{\langle\,\tr \Mat^k\,\rangle}. Similarly we may define the
generating functional for higher point amplitudes, the $n$--loop
correlators
\beq
W(p_1,\ldots,p_n)= N^{n-2}\! \sum_{k_1,\ldots ,k_n=1}^\infty 
\frac{\langle\,\tr \Mat^{k_1}\,\ldots\,\tr \Mat^{k_n}\rangle_{\mbox{
\footnotesize conn.}}}{{p_1}^{k_1+1}\, \ldots \, {p_n}^{k_n+1}},
\eeq[hermnloopcorrelator]
where ``conn.'' refers to the connected part. The last two equations may
be rewritten as 
\beq
W(p_1,\ldots,p_n)= N^{n-2} \, \Bigl\langle\, \tr \frac{1}{p_1-\Mat}\ldots
\tr \frac{1}{p_n-\Mat}\,\Bigr\rangle_{\mbox{conn.}}.
\eeq[hermnloopcorrelator2]
\par
A useful object to define is the loop insertion operator \cite{Amb93}
\beq
\dV{p} \equiv - \sum_{j=1}^\infty \frac{1}{p^{\, j+1}}\, \frac{\del}{\del g_k},
\eeq[loopinsertion]
because we may now obtain the $n$--loop correlators from the free energy
$F$ by applying the loop insertion operators:
\beq
W(p_1,\ldots,p_n)= \dV{p_1}\dV{p_2}\ldots \dV{p_n}\, F.
\eeq[hermnloopcorrelator3]
Hence once the free energy for a general potential is known, all
observables may be calculated. The same holds true for the one--loop
correlator \Math{W(p_1)}, as with
\beq
W(p_1,\ldots,p_n)= \dV{p_2}\ldots \dV{p_n}\, W(p_1)
\eeq[hermnloopcorrelator4]
all multi--loop correlators follow from it. We thus see that solving
for \Math{W(p)} with a general potential really means completely solving
the Hermitian one matrix model. We shall see in section \Sub{hermsolution}
that this is most efficiently done by considering the loop equations 
of the model.
\par
It is known since the work of t'Hooft \cite{tHo} that the free energy 
\Math{F} and all correlators admit an expansion in \Math{1/N^2}, 
which may be seen by looking at the perturbative evaluation 
of the matrix model \gl{hermmmodel}.
\Subsubkapitel{Feynman Diagrams}
The propagators of the \Math{\Mat_{ij}} fields of 
eq.\ \gl{hermmmodel} may 
be represented diagrammatically by double lines each one corresponding 
to the separate propagation of the matrix indices. For Hermitian matrices the 
lines should be oriented in opposite directions (cf.\ Figure 1).
The propagator then
simply is
\beq
\langle\, \Mat_{ij}\, \Mat_{kl}\,\rangle_0 = \frac{1}{N}\, \d_{ik}\,
\d_{jl},
\eeq[hermpropagator]
where the subscript $0$ denotes the average taken in the free
theory, i.e. \Math{g_2=1/2} and \Math{g_k=0} for \Math{k\neq 2}. 
With the general interactions
of eq.\ \gl{hermpotential} there will be three--point vertices, four--point
vertices, etc\ldots, each n--point vertex contributing a factor of
\Math{(- g_n\, N)}. Moreover each loop of internal index will yield the
factor \Math{N=\d_{ii}}. 
\par
\begin{figure}[t]
\begin{center}
\leavevmode\epsfxsize=8cm
\epsfbox{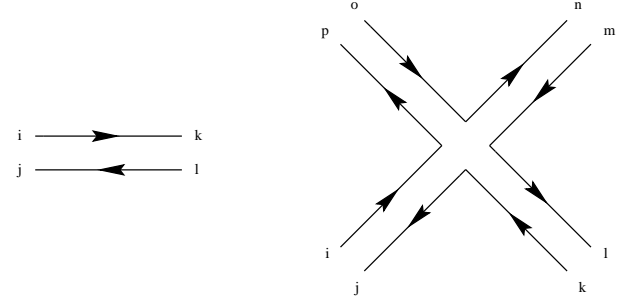}
\caption{The propagator and a four point vertex of the Hermitian matrix
model}
\end{center}
\end{figure}
Consider for instance a connected vacuum diagram built
out of \Math{P} propagators, \Math{L} closed loops of internal
index and \Math{V_3} three--point vertices, \Math{V_4} four--point vertices,
etc\ldots, and let
\beq
V= V_3 + V_4 + V_5 + \ldots .
\eeq[]
Each loop of internal index may be considered as a face of a polyhedron,
and the Euler relation then gives
\beq
V-P+L= 2-2\, g,
\eeq[Euler]
in which $g$ is the genus of the surfaces on which the polyhedron (or Feynman
diagram) is drawn (0 for a sphere, 1 for a torus, etc\ldots). The 
contribution of this diagram is
\beq
N^{V-P+L}\, (-g_3)^{V_3}\, (-g_4)^{V_4}\ldots \, = N^{2-2g}\, \prod_k\, 
(-g_k)^{V_k}.
\eeq[]
As the free energy $F$ is the sum of all connected diagrams we find the
genus expansion
\beq
N^2\, F[g_k]= \sum_{g=0}^\infty\, N^{2-2g}\, F_g [g_k],
\eeq[hermgenusexp] 
where \Math{F_g} is the sum of all diagrams that can be drawn on a surface
of genus $g$. The Hermitian one--matrix model can be solved rather easily in
the ``planar'' limit of infinite matrix size \Math{N\ra \infty} as we shall 
see in section \Sub{hermsolution}. 
Here only the planar diagrams of \Math{F_0} survive, the Hermitian
matrix model thus represents a powerful tool for planar graph counting
\cite{PlanarDiagrams}. 
\Subkapitel{Discretized Surfaces and 2d Quantum Gravity}{discret}
We learned in the previous section that the closed Feynman diagrams 
of the Hermitian matrix model may be viewed as polyhedrons. Moreover
the matrix size $N$ appeared as the parameter controlling the genus of
such a polyhedron. The nontrivial combinatorial problem of how many
inequivalent polyhedrons of genus $g$ with $V_3$ 3--point vertices,
$V_4$ 4--point vertices, etc.\ \ldots exist, may be answered by expanding
\Math{F_g[g_k]} of eq.\ \gl{hermgenusexp} in the \Math{g_k}'s. But as
a polyhedron may be interpreted as a discrete approximation to a smooth
two dimensional surface, this result may be exploited to study an entirely
different problem: The quantization of two dimensional Euclidean
gravity!
\par
In two dimensions the Einstein--Hilbert action forms a topological
invariant, the Euler characteristic of the underlying manifold. So for 
fixed topologies
only the cosmological term will be dynamic. If we consider a specific 
manifold the action of 2d Euclidean gravity is
\beq
S= \m\, \int dx^2\, \sqrt{g}\, - \frac{1}{4\p G}\, \int dx^2\, \sqrt{g}\, R \, 
 = \m \, A\, - \,\frac{2-2g}{G} 
\eeq[Class2dG]
where $G$ denotes the gravitational constant, $\m$ the cosmological
constant and $A$ the area of the surface. In a path integral 
quantization of this theory the integration over the metric may be split
up into separate integrals over topologies and areas
\beq
{\cal Z}_{\mbox{\scriptsize QG}}= \int \frac{{\cal D}
g_{\m\n}}{\mbox{\scriptsize Vol({\it Diff})}}\, e^{-S}
= \sum_{g=0}^\infty\,\int_0^\infty  dA\,\, e^{(2-2g)/G \,- \,\m \, A}\, 
\int_{\S_{g,A}}\, \frac{{\cal D}g_{\m\n}}{\mbox{\scriptsize
Vol({\it Diff})}},
\eeq[ZQG]
where we have formally divided out the volume of the group of diffeomorphisms.
In general the volume of the moduli space \Math{\mbox{Vol}(\S_{g,A})=
\int_{\S_{g,A}}\, {\cal D}g_{\m\n}/\mbox{\scriptsize Vol({\it Diff})}} is
difficult to calculate. The most progress in quantizing 2d
gravity in the continuum has been made via the Liouville approach 
\cite{KPZ-DK}. If we discretize the surface,
on the other hand, it turns out that \gl{ZQG} is much easier to calculate,
even before removing the finite cutoff. We consider in particular a
``random triangulation'', in which the surface is constructed
from equilateral triangles. \footnote{This constitutes the basic difference
to Regge calculus, where the link lengths are the geometrical degrees
of freedom.} Assign the unit area $\e$ to each triangle, so that the
total area of this simplicial manifold built out of $F$ triangles 
is given by \Math{\e\, F}. The path integral over the metric \Math{\int
{\cal D}g_{\m\n}} is now replaced by a sum over triangulations, so that
the discretized partition function \Math{{\cal Z}_{\mbox{\scriptsize QG}}} 
of eq. \gl{ZQG} reads
\beq
{\cal Z}_{\mbox{\scriptsize QG}}^{\mbox{\it\tiny discr.}}
= \sum_{g=0}^\infty\,
\sum_{F=1}^\infty\, e^{(2-2g)/G}\, e^{-\m\, \e F\,}\, {\schnorkel N}_{g,F}
\eeq[ZQGdiscrete]
where \Math{{\schnorkel N}_{g,F}} denotes the total number of inequivalent
triangulations of genus $g$ built out of $F$ triangles.
\par
In the above, triangles do not play an essential role and may be replaced
by any set of polygons leading to a similar expression as in eq. 
\gl{ZQGdiscrete}.
\par
As we saw in the previous section \Math{{\schnorkel N}_{g,F}} is calculated
by the free energy of the Hermitian matrix model. In particular to 
establish contact to
random triangulations consider the matrix model of eq. \gl{hermmmodel} 
with a cubic interaction, i.e.\ \Math{g_2=1/2}, \Math{g_3=\l} and 
\Math{g_k=0}
for \Math{k>3}. The dual to a Feynman diagram of this model (in which each
face, edge and vertex is associated respectively to a dual vertex, edge and 
face) is identical to a random triangulation of some orientable 
Riemann surface (cf.\ Figure 2).
\begin{figure}[t]
\begin{center}
\leavevmode\epsfxsize=8cm
\epsfbox{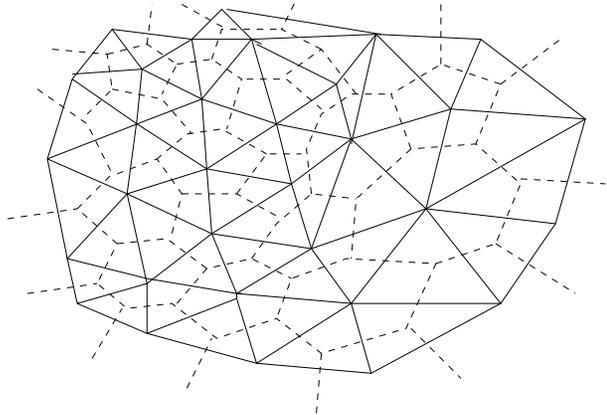}
\caption{A piece of a random triangulation of a surface. Each of the
triangular faces is dual to a three point vertex of a Hermitian matrix
model.}
\end{center}
\end{figure}
Hence there is a one--to--one correspondence between Feynman diagram 
vertices and the faces of the random triangulation. By looking at the
free energy of this cubic matrix model
\beq
N^2\, F[\l ]= \sum_{g=0}^\infty\, \sum_{V=0}^\infty \,
N^{2-2g}\, (-\l)^V\, F_{g,V},
\eeq[Cubicmm]
we see that by formally identifying
\beq
e^{1/G}=N \qquad  \mbox{and}  \qquad  e^{-\m\, \e}= -\l
\eeq[identify]
in eqs.\ \gl{ZQGdiscrete} and \gl{Cubicmm} the free energy of the matrix 
model is actually the partition function  of discrete 2d gravity
\beq
N^2\, F[g_k]= {\cal Z}_{\mbox{\scriptsize QG}}^{
\mbox{\it\tiny discr.}}.
\eeq[discrid]
\par
Needless to say that this argument goes through for general matrix
potentials of eq.\ \gl{hermpotential} as well, then describing
``random polygonulations'' of surfaces. In the continuum limit the
result should not depend of the particular discretization used. However
the different possibilities for generating vertices constitute additional
degrees of freedom that can be realized as the coupling of 2d gravity
to different varieties of matter. We shall see how this works in 
chapter III in the context of the $m$'th multicritical
points \cite{Kaz}.
\Subsubkapitel{Continuum Limit}
In order to discuss the continuum limit we have to anticipate a result
which will be derived in later chapters. The genus $g$ contributions
to the free energy of the matrix model are analytic functions of the
couplings \Math{g_k} around the origin. Singular behaviour arises at
a critical surface in the space of couplings. Specializing again to the
cubic model of eq. \gl{Cubicmm} this means that by moving the constant
$\l$ to its critical value $\l_c$ the free energy becomes singular. In the
vicinity of the critical point one finds the scaling behaviour
\beq
F_g[\l ] \sim (\l-\l_c )^{(2-\g_0)\, (1-g)} ,
\eeq[Fgscaling]
in the case of the cubic potential the critical exponent $\g_0$ takes the 
value $\g_0=-1/2$. We can extract
the continuum limit of the random triangulation by tuning \Math{\l \ra \l_c}.
This is because the average value of the area of the surface is given
by
\beq
\langle A\rangle = \e\, \langle V \rangle= \e \,\l\, \frac{\del}{\del\l}\,
\ln F_g[\l] \, \sim \, \frac{\e}{\l -\l_c}
\eeq[Area]
(recall that the area is proportional to the number of vertices $V$, appearing
as the power of $\l$ associated to each graph which has to be averaged 
with respect to \Math{F_g[\l ]}, as the free energy of the matrix models
represents the partition function of 2d gravity, cf.\ eq.\ \gl{discrid}). 
As \Math{\l \ra \l_c}, we
see that \Math{A\ra\infty} so that we may rescale the area of an individual
triangle $\e$ to zero, thus giving a continuum surface with finite area.
Intuitively this is clear, by tuning the coupling constant to the point
where the perturbation series diverges the sum becomes dominated by
diagrams with infinite number of vertices, which is precisely what we
need for the continuum limit.
\par
The above reasoning constitutes by no way a proof that the matrix model
in the continuum limit coincides with continuum 2d--gravity. However,
one is able to compare properties of the partition function and
correlation functions calculated by matrix models with results
directly computable in the continuum formulation of Liouville theory. This
gives implicit confirmation that the matrix model approach is sensible.
\par
One such quantity computable in Liouville theory is the scaling behaviour
of the Volume of $\S_{g,A}$ of eq. \gl{ZQG}
\beq
 \mbox{Vol}(\S_{g,A})  = e^{\m_c\, A}\, A^{(1-g)\, (\g_{\mbox{\tiny str.}}
  -2) -1}\, W_g,
\eeq[LVol]
where $\m_c$ and $W_h$ are undetermined constants. Here the critical
exponent \Math{\g_{\mbox{\tiny str.}}} is called the string susceptibility.
The above formula even holds true if one couples conformal field theories
with central charge $c$ to Euclidean gravity. 
The Liouville theory prediction for the string susceptibility then is
\cite{KPZ-DK}
\beq
\g_{\mbox{\tiny str.}}= \frac{1}{12}\, ( c-1-\sqrt{(c-1)\, (c-25)} \, ).
\eeq[gammastring]
By plugging eq.\ \gl{LVol} into eq.\ \gl{ZQG} and integrating over the 
area $A$ we find \footnote{Simply use the definition of the gamma function
\Math{\G (x)=\int_0^\infty dt\, e^{-t}\, t^{x-1}}.}.
\beq
{\cal Z}_{\mbox{\scriptsize QG}}
= \sum_{g=0}^\infty\, e^{(2-2g)/G }\,  
(\m -\m_c)^{(2-\g_{\mbox{\tiny str.}})
\, (1-g)}\,\,  \G[\, (g-1)\, (2-\g_{\mbox{\tiny str.}})\, ]\, W_g.
\eeq[ZQGres]
Thus we see that the scaling behaviours of the matrix model in eq.\ 
\gl{Fgscaling} and that of quantum gravity in eq.\ \gl{ZQGres} are identical 
with
\Math{\g_o=\g_{\mbox{\tiny str.}}=-1/2}. Comparing this with eq.\ 
\gl{gammastring}, shows that we in fact are describing the 
\Math{c=0} theory, i.e.\ ``pure'' 2d-Euclidean gravity. This result may be
further confirmed by comparing planar correlation functions.
\Subsubkapitel{ Double Scaling Limit }
There is yet another limit we have to discuss. As a matter of fact the
matrix model can only be solved in the \Math{N\ra\infty} limit, as
we shall see in the next chapters. But then due to eq.\ \gl{hermgenusexp}
only the planar contribution to the free energy survives. However, 
the successive contributions \Math{F_g} all diverge at the critical 
value for the couplings \Math{\l=\l_c}. This suggests that if we
take the limits \Math{N\ra\infty} and \Math{\l\ra \l_c} not independently,
but together in a correlated manner, we may compensate the large $N$
high genus suppression with a \Math{\l\ra \l_c} enhancement. This
results in a coherent contribution from all genus surfaces.
\par
We already saw the leading singular behaviour of $F_g$ in
eq.\ \gl{Fgscaling}, i.e.
\beq
F_g[\l]= f_g\, (\l-\l_c)^{(2-\g_0)(1-g)}. 
\eeq[]
Then in terms of 
\beq
\a= N^{-2}\, (\l-\l_c)^{\g_0-2}
\eeq[stringcouplingconst]
the genus expansion \gl{hermgenusexp} may be rewritten as
\beq
N^2\, F[\l ] = \sum_{g=0}^\infty \a^{g-1}\, f_g.
\eeq[DoubleScalingExp]
The double scaling \cite{DSL} limit is thus obtained by taking the limits 
\Math{N\ra\infty}, \Math{\l\ra\l_c} while keeping fixed the ``renormalized''
string coupling constant $\a$ of eq. \gl{stringcouplingconst}.
\par
The above limit may be performed in the case of general potentials
as well. In this case the additional degrees of freedom can be used 
to fine tune the couplings in such a manner as to adjust an alternative 
string susceptibility, i.e.\ at the $m$'th ``multicritical'' point we 
find \Math{\g_0=-1/m}. It may be shown that these theories then
correspond to the \Math{(2,2m-1)} non--unitary, minimal conformal 
field theories
\footnote{These are classified by the pair (p,q) with
\Math{c=1-6(p-q)^2/pq}.} coupled to quantum gravity and having
the central charge \Math{c=1-3\, (3-2m)^2/(2m-1)}.
\Subkapitel{ Virasoro Constraints and the Loop Equation}{VirLoop}
We now derive a set of constraint equations
for the Hermitian matrix model. The generators of these constraints 
obey the Virasoro algebra, alluding at the integrability of the model
as well as a conformal field theory formulation of it. This structure
sets the basis for a supersymmetric generalization of the Hermitian
one matrix model to the supereigenvalue model.
\par
Consider the eigenvalue model \gl{hermeigenvaluemodel} under the 
shift of integration variables \Math{\l_i\mapsto \l_i + \e\,{\l_i}^{n+1}},
with $\e$ infinitesimal and $n\geq -1$ free. The resulting expression 
proportional to $\e$ must equal zero. One finds
\beq
\e\,\Bigl\langle \,N\, \sum_{k\geq0} k\, g_k\, \sum_{i}{\l_i}^{k+n} +
\sum_{k=0}^n (\sum_i {\l_i}^{n-k})\, (\sum_j {\l_j}^k)\, \Bigr \rangle =0 
\qquad n\geq -1 .
\eeq[preVir]
Of course this analysis may also be performed in the matrix formulation
of eq. \gl{hermmmodel}.
\Comment{
To see this look at the following variations
\beqx
\d\, (\prod_i d\l_i)= \sum_j (n+1)\, {\l_j}^n\, \prod_i d\l_i,
\eeqx
and
\beqx
\d\, V(\l_i )= \sum_{k\geq 0} k\, g_k\, {\l_i}^{k+n}.
\eeqx
For the van der Monde determinant we have \Math{\d\,\D(\l)=\D (\l)\,
\sum_{i<j}\frac{\l_i^{n+1}-\l_j^{n+1}}{\l_i-\l_j}}, and thus
\beqx
\d\,\D ^2 (\l)=\D ^2 (\l)\,\sum_{i\neq j}\frac{\l_i^{n+1}-\l_j^{n+1}}
{\l_i-\l_j}=\D ^2 (\l)\,\sum_{k=0}^n (\sum_i \l_i^{n-k})\, (\sum_j \l_j^k)
- \D ^2 (\l)\, (n+1)\,\sum_i\l_i^n.
\eeqx
Putting these three equations together yields eq.\ \gl{preVir}.
}
The Schwinger--Dyson equation \gl{preVir} may be recast in the form
\cite{Vir}
\beq
\VL{n}\, \ZN = 0 \qquad n\geq -1,
\eeq[Virasoroconstraint]
where \Math{\VL{n}} is a differential operator in the coupling constants
\beq
\VL{n} = \sum_{k\geq  0} k\, g_k\, \del_{g_{k+n}} + 
\frac{1}{N^2}\,\sum_{k=0}^n  \del_{g_k}\,\del_{g_{n-k}}.
\eeq[LVir]
In the case \Math{n=-1} the second sum simply vanishes. The operators
\Math{\VL{n}} are generators of a closed subset of the Virasoro algebra
\beq
[\,\VL{m}\, ,\, \VL{n}\, ] = (m-n)\, \VL{m+n}
\eeq
without central extension (remember that \Math{n,m\geq -1}). 
This is the reason why 
\gl{Virasoroconstraint} are called the Virasoro constraints, 
in bosonic string theory physical states obey an analogous set of 
constraints. 
\par
\Subsubkapitel{Conformal Field Theory Formulation}
Given a complete set of constraints on a partition function 
which form a closed algebra, one might now ask the inverse
question: What is the integral representation of the partition function
obeying these constraints? In the case of the Virasoro algebra it is
natural to look for an answer to this question in form  of a correlation
function in a conformal field theory \cite{MMM91}. The methods 
introduced in this
subchapter can be easily generalized to the case of the super--Virasoro
algebra considered in chapter III and lead to the
construction of the supereigenvalue model.
\par
Just as the generators \Math{\VL{n}} of eq.\ \gl{Virasoroconstraint}
the modes $T_n$ of the energy--momentum tensor
of a free scalar field obey the Virasoro algebra. Let us thus
consider the simplest possible conformal field theory, a holomorphic
scalar field
\beq
\f (z)= \wh{q} + \wh{p}\, \ln z + \sum_{k\neq 0} \frac{\wh{J}_{-k}}{k}
\, z^k,
\eeq[phidef]
with the commutation relations
\beq
[\wh{J}_n,\wh{J}_m]= n\, \d_{n+m,0}, \qquad [\,\wh{q},\wh{p}\, ]=1.
\eeq[comrel]
Define the vacuum states 
\beql
\wh{J}_k \, |0\rangle &=& 0 \qquad \langle N| \, \wh{J}_{-k}= 0 \qquad k>0
 \zeile \wh{p}\, |0\rangle &=& 0 ,
\eeql[vacuums]
where \Math{|N\rangle \equiv \exp [ \sqrt{2}\, N \, \wh{q}\,]\, |0\rangle}
and \Math{\widehat{q}} is anti--Hermitian. The energy--momentum tensor is
given by
\beqx
T(z)= \frac{1}{2} \, : [\del \f (z) ]^2 :\, = \sum_{n\in \Z} T_n\, z^{-n-2},
\eeqx
\beq
T_n= \sum_{k>0} \wh{J}_{-k}\, \wh{J}_{k+n} + \frac{1}{2}\, \sum_{a+b=n}
\wh{J}_a\,\wh{J}_b \qquad n\geq 0 ,
\eeq[emtensor]
and we define a Hamiltonian by
\beq
H(g_k)=  \frac{1}{\sqrt{2}}\, \sum_{k>0} g_k\, \wh{J}_k=
\frac{1}{\sqrt{2}}\, \oint_{C_0}\frac{dz}{2\p i}\, V(z)\, \del\f(z),
\eeq[hamiltonian]
where \Math{V(z)=\sum_{k>0}g_k\, z^k}. With these definitions one 
shows that
\beq
\VL{n}\, \langle N| \, \exp [ \, H(g_k)\, ]\ldots = 
\langle N| \, \exp [ \, H(g_k)\, ]\, T_n\,\ldots.
\eeq[leftVL]
So any operator \Math{\schnorkel G} satisfying
\beq
[\, T_n , {\schnorkel G}\, ]=0, \qquad n\geq -1,
\eeq[TG0]
will lead to \footnote{Note that we have extended the definition of
\Math{T_n} to \Math{n=-1} by dropping the second term in eq.\ \gl{emtensor}.}
\beq
\VL{n}\, \langle N| \, \exp [ \, H(g_k)\, ]\, {\schnorkel G}\,| 0\rangle=0,
\eeq[]
as \Math{T_k\, |0\rangle = 0}. Therefore any nonvanishing correlator
of this form is a candidate for the partition function 
\beq
\ZN =
 \langle N| \, \exp [ \, H(g_k)\, ]\, {\schnorkel G}\,| 0\rangle,
\eeq[Confformmm]
obeying the Virasoro constraints.
Finding the operators \Math{\schnorkel G} satisfying eq.\ \gl{TG0} is an
internal problem of conformal field theory. The solution is given by
an arbitrary function of the screening charges \Math{{\cal Q}_\pm}
which are of conformal dimension zero
\beq
{\cal Q}_\pm = \oint_{C}d \w \, :\exp [\, \pm \sqrt{2}\, 
\f (\w)\,]:.
\eeq[Qpm]
Choose \Math{{\schnorkel G}= {{\cal Q}_+^N}} 
in order to get a nonvanishing correlator in eq. \gl{Confformmm} 
\footnote{More generally one
could take \Math{{\cal Q}_+^{N+M}\, {\cal Q}_-^M}, for
more details see \cite{MMM91}.}. We then have
\beq
\ZN= \langle N| \, :\exp [ \, \frac{1}{\sqrt{2}}\,\oint_{C_0}\,
V(z)\,\del\f(z)\,]:\, \prod_{i=1}^N \oint_{C_i} dz_i\,
:\exp [\, \sqrt{2}\, \f(z_i)\,]: \,| 0\rangle .
\eeq[ZN1]
Using the operator product expansion 
\Math{\f(z)\,\f(\tilde{z}) \sim \ln (z-\tilde{z})} this
is evaluated to
\beq
\ZN=\prod_{i=1}^N \oint_{C_i} dz_i\, \exp[\sum_i V(z_i)\, ]\, \prod_{i<j}
(z_i -z_j)^2.
\eeq[ZN2]
After deforming the contour integrals to integrals on the real line we thus
recover the Hermitian matrix model in the eigenvalue representation
of eq. \gl{hermeigenvaluemodel}.
\Comment{
To obtain eq.\ \gl{ZN2} from eq.\ \gl{ZN1} use \Math{:e^A:\, :e^B:=
e^{\AB}\, :e^{A+B}:\, }, therefore
\beql
\ZN &=&\Bigl( \prod_{i=1}^N \oint_{C_i} dz_i\Bigr )\,
\langle N| \, :\exp [ \, \frac{1}{\sqrt{2}}\,\oint_{C_0}
V(z)\,\del\f(z)\, + \sqrt{2}\, \f(z_1)\,]:\, \prod_{i=2}^N
:\exp [\, \sqrt{2}\, \f (z_i)\,]: \,| 0\rangle \zeile
&&\qquad \cdot\: e^{V(z_1)} \zeile
&=& \prod_{i=1}^N \oint_{C_i} dz_i\,
\langle N| \, :\exp [ \, \frac{1}{\sqrt{2}}\,\oint_{C_0}
V(z)\,\del\f(z)\, + \sqrt{2} \{ \f(z_1)+\f(z_2)\} \,]: \prod_{i=3}^N
:\exp [\, \sqrt{2}\, \f (z_i)\,]: \,| 0\rangle \zeile
&&\qquad \cdot\: e^{V(z_1)+V(z_2)}\, (z_1-z_2)^2 \zeile
&\vdots& \zeile
&=&\prod_{i=1}^N \oint_{C_i} dz_i\,
\langle N| \, :\exp [ \, \frac{1}{\sqrt{2}}\,\oint_{C_0}\,
V(z)\,\del\f(z)\, + \sqrt{2}\,\sum_i \f(z_i)\,]:\,| 0\rangle\zeile
&&\qquad \cdot \:
 \exp[\,{\sum_i V(z_i)}\,]\, \prod_{i<j} (z_i-z_j)^2 \zeile
&=& \prod_{i=1}^N \oint_{C_i} dz_i\,
\langle N| \, :\exp [ \, \sqrt{2}\, N\, \wh{q}\,\, ]:\,| 0\rangle\,
 \exp[\,{\sum_i V(z_i)\,]}\, \prod_{i<j} (z_i-z_j)^2 ,\nonumber
\eeql
and as \Math{\langle N|\, :\exp [ \, \sqrt{2}\, N\, \wh{q}\,\, ]:\,| 0\rangle 
=1} we have proven eq.\ \gl{ZN2}.
}
\Subsubkapitel{Loop Equations}
An immediate consequence of the Virasoro constraints is the loop
equation, an integral equation for the one--loop correlator \Math{W(p)}
of eq.\ \gl{herm1loopcorrelator}. Multiply 
eq.\ \gl{preVir} for $n$ with \Math{1/p^{n+2}} and sum up the obtained
equations. The resulting geometric series \footnote{The subtlety of
$p$ larger or smaller than $\l_i$ may be removed by directly considering
a shift \Math{\l_i\mapsto \l_i + \frac{\e}{p-\l_i}}.} may be
performed to give
\beq
\e\,\Bigl\langle \,N\, \sum_{i}\frac{V^\prime(\l_i )}{p-\l_i}\,\, -\,\,
\Bigr [ \, \sum_i \frac{1}{p-\l_i}\,\Bigr ]^2 \, \Bigr \rangle =0 .
\eeq[Loop1]
Note that generally \Math{\langle\sum_i {\cal O}(\l_i)^2\,
\rangle_{\mbox{\footnotesize conn.}}
=\langle\,\sum_i {\cal O}(\l_i)^2\,\rangle - 
\langle\,\sum_i {\cal O}(\l_i)\,\rangle ^2} and hence
\beq
\Bigr \langle\, \Bigl [\sum_i \frac{1}{p-\l_i}\,\Bigr ]^2 \, \Bigr \rangle =
W(p,p) + {N^2}\, W(p)^2,
\eeq[Loop2]
with the definition of the loop--correlators of eq.\ 
\gl{hermnloopcorrelator2}. In order to rewrite the first term of eq.\ 
\gl{Loop1} in an integral form introduce the eigenvalue density
\beq
\r (\l) = \frac{1}{N}\, \sum_i \langle \,\d (\l -\l_i)\,\rangle.
\eeq[evdensity]
The eigenvalue density has the remarkable property that 
it vanishes outside a support of \Math{[y,x]} on the real line in the limit of
infinite matrix size, \Math{N\ra \infty}. This can be seen by studying the  
saddlepoint evaluation of the eigenvalue model \gl{hermeigenvaluemodel}
in this limit \cite{PlanarDiagrams}. It is precisely this limit which we
consider in the following, keeping all subleading terms in
\Math{1/N^2} of the genus expansion. In this sense we will be able to solve
the matrix model perturbatively by expanding around the point 
\Math{N=\infty}. Each term of this expansion will be some function of
the couplings \Math{g_k}, which can be determined exactly. So the Hermitian
matrix model will turn out to be solvable nonperturbatively in couplings
but perturbatively in \Math{1/N^2}.
\par
By using the eigenvalue density of eq.\ \gl{evdensity} the first 
term of eq.\ \gl{Loop1} becomes
\beql
\langle\, N\, \sum_i  \frac{V^\prime(\l_i)}{p-\l_i}\,  \rangle&=&
N^2\,\int d\l \, \r (\l)\,\frac{V^\prime(\l)}{p-\l} \zeile
&=& N^2\,\int d\l \, \r (\l)\, \oint_C \frac{d\w}{2\p i}\, \frac{1}{\w -\l}
\, \frac{V^\prime(\w)}{p-\w }.
\eeql[Loop3]
Here the curve $C$ of the contour integral runs around the cut \Math{[x,y]}.
Moreover $p$ lies outside the curve. The integral over $\l$ may now be
performed in eq.\  \gl{Loop3} yielding the one--loop correlator \Math{W(\w )}.
Combining this with the result of eq.\ \gl{Loop2} finally leads us to the 
loop equation of the Hermitian one--matrix model \cite{Mig83,loop,Mak}
\beq
\cint{\w}\, \frac{V^\prime(\w)}{p-\w}\, W(\w) - W(p)^2 = \frac{1}{N^2}\,
W(p,p).
\eeq[LoopEquation]
This integral equation basically captures the whole Hermitian matrix 
model. Due to eq.\ \gl{hermnloopcorrelator4} we have \Math{W(p,p)=
\d/\d V(p)\, W(p)}. So the loop equation \gl{LoopEquation} is a closed
equation for the one--loop correlator \Math{W(p)}, determining
this quantity. As explained in section \Sub{model} knowing \Math{W(p)} 
then means knowing all correlators of the model. 
\Subkapitel{The Solution}{hermsolution}
There are various methods to solve the Hermitian one matrix model. Two
of the most prominent ones are the saddlepoint evaluation 
\cite{PlanarDiagrams} and the method of orthogonal polynomials 
\cite{OrthPol}. The most effective solution, however, 
is based on an iterative
procedure to solve the loop equation \gl{LoopEquation} genus by
genus \cite{Amb93}. This method is the one which has found a
generalization in the supereigenvalue model, so let us briefly review
its basic concepts. 
\par 
The loop equation \gl{LoopEquation} may be solved by making use of the
genus expansion of \Math{W(p)}
\beq
W(p)= \sum_{g=0}^\infty \frac{1}{N^{2g}}\, W_g(p).
\eeq[Wgdef]
Plugging this expansion into the loop equation \gl{LoopEquation} and
comparing terms of common order in \Math{1/N^2} leads to a 
coupled hierarchy of equations for the \Math{W_g(p)}.
To leading order in \Math{1/N^2} one simply has
\beq
\cint{\w}\, \frac{V^\prime(\w)}{p-\w}\, W_0(\w) =  W_0(p)^2.
\eeq[LoopEquation0]
From the definition of \Math{W(p)} in eq.\ \gl{herm1loopcorrelator} we
know that asymptotically \Math{W_0(p)=1/p + {\cal O}(p^{-2})}  for 
\Math{p\ra\infty}.
If one additionally assumes that the singularities of \Math{W_0(p)}
consist of only one cut on the real axis \footnote{This follows from
the relation of \Math{W_0(p)} to the eigenvalue density of eq. \gl{evdensity}:
\Math{\r(z)= \lim_{\e\ra 0}\frac{1}{2\p i}\,[ W_0(z-i\e)
- W_0(z+i\e )\, ]}.} one finds \cite{Mig83}
\beq 
W_0(p)= \frac{1}{2}\, \cint{\w} \frac{V^\prime (\w)}{p-\w}\,
\Bigl [ \frac{(p-x)\, (p-y)}{(\w-x)\, (\w-y)}\Bigr ] ^{1/2} \,\, ,
\eeq[W0]
where the endpoints of the eigenvalue distribution $x$ and $y$ are
determined through the matrix potential in the following way:
\beq
\cint{\w} \frac{V^\prime (\w)}{\sqrt{(\w -x)\, (\w-y)}}=0,
\eeq[BoundCon0]
\beq
\cint{\w} \frac{\w\, V^\prime (\w)}{\sqrt{(\w -x)\, (\w-y)}}=2,
\eeq[BoundCon2]
which are a direct consequence of \Math{W(p)=1/p + {\cal O}(p^{-2})}.
\Comment{
Let us verify the planar solution \gl{W0}:
\beqx
W_0(p)^2=\frac{1}{4}\,\ccint{C_1}{\w}\, \ccint{C_2}{z}\, 
\frac{V^\prime (\w)
\, V^\prime (z)}{(p-\w)\, (p-z)}\, \frac{ (p-x)\, (p-y)}{[\, (\w -x)\, (\w-y)\,
(z-x)\, (z-y)\, ]^{1/2}}.
\eeqx
Now rewrite \Math{(p-x)=(p-\w) + (\w-x)} in the numerator. By eq.\ 
\gl{BoundCon0} only
the \Math{(\w-x)} terms survives. Doing the same for \Math{(p-y)}
gives
\beq
W_0(p)^2=\frac{1}{4}\,\ccint{C_1}{\w}\, \ccint{C_2}{z}\, 
\frac{V^\prime (\w)
\, V^\prime (z)}{(p-\w)\, (p-z)}\, \frac{ (\w-x)\, (z-y)}{[\, (\w -x)\, (\w-y)\,
(z-x)\, (z-y)\, ]^{1/2}}.
\eeq[W0calc]
Using
\beqx
\frac{1}{p-\w}\,\frac{1}{p-z}=\frac{1}{\w-z}\, ( \frac{1}{p-\w}-
\frac{1}{p-z })
\eeqx
in eq.\ \gl{W0calc} and renaming \Math{\w \leftrightarrow z} in the
resulting second term one finds
\beqx
W_0(p)^2=\frac{1}{4}\,\ccint{C_1}{\w}\, \ccint{C_2}{z}\, 
\frac{V^\prime (\w)
\, V^\prime (z)}{(p-\w)\, (\w-z)}\, \frac{ (\w-x)\, (z-y)+(\w-y)\, (z-x)}{[\, 
(\w -x)\, (\w-y)\,(z-x)\, (z-y)\, ]^{1/2}},
\eeqx
(The process of pulling the curve $C_2$ over $C_1$ gives no contribution).
Now we do the same trick again and write \Math{(z-y)=(z-\w)+(\w-y)}
as well as  \Math{(z-x)=(z-\w)+(\w-x)} to get
\beqx
W_0(p)^2=\frac{1}{2}\,\ccint{C_1}{\w}\, \ccint{C_2}{z}\, 
\frac{V^\prime (\w)
\, V^\prime (z)}{(p-\w)\, (\w-z)}\, \Bigl [
\frac{ (\w-x)\, (\w-y)}{(z-x)\, (z-y)} \Bigr ]^{1/2}= 
\ccint{C_1}{\w}\, \frac{V^\prime (\w)}{p-\w}\, W(\w),
\eeqx
which proves eq.\ \gl{LoopEquation0}.
}
An interesting consequence of the planar solution \gl{W0} is the 
rather universal form of the
planar two--loop correlator \Math{W_0(p,p)}
\beq
W_0(p,p)= \frac{(x-y)^2}{16\, (p-x)^2\, (p-y)^2},
\eeq[W0PP]
depending on the matrix potential only through the endpoints of the
eigenvalue distribution.
\par
Let us now turn to the higher genus contributions of \Math{W(p)}.
By plugging the genus expansion \gl{Wgdef} into the loop equation 
\gl{LoopEquation} it appears that the \Math{W_g(p)} for \Math{g\geq 1}
 obey the equation \cite{Amb93}
\beq
{\schnorkel\widehat{V^\prime}}\,\circ\, W_g(p) =
\sum_{g^\prime =1}^{g-1} W_{g^\prime}(p)\, W_{g-g^\prime}(p) +
\dV{p}\, W_{g-1}(p),
\eeq[Wgeq]
where we have introduced the linear operator 
\Math{\schnorkel\widehat{V^\prime}} by
\beq
{\schnorkel\widehat{V^\prime}}\, \circ\, f(p) = \cint{\w} \frac{V^\prime 
(\w)}{p-\w}\, f(\w) - 2\, W_0(p)\, f(p).
\eeq
In eq.\ \gl{Wgeq} \Math{W_g(p)} is expressed entirely in terms of the
\Math{W_{g^\prime}}, \Math{g^\prime < g}. This makes it possible to
develop an iterative procedure to solve for \Math{W_g(p)} as done by
Ambj\o rn, Chekhov, Kristjansen and Makeenko \cite{Amb93}. One
basically has to find a way to invert the operator 
\Math{{\schnorkel\widehat{V^\prime}}} 
of the left hand side of eq.\ \gl{Wgeq}. It turns out that this is effectively
done by expanding the \Math{W_g(p)} in a set of basis functions of
\Math{\schnorkel\widehat{V^\prime}} augmented by a change of variables
from coupling constants to moments of the potential. As we shall encounter
precisely the same problem in our discussion of the iterative solution
of the superloop equations for the supereigenvalue model in 
chapter III, let us postpone the analysis of this point.

%% file: c2mm
%
%
%
To construct supersymmetric versions of matrix models it is necessary to
augment the one matrix model of chapter I by fermionic degrees of
freedom. There are various ways to do this. On the level of matrices
one could think of replacing the Hermitian matrices $\Mat$ by supermatrices
$\schnorkel M$ of type \Math{(N|M)}
\beq
{\schnorkel M}_{AB} = \left ( \matrix{ X_{ij} & \Y_{i\a} \cr
  \bar{\Y}_{\a i} & Y_{\a\b} } \right ) \qquad \matrix{ i,j=1,\ldots,N \cr
\a ,\b = N+1, \ldots , N+M ,}
\eeq[Supermatrix]
where the $X_{ij}$ and $Y_{\a\b}$ are Grassmann even and 
$\Y_{i\a}$ and $\bar{\Y}_{\a i}$ are Grassmann odd quantities.
A supermatrix model is then naturally defined by the
partition function of eq.\ \gl{hermmmodel} where one replaces the trace by
a supertrace. This model has been studied in refs.\ \cite{AlvYost}, but the
outcome is rather disappointing: The partition function \Math{{\cal Z}_S
(N|M)} of such a supermatrix model of type \Math{(N|M)} turns out to be
proportional to the partition function \Math{{\cal Z}_S(N-M|0)}, which is
nothing but an ordinary matrix model built out of \Math{(N-M)\times(N-M)}
matrices \footnote{ A negative \Math{(N-M)} corresponds to a change of sign 
in the exponent of eq.\ \gl{hermmmodel}.}.
\par
In this chapter we shall address an alternative approach to this problem. 
Guided by the superspace formulation of supersymmetric field theories,
one augments the zero dimensional space of the Hermitian matrix model
by two anticommuting coordinates. The fields living in this space will
be matrix valued superfields, whose component expansion consists of two
bosonic and two fermionic matrices. Thinking of the interpretation of
anticommuting coordinates as negative dimensions, this model should be
considered as a \Math{d=-2} matrix model. The planar triangulations of
closed surfaces in \Math{d=-2} dimensions was first studied on combinatorial
grounds by Kazakov, Kostov and Mehta \cite{KazKosMig}. The matrix model
approach to this problem was introduced by David \cite{Dav85} and
further studied in refs.\ \cite{Kos87,KlebWilk}. And in fact in its 
scaling limit the matrix model does describe the coupling of \Math{c=-2}
matter to 2d gravity.
\par
In our analysis of the model we shall stress its supersymmetric structure
and discuss the emergence of a Nicolai--map to a Gaussian \Math{d=0} Hermitian
matrix model. The diagrammatic interpretation of the model is investigated,
leading us to its equivalence to random surfaces decorated by fermion loops.
The one--, two-- and three--point functions are computed in the case of  
a quartic potential, which has
not appeared in literature so far. In the critical regime of this solution 
we recover the scaling exponents of \Math{c=-2} Liouville theory. Finally
as a nice application of the solution we present the counting of the first few
supersymmetric graphs, similar in fashion to the counting of bosonic
planar diagrams in ref. \cite{PlanarDiagrams}.
\Subkapitel{The Model}{c2model}
In chapter I we studied Hermitian matrices living in zero 
dimensional space. As a method of supersymmetrization of matrix
models, let us consider matrices living in superspace. In order to maintain
exact solvability we introduce two fermionic 
coordinates \Math{\qb} and \Math{\q}, but keep the zero dimensional
bosonic space. A matrix valued superfield \Math{\SF} will then have the
following component expansion
\beq
\SF= \Mvf + \MYb\, \q +  \bar{\q}\, \MY + \q\bar{\q}\,
\MF,
\eeq[superfield]
where $\Mvf$ and $\MF$ are Hermitian \Math{N\times N} matrices and
the $\MY$ and $\MYb$ are \Math{N\times N} matrices with 
complex Grassmann odd entries. We have \Math{\MY^\dagger=\MYb} and
\Math{\q^\ast=\qb}. Note that $\SF$ is Hermitian.
\par
The matrix model built out of $\SF$ is given by the partition 
function
\beq
{\cal Z}_{\SF} [\, g_k \, ]= \int_{N\times N} {\cal D}\SF \, 
\exp \Bigl [ \, -N\, \tr S[\,\SF \, ]\, \Bigr ].
\eeq[c2model]
As measure we take \Math{{\cal D}\SF= {\cal D}\Mvf\,
{\cal D}\MF\, {\cal D}\MYb\,{\cal D}\MY}, where the
measure of the Hermitian matrices $\Mvf$ and $\MF$ are as in eq.\ 
\gl{hermeasure}  and for the fermionic matrices we take
\beq
{\cal D}\MYb\,{\cal D}\MY= \prod_{\a,\b=1}^N\,
d\MYb_{\a\b}\, d\MY_{\b\a}.
\eeq[grassmeasure]
The action of the $\SF$--matrix model of eq.\ \gl{c2model} reads
\beq
S(\SF)= \int d\q d\bar{\q}\, \Bigl \{ -{\schnorkel D}\SF\, \bar{{\schnorkel D}}
\SF \, +\, \sum_{k=0}^\infty g_k\, \SF^k\, \Bigr \},
\eeq[c2action]
with the ``superspace'' derivatives \Math{{\schnorkel D}=\del /\del \q}
and \Math{\bar{{\schnorkel D}}=\del/\del \bar{\q}}. Bearing in mind the
interpretation of anticommuting coordinates as negative dimensions it 
should be clear why eq.\ \gl{c2model} is considered as a \Math{d=-2} matrix
model. 
\par
The obvious extension
of this model to one bosonic dimension, i.e.\ considering component
fields $\Mvf(x)$, $\MYb(x)$, $\MY(x)$ and $\MF(x)$ depending on one
variable $x$, is known as the Marinari--Parisi superstring \cite{MarPar}
\footnote{The question whether the continuum limit of this
model describes a theory with target space supersymmetry is still
unsettled \cite{MarPar2}.} .
\par
After performing the integrals over $\q$ and $\bar{\q}$ in the action 
\gl{c2action} and using cyclicity under the trace one finds
\beq
\tr S[ \, \SF\, ]= - \tr \MF^2 + \tr V^\prime (\Mvf )\, \MF +
\sum_{k=0}^\infty k\, g_k \sum_{a+b=k-2} \tr \Mvf^a\, \MYb\, 
\Mvf^b\, \MY,
\eeq[glg3]
using the general matrix potential
\beq
V(\Mvf)= \sum_{g=0}^\infty g_k\, \Mvf^k.
\eeq[c2potential]
As the auxiliary matrix $\MF$ enters in the action \Math{\tr S [\, \SF\,]}
of eq. \gl{glg3} only quadratically, a shift in integration variables 
\Math{\MF_{\a\b}\mapsto \MF_{\a\b}+ 1/2\, V^\prime(\Mvf)_{\a\b}} lets the
integral over $\MF$ decouple. The Jacobian associated with this shift is
unity, hence the integral over the auxiliary matrix \Math{\MF} can be
performed yielding an $N$ dependent constant $c_N$. We are
thus led to the effective action
\beq
\tr S_{\mbox{\footnotesize eff}}[\, \SF\, ]=  \frac{1}{4}\, 
\tr V^\prime (\Mvf )^2\, 
+\, \sum_{k=0}^\infty k\, g_k \sum_{a+b=n-2} \tr \Mvf^a\, \MYb\, 
\Mvf^b\, \MY ,
\eeq[Seff]
and the partition function now has the form
\beq
{\cal Z}_{\SF} [\, g_k \, ]= c_N\, \int_{N\times N} {\cal D}\Mvf \, 
{\cal D}\MYb\,
{\cal D}\MY\, \exp \Bigl [ \, -N\, \tr
S_{\mbox{\footnotesize eff}}[\, \SF \, ]\, \Bigr ].
\eeq[Zc2eff]
The next obvious thing to do is to integrate out the fermionic matrices.
Let us, however, first investigate the symmetries of this effective
action.
\Subsubkapitel{Supersymmetry Transformations}
After integrating out the auxiliary matrix $\MF$ we find that the effective 
action of eq. \gl{Seff} is invariant under the following set of (global)
supersymmetry transformations
\beql
\d\Mvf_{\a\b}&=& \bar{\e}\, \MY_{\a\b} + \MYb_{\a\b}\, \e \zeile
\d\MY_{\a\b}&=& -\frac{1}{2}\, \e\, V^\prime(\Mvf)_{\a\b}\zeile
\d\MYb_{\a\b}&=& -\frac{1}{2}\, \bar{\e}\, V^\prime(\Mvf)_{\a\b},
\eeql[SusyTrafo]
where $\e$ and $\bar{\e}$ are anti--commuting parameters. 
\Comment{
This is easily verified. Consider the variation of eq. \gl{Seff} under
the transformations \gl{SusyTrafo}, then
\beqx
\d \Bigl [ \, \frac{1}{4}\,\tr [\, V^\prime(\Mvf )\, ] ^2\, \Bigr ] =
\frac{1}{2}\,\tr V^\prime (\Mvf)\, V^{\prime\prime}(\Mvf)\, \Bigl (\, 
\bar{\e}\, \MY + \MYb\, \e\, \Bigr )
\eeqx
cancels with
\beqx
\d \Bigl [ \, \sum_k g_k\, k\, \sum_{a+b=n-2} \tr \Mvf^a\,\MYb\, \Mvf^b\,
\MY\, \Bigr ]= -\frac{1}{2}\,\tr \bar{\e}\, \MY\, V^\prime(\Mvf)\, 
V^{\prime\prime}(\Mvf) - \frac{1}{2}\, \tr V^\prime(\Mvf)\, 
V^{\prime\prime}(\Mvf)\, \MYb\, \e + \{ \, \d\Mvf\mbox{`s}\,\}.
\eeqx
except for the variations \Math{\{ \, \d\Mvf\mbox{`s}\,\}}, which cancel 
by themselves. See this by looking at the  part
of the transformations \gl{SusyTrafo} proportional to $\bar{\e}$
\beql
\{ \, \d\Mvf\mbox{`s}\,\}_{\bar{\e}}
&=& \sum_{a+b=n-2} \tr \d_{\bar{\e}}(\Mvf^a)\,\MYb\,\Mvf^b\,
\MY+ \tr \Mvf^a\,\MYb\,\d_{\bar{\e}} (\Mvf^b)\, \MY\zeile
&=& \sum_{a+b+c=n-3} \tr \Mvf^a\,\bar{\e}\,\MY\,\Mvf^b\,\MYb\,\Mvf^c\,
\MY + \tr \Mvf^a\,\MYb\,\Mvf^b\,\bar{\e}\,\MY\,\Mvf^c\,\MY =0,
\nonumber
\eeql[]
the calculation for the part proportional to $\e$  works analogously.}
If one goes back to the form of the action including the auxiliary matrix 
$\MF$ of eq. \gl{glg3}, the supersymmetry transformations become
linear in the fields:
\beql
\d\Mvf_{\a\b}&=& \bar{\e}\, \MY_{\a\b} + \MYb_{\a\b}\, \e \zeile
\d\MY_{\a\b}&=& -\e\, \MF_{\a\b}\zeile
\d\MYb_{\a\b}&=& -\bar{\e}\,\MF_{\a\b}\zeile
\d\MF_{\a\b} &=& 0.
\eeql[SusyLinearTrafo]
\par
Note that the measure 
\Math{{\cal D}\Mvf\,\, {\cal D}\MYb\,{\cal D}\MY}
is invariant under the supersymmetry transformation \gl{SusyTrafo}.
This immediately follows from the form of the Jacobian of the 
transformations \gl{SusyTrafo}
which is zero in linear order of $\e$ and $\bar{\e}$.
\Subsubkapitel{Integrating out the Fermions}
Just as in the Hermitian matrix model the action 
\Math{\tr S_{\mbox{\footnotesize eff}}} of eq.\ \gl{Seff} 
is invariant under the simultaneous
$U(N)$ transformations \Math{\Mvf\mapsto U^\dagger \Mvf U}, 
\Math{\MY\mapsto U^\dagger \MY U} and 
\Math{\MYb\mapsto U^\dagger \MYb U}. This may be employed to diagonalize
the Hermitian matrix \Math{\Mvf= U^\dagger \, \Mvf_{\fett D}\, U},
with \Math{\Mvf_{\fett D}= \mbox{diag}\, (\vf_1,\ldots ,\vf_N)}. We
again pick up the van der Monde determinant
\Math{\prod_{\a<\b} (\vf_\a -\vf_\b )^2} as the Jacobian of this 
transformation. As the
transformation matrix $U$ does not depend on the fermionic 
matrices $\MY$ and
$\MYb$, the Jacobian of their transformation is unity. We thus obtain
the semi--eigenvalue picture of the $\SF$--model 
\beqy
{\cal Z}_{\SF} [\, g_k \, ]= 
\eeqy
\beq
\tilde{c}_N\, \int 
\prod_{\a=1}^\infty  d\vf_\a \, \prod_{\a,\b=1}^N\,
d\MYb_{\a\b}\, d\MY_{\b\a}\, \prod_{\a<\b}(\vf_\a -\vf_\b )^2\,
\exp \Bigl [ \, -N\, 
S_{\mbox{\footnotesize eff}}[\, g_k\, ]\, \Bigr ],
\eeq[diagc2mm]
where the constant \Math{\tilde{c}_N} now contains the integral over
$U(N)$ and the constant $c_N$ of eq.\ \gl{Zc2eff}. 
The partially diagonalized effective action reads

\beq
S_{\mbox{\footnotesize eff}}[\, g_k\, ] = \frac{1}{4}\, \sum_\a V^\prime
(\vf_\a)^2\, +\sum_k k\, g_k \sum_{a+b=k-2} \sum_{\a\b}\,
\vf_a^a\, \MYb_{\a\b}\, \vf_\b^b\, \MY_{\b\a}.
\eeq[Seffdiag]
\par
In this form the integration over the fermions may be performed rather
easily. To do this introduce the abbreviation
\beq
C_{\a\b}= N\, \sum_k k\, g_k \sum_{a+b=k-2} \,
\vf_a^a\, \vf_\b^b = \cases { N\,\frac{V^\prime(\vf_\a) - V^\prime
(\vf_\b)}{\vf_\a-\vf_\b} & if $\a\neq\b$ \cr & 
\cr N\,V^{\prime\prime}(\vf_\a)&
if $\a=\b$} .
\eeq[shorthand]
We then have
\beqy
\int \prod_{\a,\b=1}^N\, d\MYb_{\a\b}\, d\MY_{\a\b}\, \exp \Bigl [
-\sum_{\a,\b}\, C_{\a\b}\, \MYb_{\a\b}\, \MY_{\b\a}\, \Bigr ] 
\eeqy
\beql
&=& \int \prod_{\a,\b=1}^N\, d\MYb_{\a\b}\, d\MY_{\a\b}\, \prod_{\a,\b}
\Bigl [ \, 1+ 
\, C_{\a\b}\, \MY_{\a\b}\, \MYb_{\b\a}\, \Bigr ] \zeile
&=& \, \prod_{\a,\b} \, \Bigl [ \, C_{\a\b}\, \Bigr ] \zeile
&=& N^{N^2}\, \prod_\a [ \, V^{\prime\prime}(\vf_\a)\, ]\, \prod_{\a<\b}\,
\Bigl [ \frac{V^\prime(\vf_\a) - V^\prime (\vf_b)}{\vf_\a-\vf_\b}\,
\Bigr ]^2.
\eeql[2mmcalc]
This is a rather astonishing result, as the van der Monde determinant
will now drop out of the partition function. Plugging eq.\ \gl{2mmcalc} 
into eq.\ \gl{diagc2mm} yields the purely bosonic partition
function
\beql
{\cal Z}_{\SF} [\, g_k \, ]&=& \hat{c}_N\, 
\prod_{\a=1}^N  \Bigl (\,\int_{-\infty}^\infty 
d\vf_\a \, V^{\prime\prime}(\vf_\a)\,\Bigr )\, 
\prod_{\a<\b}\Bigl [ \, V^\prime (\vf_\a) -V^\prime (\vf_\b )\,\Bigr ]^2
\zeile && \qquad
\exp \Bigl [ \, -\frac{N}{4}\,  \sum_\a V^\prime(\vf_a)^2\, \Bigr ].
\eeql[ZSF6]
So by making the $N$ independent substitutions
\beq
\l_\a = V^\prime(\vf_\a) \quad \Rightarrow \quad d\l_\a = 
V^{\prime\prime}(\vf_\a)\, d\vf_\a \qquad \a=1,\ldots, N,
\eeq[c2mmsubst]
the  partition function \Math{{\cal Z}_{\SF} [\, g_k \, ]} 
takes the Gaussian form
\beq
{\cal Z}_{\SF} [\, g_k \, ] = \hat{c}_N\, \prod_{\a=1}^N \, \Bigl ( \,
 \int_{V^\prime(-\infty)} ^{V^\prime(\infty )} d\l_\a\, \Bigr )\, 
\prod_{\a<\b} ( \, \l_\a -\l_\b\, )^2 \,  e^{\, -\frac{N}{4}\, 
\sum_{\a=1}^N \l_a^2\, },
\eeq[ZSFgauss]
note that this is true for any continuous function $V^\prime(\vf_\a)$.
\par
We conclude that in the case of unbounded potentials \Math{V^\prime
(-\infty)=V^\prime(\infty)} the partition function ${\cal Z}_{\SF}$ 
vanishes. On the other hand for potentials with a lower bound, i.e.\
\Math{V^\prime(-\infty)=-\infty} and \Math{V^\prime(\infty)=\infty},
the $\SF$--matrix model is proportional to a pure Gaussian Hermitian matrix
model
\beq
{\cal Z}_{\SF} [\, g_k \, ] = \breve{c}_N\, \int {\cal D}{\fett \L}\, \exp \Bigl [
\, -\frac{N}{4}\, \tr {\fett \L}^2\, \Bigr ] = 2^N\, \p^{\, N^2},
\eeq[Gaussian]
and is completely independent of the coupling constants $g_k$! 
The result \Math{2^N\, \p^{N^2}} is obtained by collecting all $N$ dependent
constants and making use of the result for the Gaussian partition function
\cite{OrthPol}. 
\par
But this means that the partition function can obviously not have any
critical behaviour.
Moreover any integrated supersymmetric expectation value vanishes
\beq
\Bigl \langle \int d\q \, d\qb\, \tr \SF^n\, \Bigr \rangle =0 ,
\eeq[SFn0]
which is a direct consequence of differentiating \Math{{\cal Z}_{\SF}}
of eq. \gl{c2model} with respect to $g_n$. On the level of the effective
action \gl{Seff} this Ward identity reads
\beq
\frac{1}{2}\, \Bigl \langle\, \tr V^\prime(\Mvf)\, \Mvf^{n-1}\, \Bigr \rangle
= \sum_{a+b=n-2} \Bigl \langle \, \tr \Mvf^a\, \MY\, \Mvf^b\, \MYb\, 
\Bigr \rangle.
\eeq[WardId]
Nevertheless non--supersymmetric correlation functions like 
\Math{\langle\, \tr \f^k\, \rangle} are non--trivial and display
critical behaviour as we shall see in the subsequent sections. 
\Subsubkapitel{Nicolai--Map}
The observed situation is very reminiscent to the
Nicolai--map \cite{NicolMap} of globally supersymmetric field theories: 
Integrating
out the fermions in a supersymmetric field theory yields a determinant
and an effective bosonic action. The vanishing of the vacuum energy
of these theories alludes at their equivalence to a free theory. As a 
matter of fact one can show that there always exists
a map from a free, Gaussian action to the effective bosonic action of
the supersymmetric theory, whose Jacobian is identical to the determinant 
of the fermionic integration. On the level of correlation 
functions this implies that an operator
expectation value of the supersymmetric theory may be calculated
as a generically rather complicated operator expectation value in a simple
free theory.
\par
This scenario directly translates to the $\SF$--matrix
model. The correlator
\Math{\langle\, \tr \Mvf^k\, \rangle} may be evaluated in the Hermitian
matrix model with Gaussian measure \Math{S_0=-N/4\, \tr {\fett \L}^2}
by using the inverse map of eq.\ \gl{c2mmsubst}
\beq
\Bigl \langle \, \tr \Mvf^k\, \Bigr \rangle = \Bdangle{
\tr [\, V^{\prime\, -1}({\fett \L})\,]^k},
\eeq[NicMap]
where \Math{\dangle{\ldots}} denotes the expectation value in the free theory 
\beq
\Bdangle{ {\cal O}({\fett \L})} =
\frac{1}{{\cal Z}_0}\, \int_{N\times N} {\cal D}{\fett \L}\, 
{\cal O}({\fett \L})\, \exp [\,  -\frac{N}{4}\, \tr {\fett \L}^2\, ] .
\eeq[GaussNic]
We shall exploit this relation in sections \Sub{bos1PFunct} and
\Sub{23PF}
for the calculation of bosonic one--, two-- and three--point correlators.
\Subkapitel{Feynman Diagrams}{FeynDiagr}
Let us derive the Feynman rules for the $\SF$-matrix model. This will
lead to a geometrical interpretation in terms of ``fat'' graphs, 
as we saw in the Hermitian matrix model
in chapter I. 
\par
To be specific we consider a quartic potential
of the form
\beq
V(\SF)= \frac{1}{2}\, \SF^2 + \frac{1}{4}\, g\, \SF^4.
\eeq[SF4Pot]
Note that the quartic potential is the simplest non--trivial bounded
potential to consider, i.e.\ satisfying \Math{V^\prime(-\infty)=-\infty}
and \Math{V^\prime(\infty)=\infty} for $g$ positive. To our mind this point
has not found adequate attention in the literature 
\cite{Dav85,Kos87,KlebWilk}
where a cubic potential is assumed and the boundaries of integration in eq.\
\gl{ZSFgauss} are set by hand to \Math{-\infty} and \Math{\infty}.
\par
With this quartic potential  we are led to the effective action for the
matrices $\Mvf$, $\MYb$ and $\MY$ via eq.\ \gl{Seff}
\beql
\tr S_{\mbox{\footnotesize eff.}} &=& \frac{1}{4}\,\tr \Mvf^2 + \frac{1}{2}\,
g\,\tr \Mvf^4 + \frac{1}{4}\, g^2\, \tr \Mvf^6  \zeile
&& \quad + \tr \MYb\,\MY + g\, \Bigl \{ \,\tr \Mvf^2\,\MYb\,\MY +
\tr \Mvf\,\MYb\,\Mvf\,\MY + \tr \MYb\,\Mvf^2\,\MY\, \Bigr \}.
\eeql[Seff4]
\par
\begin{figure}[t]
\begin{center}
\leavevmode\epsfxsize=12cm
\epsfbox{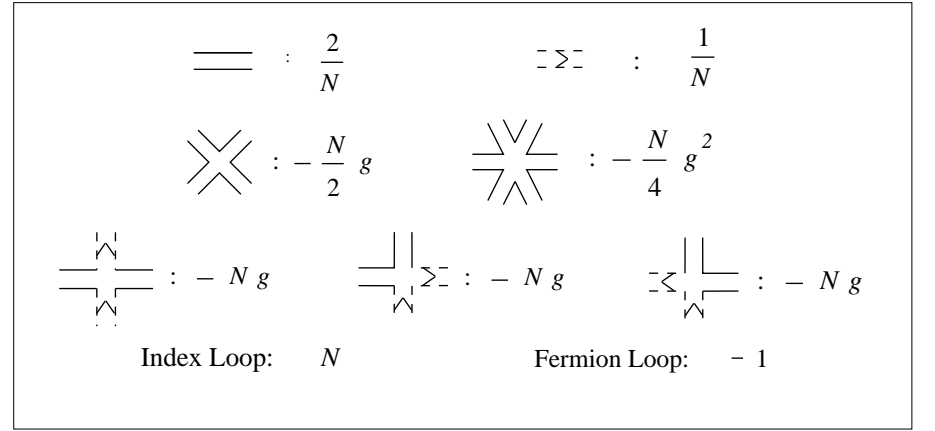}
\caption{The Feynman rules for the quartic $\SF$--matrix model.}
\end{center}
\end{figure}
The bosonic and fermionic propagators read
\beq
\langle\, \Mvf_{\a\b}\,\Mvf_{\g\d}\,\rangle_{g=0} = 
\frac{2}{N}\, \d_{\a\d}\,\d_{\b\g},
\qquad\qquad
\langle\, \MY_{\a\b}\,\MYb_{\g\d}\,\rangle_{g=0}=
 \frac{1}{N}\, \d_{\a\d}\,\d_{\b\g},
\eeq[propagators]
and again they are represented by double
lines each one corresponding to the separate propagation of matrix
indices. Moreover one has to assign a direction of propagation to the
fermions due to a necessary canonical ordering of the matrices 
\Math{\MY} and \Math{\MYb}.
\Comment{
The form of the propagators in eq.\ \gl{propagators} is directly derived by
\beql
\langle\, \Mvf_{\a\b}\,\Mvf_{\g\d}\,\rangle_{g=0} &=&
{\cal Z}_\SF^{-1}\, \int {\cal D}\Mvf\, {\cal D}\MYb\, {\cal D}\MY\,\,
\Bigl(-\frac{2}{N}\Bigr )\,\Mvf_{\a\b}\,\frac{\d}{\d\Mvf_{\d\g}}\, 
\exp\Bigl [\, -\frac{N}{4}\, 
\Mvf_{ij}\,\Mvf_{fi}  -N \, \MYb_{ij}\, \MY_{ji}\, \Bigr ] \zeile
&=& \frac{2}{N}\, \d_{\a\d}\,\d_{\b\g} \nonumber
\eeql
and
\beql
\langle\, \MY_{\a\b}\,\MYb_{\g\d}\,\rangle_{g=0}&=&
{\cal Z}_\SF^{-1}\, \int {\cal D}\Mvf\, {\cal D}\MYb\, {\cal D}\MY\,\,
\Bigl(\frac{1}{N}\Bigr )\,\MY_{\a\b}\,\frac{\d}{\d\MY_{\d\g}}\, 
\exp\Bigl [\, -\frac{N}{4}\, 
\Mvf_{ij}\,\Mvf_{fi}  -N \, \MYb_{ij}\, \MY_{ji}\, \Bigr ] \zeile
&=& \frac{1}{N}\, \d_{\a\d}\,\d_{\b\g} \nonumber
\eeql
via partial integration.}
With the interactions of eq.\ \gl{Seff4} there will be a bosonic
four--point vertex contributing \Math{(-N/2\, g)}, a bosonic 
six--point vertex contributing \Math{(-N/4\, g^2)} and four--point 
Yukawa vertices contributing \Math{(-N\, g)}. Again each loop of
internal index will yield a factor of \Math{N=\d_{\a\a}}. Moreover
there is a factor of \Math{(-1)} for every closed fermionic propagator
loop. To see this consider a fermionic loop in an arbitrary diagram
built out of $n$ Yukawa vertices. We then have to evaluate
the following expectation value, where the dots stand for  
bosonic propagators running out of the Yukawa vertices and all matrix
indices are suppressed
\par
\bigskip
\parbox[l]{2cm}{
\leavevmode\epsfxsize=2cm
\epsfbox{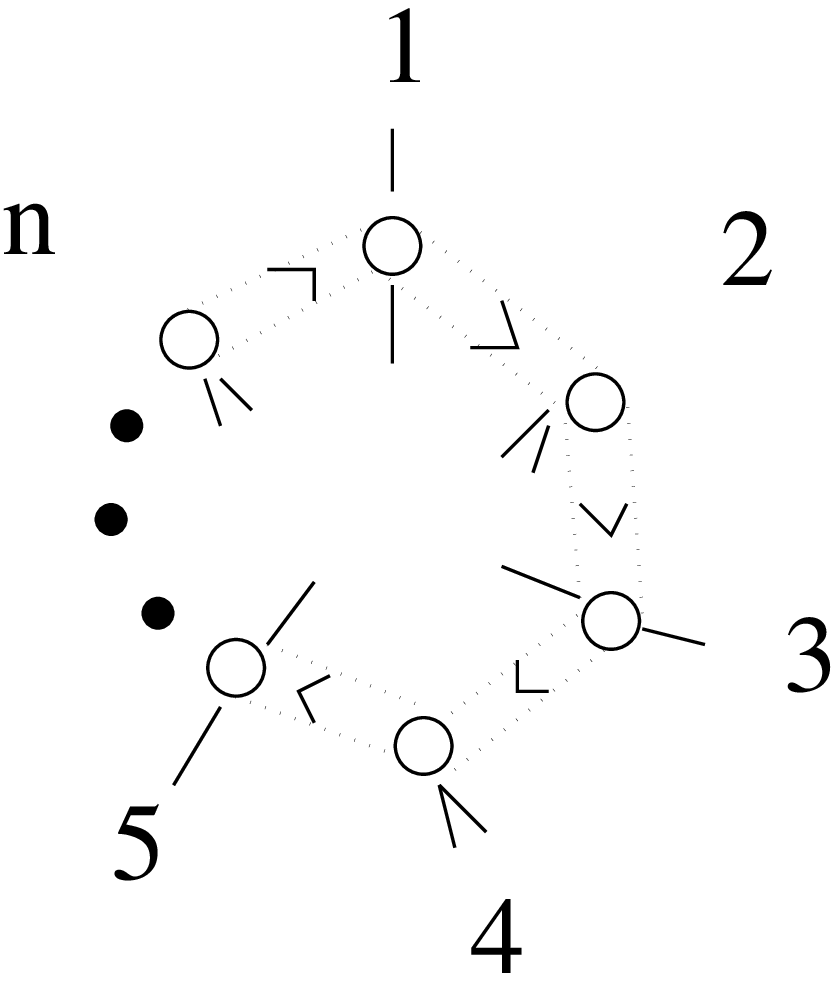} }
\parbox[r]{8cm}{\footnotesize
$$
\Bigl\{ \ldots \MYb_1\ldots \Wick{1}{2}
\ldots \Wick{2}{3}\ldots \Wick{3}{4}\ldots \Wicke{4}\ldots \eWick{n}
\ldots \ldots \MY_n \put(-3,-4){\line(-1,0){100}}
                        \put(-3,-4){\line(0,1){3}}
                        \put(-103,-4){\line(0,1){3}}
\Bigr \}
$$
$$
= (-) \, \Bigl\{ \ldots \Wick{n}{1}\ldots \Wick{1}{2}
\ldots \Wick{2}{3}\ldots \Wicke{3}\ldots \eWick{n}
\ldots \ldots\Bigr \}.
$$}
\bigskip
\newline
One thus gets a factor of 
$(-1)$ for every closed fermionic propagator loop.
\par
Hence a diagram consisting of $n_4$ 4--point and $n_6$ 
6--point bosonic vertices, $n_F$ Yukawa--vertices, $E_B$ boson propagators,
$E_F$ fermion propagators, $L$ index loops and $l$ fermion propagator 
loops contributes the factor
\beq
\mbox{Diag} =
(-)^l\, \Bigl(-\frac{g}{2}\Bigr)^{n_4}\, \Bigl( -\frac{g^2}{4}
\Bigr ) ^{n_6}\, (-g)^{n_F}\,
N^{n_4+n_6+n_F}\, \Bigl (\frac{1}{N}\Bigr )^{E_B+E_F}\, 2^{E_B}\, N^L
\eeq[SusyDiagram]
to the partition function. Using the Euler relation one again has the
overall genus dependent \Math{N^{2-2g}} weight for a graph of genus $g$.
One may envisage the partition function \Math{{\cal Z}_\SF}
as coming from a Hermitian matrix model with a \Math{(4,6)} vertex
interaction, where one draws $l$ loops running over $n_F$ 4--point vertices 
on the diagrams and weights each resulting
 diagram by a factor of \Math{(-1)^l}.
\begin{figure}[t]
\begin{center}
\leavevmode\epsfxsize=8cm
\epsfbox{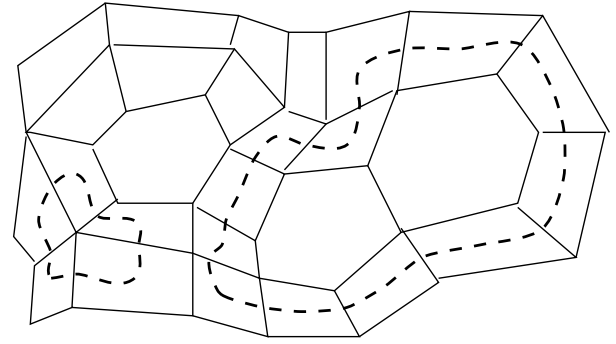}
\caption{A dual graph of the \Math{\SF}--matrix model.}
\end{center}
\end{figure}
(This is true as $n_F$ boson propagators are turned into fermion propagators
yielding a factor of \Math{2^{n_F}}, which is exactly the factor needed
to convert $n_F$ bosonic 4--point vertices into Yukawa--vertices). 
Note that the loops
drawn on the graph are not allowed to intersect or touch. Hence
\beq
{\cal Z}_\SF = \sum_{l=0}^\infty (-)^l\, {\cal Z}_{\mbox{\footnotesize
herm}}^{(l)}(g_2=1/4, g_4=g/2, g_6=g^2/4)
\eeq[ZidSFB]
in the notation of eq.\ \gl{hermpotential}. Interestingly enough the 
loop drawing lets the sum in eq.\ \gl{ZidSFB} trivialize. On the level of
the dual graphs we see that each diagram of \Math{{\cal Z}_\SF}
represents a discretized surface built from squares and hexagons with
loops painted across squares.
\Comment{
As a nice check of the triviality of the partition function \Math{{\cal
Z}_\SF} of eq.\ \gl{Gaussian} let us evaluate the \Math{{\cal O}(g)}
vacuum graphs:
\beqx
\begin{array}{ccccc}
\FGraph{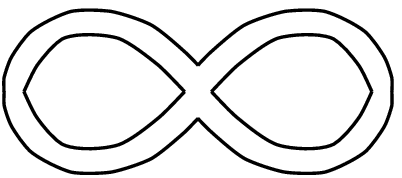}&\quad & : & \qquad &
      [2]\, \Bigl( - \frac{N}{2}\, g\Bigr )\, 
     \Bigl ( \frac{2}{N} \Bigr )^2\, N^3 = -4\, g\, N^2 \\ \\
\FGraph{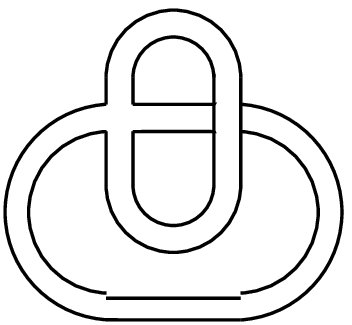}& & : & \qquad &
     [1]\, \Bigl( - \frac{N}{2}\, g\Bigr )\, 
     \Bigl ( \frac{2}{N} \Bigr )^2\, N= -2\,g  \\ \\
\FGraph{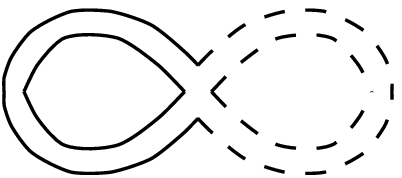}& & : & \qquad & 
     - [2]\, \Bigl( - N\, g\Bigr )\, 
     \Bigl ( \frac{2}{N} \Bigr )\,\Bigl(\frac{1}{N}\Bigr )
     N^3 =  4\, g\, N^2 \\ \\
\FGraph{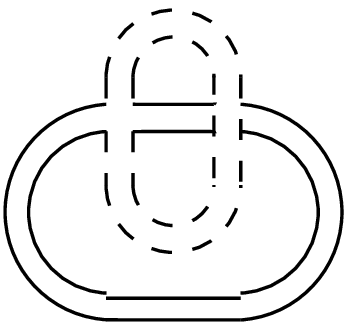}& & : & \qquad &
     - [1]\, \Bigl( - N\, g\Bigr )\, 
     \Bigl ( \frac{2}{N} \Bigr )\,\Bigl(\frac{1}{N}\Bigr )
     N =  2\, g \\
\end{array}
\eeqx
Keep in mind that the six--point vertex is of order \Math{g^2}.
The numbers in \Math{[\ldots]} denote the combinatorial factors. We see
that the sum of all vacuum diagrams to first order $g$ vanishes, as they
should. Note the genus controlling factor of $N^2$ in the above graphs.}
\Subkapitel{Bosonic One--Point Functions}{bos1PFunct}
As mentioned in section
\Sub{c2model} bosonic one--point functions of
the form \Math{\langle\, \tr\Mvf^{k}\,\rangle} might display critical
behaviour despite of the triviality of the partition function 
\Math{{\cal Z}_\SF}. To calculate these quantities one makes use of the
 Nicolai--map of eq.\ \gl{NicMap} to map these correlators onto correlators
 of a Hermitian matrix model with Gaussian potential.
\par
Let us again assume the quartic matrix potential of eqs.\ \gl{SF4Pot}
and \gl{Seff4}. As it is symmetric all odd bosonic one--point correlators
\Math{\langle\, \tr\Mvf^{2n+1}\,\rangle} vanish. For the even correlators
one has to solve the equation
\beq
V^\prime(\vf)= \vf + g\, \vf ^3 = \l 
\eeq[cubic]
for \Math{\vf(\l)=V^{\prime\,-1}(\l)}. For \Math{g>0} there is only one real
solution to eq.\ \gl{cubic}, whereas for \Math{g<0} we have three. Picking
the branch selected by smooth continuation of \Math{g\ra -g} we find
\beqy
V^{\prime\,-1}(\l)= 
\eeqy
\beq
\frac{1}{\sqrt{|g|}}\, \Bigl [ \,\sqrt[3]{ \,
\frac{s\,\sqrt{|g|}}
{2}\,\l + \sqrt{ \frac{|g|}{4}\,\l^2 + \frac{s}{27}}\,} + 
\sqrt[3]{\, \frac{s\,\sqrt{|g|}}{2}\,\l - 
\sqrt{ \frac{|g|}{4}\,\l^2 + \frac{s}{27}}\,}\, \Bigr ],
\eeq[Root]
where we have introduced \Math{s=\mbox{sign}(g)}. 
\par
As all the correlators \Math{\dangle{\tr {\fett \L}^{2k}}} in the 
free theory are computable, 
all one has to do in order to calculate the expectation values
\Math{\langle\, \tr\Mvf^{2n}\,\rangle} is to take the $2n$'th power of the
result \gl{Root} and expand the resulting expression in $\l$. The
outcome of this straightforward calculation is
\beql
\dangle{\tr V^{\prime\, -1}({\fett \L})^{2n}}&=& \sum_{k=1}^n\, (-)^k \, 
\binom{2n}{n-k} 
\sum_{i=n-1}^\infty\,\left \{ \prod_{l=0}^i\, [\, (3l)^2-k^2\, ] 
\right \}\,\zeile
& & \qquad \cdot \: 2\:
\frac{(-3g)^{i+1-n}}{(2i+2)!}\, \dangle{\tr {\fett \L}^{2i+2}},
\eeql[Vprimemin1]
which is true for positive and negative $g$.
\Comment{
In order to obtain this result, introduce the abbreviations
\beql
(+) &\equiv &\sqrt[3]{\, \frac{s\,\sqrt{|g|}}{2}\,\l + 
\sqrt{ \frac{|g|}{4}\,\l^2 + \frac{s}{27}}\,} \zeile
(-) &\equiv & \sqrt[3]{ \,\frac{s\,\sqrt{|g|}}
{2}\,\l - \sqrt{ \frac{|g|}{4}\,\l^2 + \frac{s}{27}}\,}.\nonumber
\eeql
Then
\Math{V^{\prime\, -1}(\l)^{2n}= [\,(+) + (-)\, ]^{2n}/|g|^n}.
Using \Math{(+)\,\cdot \,  (-) = -s/3} we have
\beq
\Bigl [\, (+) + (-)\, \Bigr ]^{2n} = \sum_{k=1}^n \binom{2n}{n+k}\, 
\Bigl ( - \frac{s}{3}\, \Bigr )^{n-k}\, 
\Bigl [\, (+)^{2k}+ (-)^{2k}\, \Bigr ]
+ \binom{2n}{n}\, \Bigl ( - \frac{s}{3}\, \Bigr )^{n}.
\eeq[plusmin]
Consider the case \Math{s=1}. With the expansion \cite{Grad}
\beqx
(x+\sqrt{1+x^2})^q + (x-\sqrt{1+x^2})^q = 2 ( 1 + \sum_{i=0}^\infty
\, \frac{q^2\, [\, q^2-2^2\, ]
\ldots [\, q^2-(2i)^2\,]}{(2i+2)!}\, x^{2i+2}),
\eeqx
and eq.\ \gl{plusmin} one arrives at
\beql
V^{\prime\, -1}(\l)^{2n}&=& \sum_{k=1}^n  2\, (-)^{n-k}\,\binom{2n}{n+k}\, 
\Bigl [ \, (3g)^{-n} + \sum_{i=0}^\infty\left\{
 \prod_{l=0}^i [\, k^2-(3l)^2\, ] \right \} \,
\frac{(3g)^{i+1-n}}{(2i+2)!}\, \l^{2i+2}\, \Bigr ] \zeile
& & \qquad + \binom{2n}{n}\,(-)^n\, (3g)^{-n}.
\eeql[zsx]
With the help of the identities
\beqx
\sum_{k=1}^n \binom{2n}{n+k} \, (-)^k = -\frac{1}{2}\, \binom{2n}{n}
\eeqx
and
\beqx
\sum_{k=1}^n \binom{2n}{n+k}\, (-)^k\, k^{2q}=0 \qquad \mbox{for}
\qquad n>q>0,
\eeqx
which one may prove, one recovers the result of eq.\ \gl{Vprimemin1} from
eq.\ \gl{zsx}. A similar analysis is possible for \Math{s=-1} yielding
the same result as eq.\ \gl{Vprimemin1}.}
\Subsubkapitel{Gaussian One--Point Functions}
In order to calculate the one--point functions of the Gaussian matrix
model \Math{\dangle{\tr {\fett \L}^{2k}}}
one can use the iterative solution of the loop equations
of ref.\ \cite{Amb93} sketched
in chapter I. However, due to the simplicity of the Gaussian
model it is more convenient to employ the method of orthogonal polynomials
\cite{OrthPol}. Here the evaluation of \Math{\dangle{\tr {\fett \L}^{2k}}}
for all genera reduces to an integral involving Hermite polynomials. This
problem was solved by Kostov and Mehta \cite{Kos87} and we simply cite
their result. 
\par
For the matrix potential \Math{1/4\, {\fett \L}^2} of eq.\ \gl{GaussNic}
the result reads
\beq
\frac{1}{N}\, \Bdangle{ \tr {\fett \L}^{2i+2}} =
2^{i+1}\, \frac{(2i+2)!}{(i+1)!\, (i+2)!}\, {\schnorkel P}_{i+1}(2N),
\eeq[Gauss1PF]
where \Math{{\schnorkel P}_n(2N)} is a polynomial in \Math{(2N)^{-2}}
\beq
{\schnorkel P}_n(2N)= \sum_{j=0}^{[n/2]} a_{nj}\, \frac{1}{(2N)^{2j}},
\eeq[PPolynomial]
whose coefficients \Math{a_{nj}} are defined by the recursion relation
\beq
a_{n+1,\, j}= \sum_{k=2j-1}^n k\, (k+1)\, a_{k-1,\, j-1},
\eeq[RecRels]
with \Math{a_{n0}=1}. Note the genus expansion recovered in eq.\ 
\gl{PPolynomial}. The fact that one has a recursive solution for higher genera 
contributions is not very astounding if one recalls the general iterative
solution sketched in chapter I. 
\Comment{
The first few \Math{a_{nj}} read
\beql
a_{n0} &=& 1 \zeile
a_{n1} &=& \frac{1}{3}\, (n+1)\, n\, (n-1) \zeile
a_{n2} &=& \frac{1}{90}\, (n+1)\, n\, (n-1)\, (n-2)\, (n-3)\, [5n-2] \zeile
a_{n3} &=& \frac{1}{5670}\, (n+1)\, n\, \ldots\, (n-5)\, [35n^2-77n+12]\zeile
a_{n4} &=& \frac{1}{340200}\,  (n+1)\, n\, \ldots\, (n-7)\, 
 [175n^3-945n^2+1094n-72]\zeile 
 &\vdots & \zeile
a_{nj} &=& (n+1)\, n \, \ldots \, (n-2j+1)\, [\, \mbox{Polynomial of degree}
\: n^{j-1}\,].
\eeql[anjs]
}
\Subsubkapitel{Results for all Genera}
Putting the results of the previous subsections together, i.e.\ plugging eq.\
\gl{Gauss1PF} into eq.\ \gl{Vprimemin1}, yields the bosonic one--point
correlators of the \Math{\SF}--matrix model with quartic potential 
for all genera
\beq
G_n = \frac{1}{N}\, \Bigl \langle \, \tr \Mvf^{2n}\, \Bigr \rangle =
\sum_{h=0}^\infty G^{h}_n\, \frac{1}{N^{2h}}.
\eeq[Ggndef]
Let us define
\beq
A^n_j= \sum_{k=1}^n (-)^k\, \binom{2n}{n-k}\, \prod_{l=0}^{j+n-1}
[\, (3l)^2-k^2\,].
\eeq[Anjdef]
Note that \Math{A^1_j} simplifies to \Math{A^1_j=3^{-j}\, (3j+1)!/(j!)}.
Then we have the genus \Math{h} correlators \Math{G^{h}_n}(g):
\beql
G^0_n &=& \sum_{j=0}^\infty A_j^n\, \frac{2^{\, j+n+1}}{(j+n)!\, (j+n+1)!}
 \: (-3g)^j \zeile
G^1_n &=& \sum_{j=\k(n,{h})}^\infty A_j^n\, 
\frac{2^{\, j+n-1}}{(j+n)!\, (j+n-2)!}
 \, \,\frac{1}{3}\:(-3g)^j \zeile
G^2_n &=& \sum_{j=\k(n,{h})}^\infty A_j^n\, 
\frac{2^{\, j+n-3}}{(j+n)!\, (j+n-4)!}\,
 \, \frac{[\, 5(j+n)-2\, ]}{90}\: (-3g)^j \zeile
G^3_n &=& \sum_{j=\k(n,{h})}^\infty A_j^n\, 
\frac{2^{\, j+n-5}}{(j+n)!\, (j+n-6)!}\,
 \, \frac{[\, 35(j+n)^2-77(j+n)+12\, ]}{5670}\: (-3g)^j \zeile
 G^4_n &=& \sum_{j=\k(n,{h})}^\infty A_j^n\, 
\frac{2^{\, j+n-7}}{(j+n)!\, (j+n-8)!}
 \zeile && \qquad\cdot\:
 \frac{[\, 175(j+n)^3-945(j+n)^2+1094(j+n)-72\, ]}{340200}
 \: (-3g)^j \zeile
 &\vdots&\zeile
 G^{h}_n &=& \sum_{j=\k(n,{h})}^\infty A_j^n\, 
\frac{2^{\, j+n-(2{h}-1)}}{(j+n)!\, (j+n+1)!}
 \, a_{j+n,{h}}\: (-3g)^j ,
 \eeql[Ggnresults]
where we have introduced \Math{\k= \mbox{max}\, (0,2{h}-n)} as
the lower bound of summations. The correlators \Math{G^{h}_n}
all have the same radius of
convergence of \Math{|g_c|=1/54}. Moreover the planar 
\Math{G^0_n(g=g_c)} are seen to converge at the critical point, 
whereas the higher genus correlators \Math{G^{{h}>0}_n(g=g_c)} diverge at the
critical value of the coupling constant.
\par
Let us now compute the asymptotic form of the one--point correlators
\Math{G^h_n(g)}. The factors \Math{A^n_j} of eq.\ \gl{Anjdef} behave
as
\beq
A^n_j \sim 9^{\, j+n-1}\, \Bigl [ \, (j+n-1)!\,\Bigr ]^2, 
\eeq[Anjass]
for \Math{j\gg 1}. Using this we compute the scaling behaviour of the
one--point functions
\beql
G^h_n &\simeq & \sum_{j\gg1} A^n_j\, \frac{2^{j+n-2h+1}}{(j+n)!\, (j+n-2h)!}
\, j^{h-1}\, (-3g)^j \zeile
&\simeq& 2^{n-2h+1}\, 9^{n-1}\, \sum_{j\gg1}\,
\frac{(j+n-1)\, (j+n-2)\, \ldots\, (j+n-2h+1)}{(j+n)}\, j^{h-1}\, 
( -54 g\, )^j \zeile
& \sim & \sum_{j\gg 1} j^{3h-3}\, \Bigl ( - \frac{g}{g_c}\, \Bigr )^j ,
\eeql[Gnjass]
with \Math{g_c=1/54}. Hence the correlators scale like
\beq
G^h_n(g) \sim (\, g_c - g\, )^{2-3h},
\eeq[Gscale]
and are independent of $n$. The bosonic one--point correlators $G^h_n$
are to be interpreted as sum over random surfaces of genus $h$ with
an $n$--gon hole as a boundary. In fact the form of the boundary becomes
irrelevant at the critical point of eq.\ \gl{Gscale}. One now argues that
the sum over surfaces without boundaries is obtained by integrating
eq.\ \gl{Gscale} over $g$. The resulting scaling behaviour
\beq
\int dg\, G^h_n(g) \sim   (\, g_c - g\, )^{[\, 2-(-1)\,]\, (1-h)}
\eeq[dGscale]
may then with the help of eq.\ \gl{Fgscaling} be associated with
a string susceptibility of \Math{\g_{\mbox{\footnotesize str.}}=-1},
which via eq.\ \gl{gammastring} corresponds to a central charge of
\Math{c=-2}. In this sense we have verified the claim that the 
$\SF$--matrix model of eq.\ \gl{c2model} describes random surfaces 
embedded in \Math{-2} dimensions. The double scaling limit of this
theory and the connection to the Liouville approach of this problem 
is studied in refs.\ \cite{KlebWilk}. Let us remark that
despite the supersymmetric
structure of this model it does not show any correspondence to two
dimensional supergravity. 
\par
The results for \Math{G^h_n} encode as well all the fermionic 
one--point correlators \Math{\sum_{a+b=n}\, \tr \Mvf^a\MY\Mvf^b\MYb}
if one makes use of the Ward identities presented in eq.\ \gl{WardId}.
\Subkapitel{Planar Two-- and Three--Point Functions}{23PF}
The calculation of higher point functions goes along the same lines.
However, in this section we shall be less ambitious and study only
planar contributions. We keep the quartic potential of eq.\ \gl{SF4Pot}.
\par
For the two--point functions there are now two nonvanishing
types of correlators possible: The doubly even powers in  \Math{
\langle \tr \Mvf^{2k}\,\Mvf^{2l}\rangle} and the doubly odd ones
in \Math{\langle \tr \Mvf^{2k-1}\,\Mvf^{2l-1}\rangle}. For the computation
of the last type we will have to know the power 
expansion of \Math{V^{\prime -1}
(\l)^{2n+1}} in order to apply the Nicolai--map. This is just
another exercise in elementary algebra, take the \Math{(2n+1)}'st power
of eq.\ \gl{Root} and expand the result in $\l$. The outcome of this
calculation is
\beqy
\tr V^{\prime\, -1}({\fett \L})^{2n+1}=
\eeqy
\beql
&& \sum_{k=0}^n\, (-)^k \, (2k+1)\, 
\binom{2n+1}{n-k}\sum_{l=n}^\infty\,
\left \{ \prod_{i=1}^l\, [\, 3i-k-2\, ]\,
[\, 3i+k-1\, ]\right \}\,  \zeile  &&\qquad \cdot \: 2\: 
\frac{(-3g)^{l-n}}{(2l+1)!}\, \tr {\fett \L}^{2l+1}.
\eeql[oddpowers]
The next ingredient needed are the two point functions of the Gaussian
matrix model. Here there is no nice closed form known in the
literature. The planar result, however, may
be obtained rather easily from the planar one--loop correlator
of eq.\ \gl{W0} through application of the loop insertion operator as
discussed in eq.\ \gl{hermnloopcorrelator4} of chapter I. 
Specializing to the potential \Math{1/4\, \tr {\fett \L}^2} one 
finds the connected, planar two point functions
\beql
\Bdangle{ \tr {\fett \L}^{2i}\, \tr {\fett \L}^{2l}}_{\mbox{\footnotesize
conn.}} &=& 2^{\, i+l}\, \frac{i \cdot l}{i+l}\, \binom{2i}{i}\,
\binom{2l}{l} \zeile
\Bdangle{ \tr {\fett \L}^{2i-1}\, \tr {\fett \L}^{2l-1}}_{\mbox{\footnotesize
conn.}} &=& 2^{\, i+l-3}\, \frac{i \cdot l}{i+l-1}\, \binom{2i}{i}\,
\binom{2l}{l} ,
\eeql[Gauss2P]
through an expansion of \Math{W_0(p,q)} in the loop parameters
$p$ and $q$. From this one may now 
compose the planar bosonic--two point functions
\beql
\Bigl \langle \tr \Mvf^{2m}\,\tr\Mvf^{2n} \Bigl \rangle_{\mbox{\footnotesize
conn.}} &=& \Bdangle{ \tr V^{\prime -1}({\fett \L})^{2m} \, 
\tr V^{\prime -1}({\fett \L})^{2n}}_{\mbox{\footnotesize conn.}} \zeile
\Bigl \langle \tr \Mvf^{2m-1}\,\tr\Mvf^{2n-1} \Bigl 
\rangle_{\mbox{\footnotesize conn.}} &=& 
\Bdangle{ \tr V^{\prime -1}({\fett \L})^{2m-1} \, 
\tr V^{\prime -1}({\fett \L})^{2n-1}}_{\mbox{\footnotesize conn.}},
\eeql[2PFunctions]
by plugging eqs.\ \gl{Vprimemin1}, \gl{oddpowers} and \gl{Gauss2P}
into the right hand sides of eq.\ \gl{2PFunctions}. We do not write
out the results explicitly, as they are lengthy and not too instructive.
\par
For the evaluation of the planar three--point functions 
take \Math{W_0(p,q,r)} and
expand in $p$, $q$ and $r$ to read off the planar, connected
three--point functions of the Gaussian model \cite{Peter}
\beql
\Bdangle{ \tr {\fett \L}^{2i}\,\tr {\fett \L}^{2j-1}\,\tr 
{\fett \L}^{2k-1}}_{\mbox{\footnotesize conn.}} &=& 2^{\, i+j+k-3}
\, i\cdot j\cdot k\, \binom{2i}{i}\binom{2j}{j}\binom{2k}{k}
\phantom{\qquad p}
\zeile
\Bdangle{ \tr {\fett \L}^{2i}\,\tr {\fett \L}^{2j}\,\tr 
{\fett \L}^{2k}}_{\mbox{\footnotesize conn.}} &=& 2^{\, i+j+k}
\, i\cdot j\cdot k\, \binom{2i}{i}\binom{2j}{j}\binom{2k}{k}.
\eeql[Gauss3PF]
And we may again compose the three--point functions of the $\SF$--matrix
model by plugging these results into the appropriate combinations of
eqs.\ \gl{Vprimemin1} and \gl{oddpowers}. 
\par
Needless to mention that these results may be employed for the
calculation of higher--point amplitudes involving fermions by making
use of higher--point Ward identities. 
These are obtained from the one--point
Ward identity of eq.\ \gl{WardId} by appropriate differentiation
with respect to $g_m$, e.g.
\beqx
\frac{1}{2}\, \Bangle{\tr \Mvf^{n+m-2}} + \frac{1}{4}\, \Bangle{\tr V^{\prime
}(\Mvf)\, \Mvf^{m-1}\,\tr V^{\prime}(\Mvf)\, \Mvf^{n-1}} =
\eeqx
\beqx 
\frac{1}{2}
\sum_{a+b=n-2}\Bangle{\tr V^\prime(\Mvf)\, \Mvf^{m-1}\, 
\tr \Mvf^a\,\MY\,\Mvf^b\,\MYb }\eeqx\beqx +
\frac{1}{2} \sum_{a+b=m-2}\Bangle{\tr V^\prime(\Mvf)\, \Mvf^{n-1}\, 
\tr \Mvf^a\,\MY\,\Mvf^b\,\MYb}
\eeqx 
\beq 
-\sum_{a+b=m-2\atop c+d=n-2}
\Bangle{ \tr \Mvf^a\,\MY\,\Mvf^b\,\MYb\, \tr \Mvf^c\,\MY\,\Mvf^d\,\MYb}.
\eeq[WIF2]
\par
\vfill
\Subkapitel{Supersymmetric Graph Counting}{SugraphCount}
Quite similar to the classical paper on planar diagram counting
of Br\'ezin, Itzykson, Parisi and Zuber \cite{PlanarDiagrams} the results
of eq.\ \gl{Ggnresults} may be viewed as a solution to the combinatorial
problem of supersymmetric graph counting with the Feynman rules derived
in section \Sub{FeynDiagr}. 
\par
Let us first consider the planar diagrams of this theory. For the 
counting of all diagrams with
2--bosonic legs we have to write down \Math{G^0_1(g)}
of eq.\ \gl{Ggnresults} in the first few orders of $g$
\beq
G^0_1(g)= \lim_{N\ra\infty}\,\frac{1}{N}\, \langle\, \tr \Mvf^2\,\rangle=
2\, -\, 16\, g\, + 280\, g^2\, + \,\ldots.
\eeq[G01]
These terms correspond to the planar diagrams
\beq
\begin{array}{ccccccccccr}
\Graph{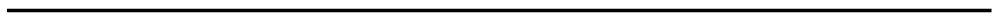} &\, 2 & & & & & & &\quad &\,{\fett \S:} & \: 2  \\ \\ \\
\Graph{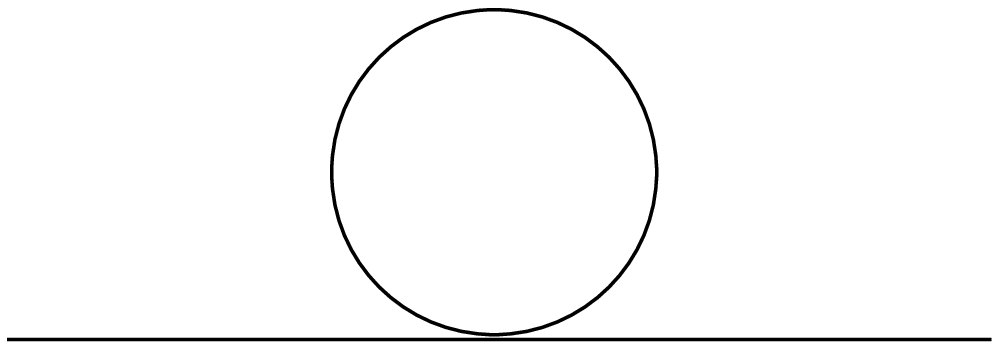} &\, -32g & & \Graph{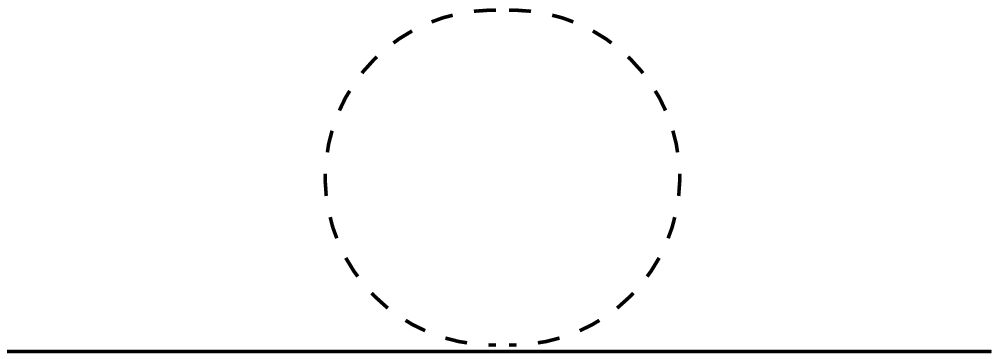} & \, 16g 
 & & & & & \, {\fett \S :} &\:  -16g \\
 \\ \\
\Graph{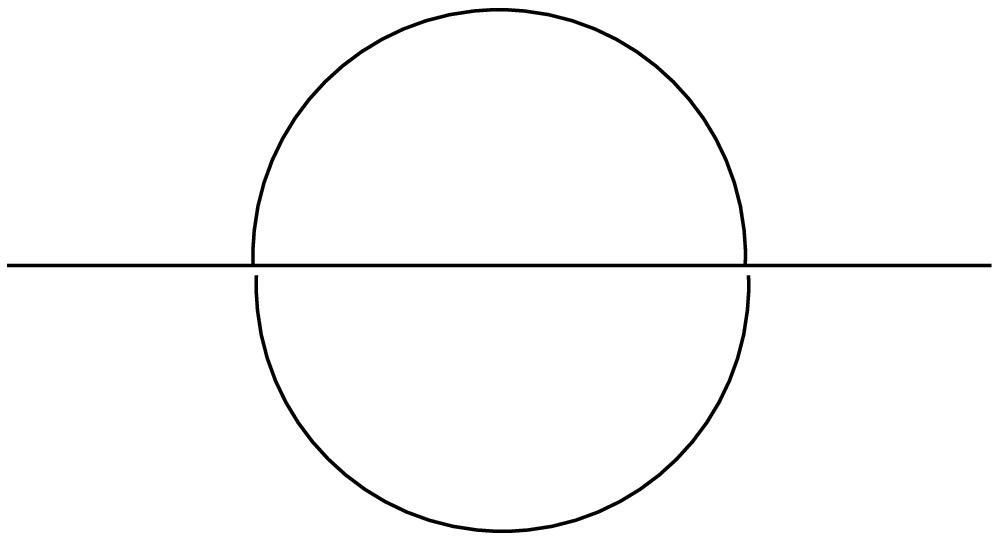} & \, 128 g^2 & \quad & \Graph{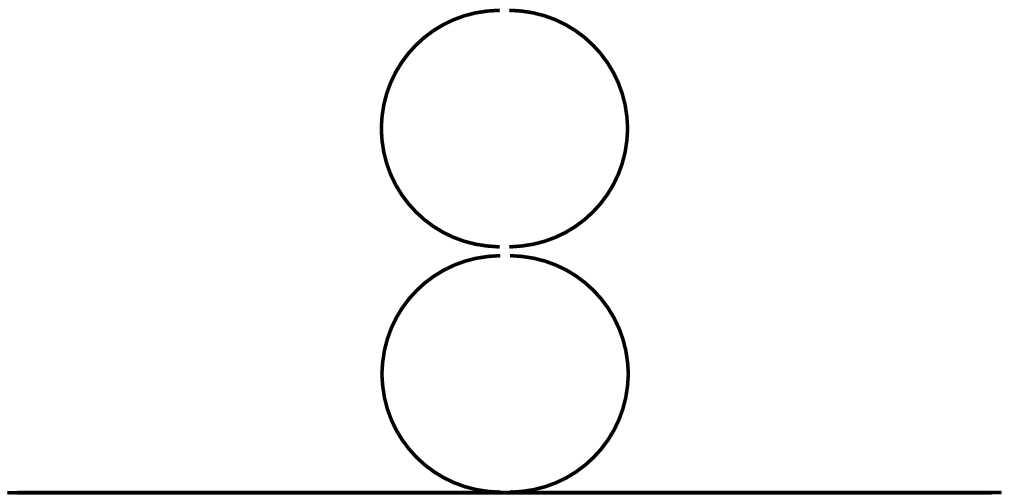} & \, 512g^2 & \quad 
   & \Graph{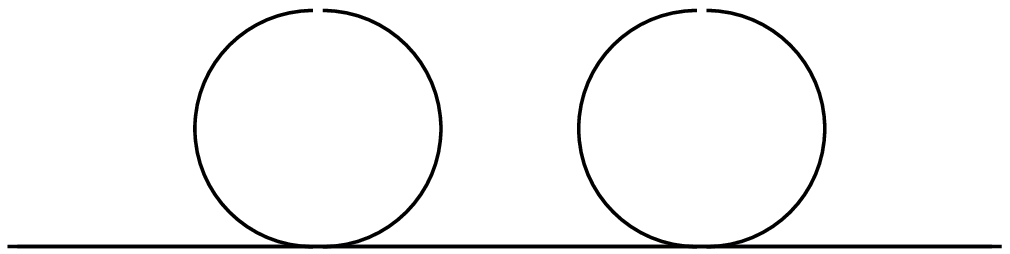}
      & \, 512 g^2 &\quad  && \\ \\
\Graph{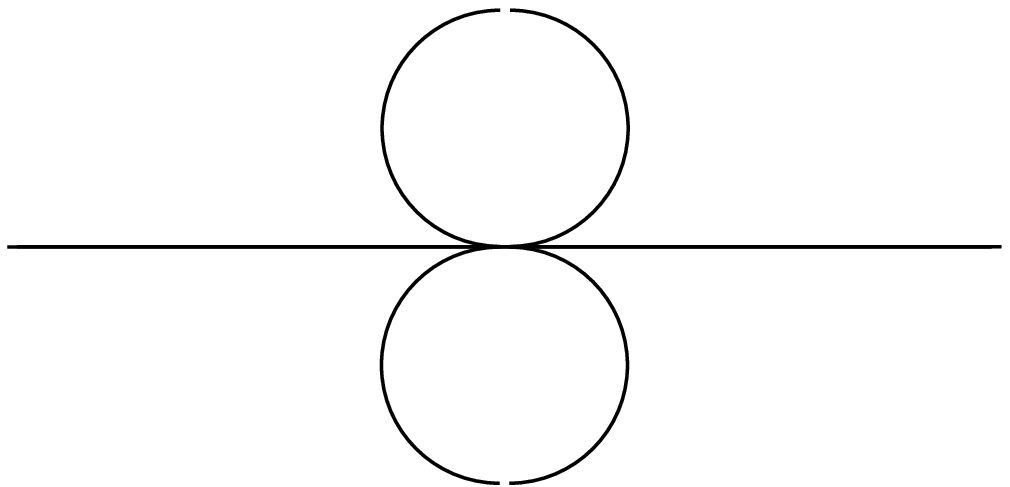} &\, -24g^2 & & \Graph{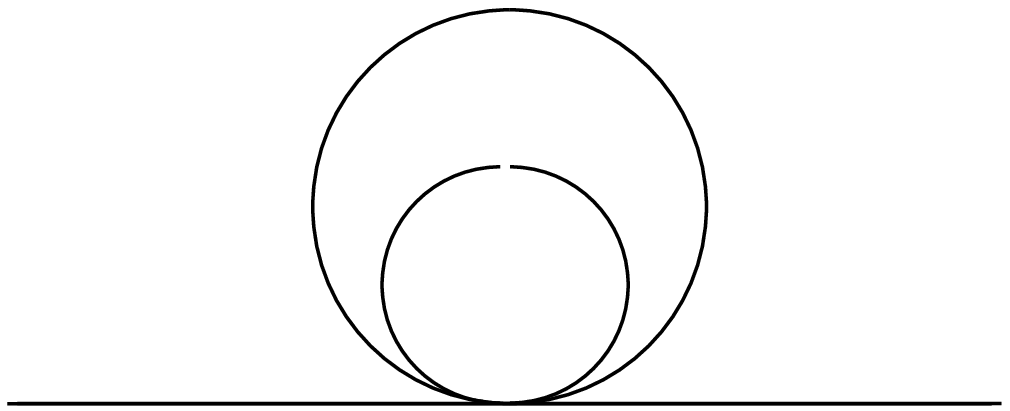} & \, -48g^2 & 
    & \Graph{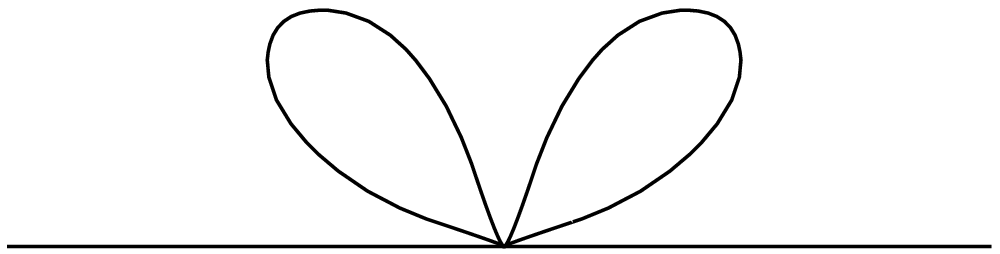}
      & \, -48g^2 & && \\ \\
\Graph{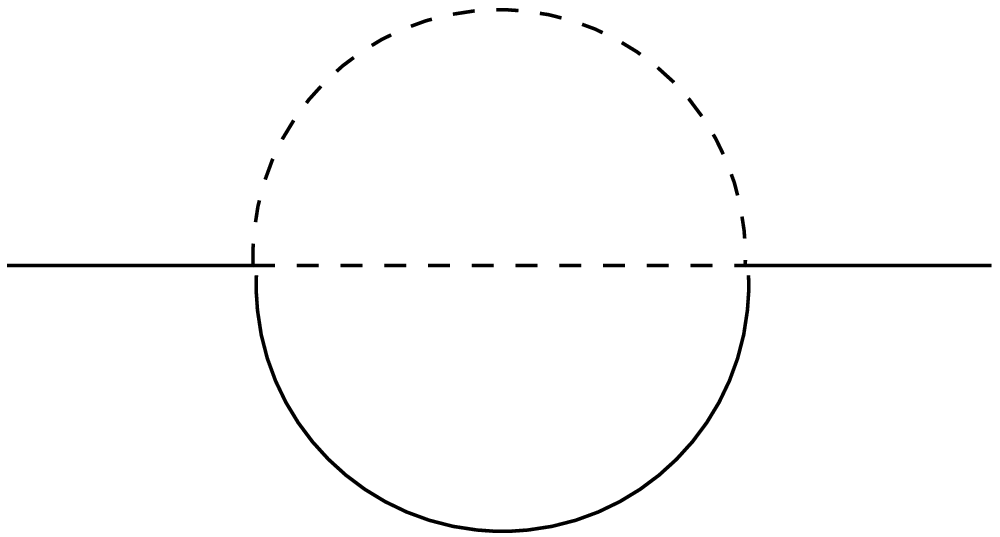} & \, -32g^2 & & \Graph{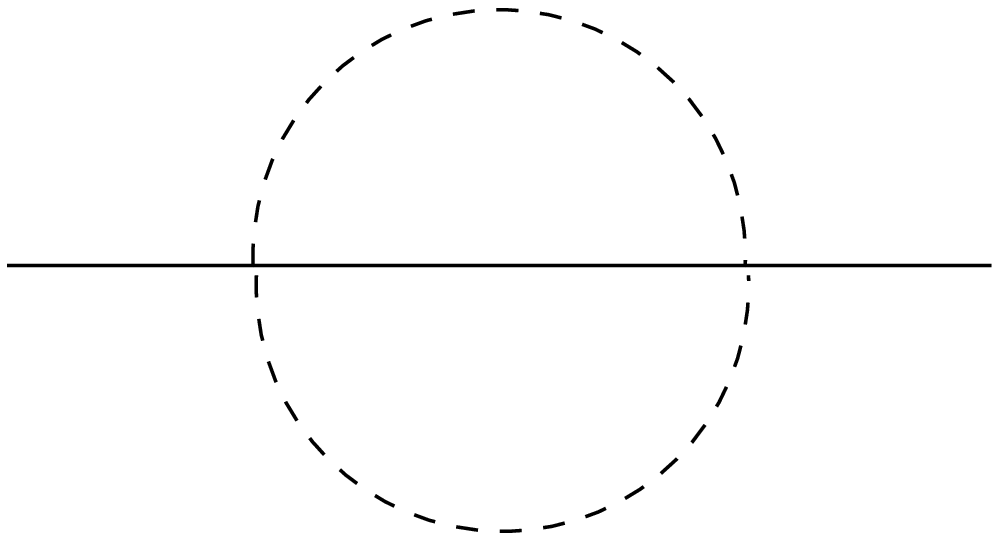} &\, -16g^2 & &
     \Graph{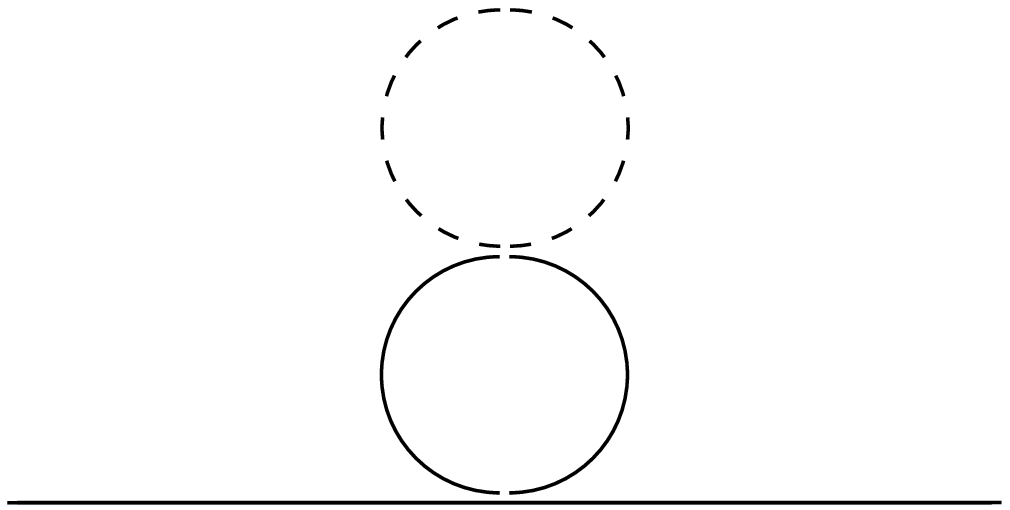} & -256g^2 & & \, {\fett \S:} &\:  280g^2  \\ \\
\Graph{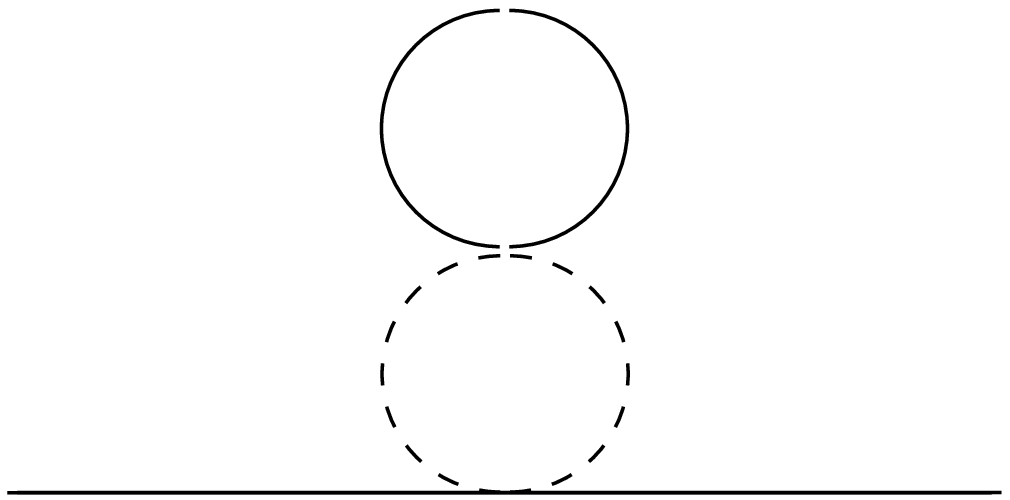} & \, -64g^2 & & \Graph{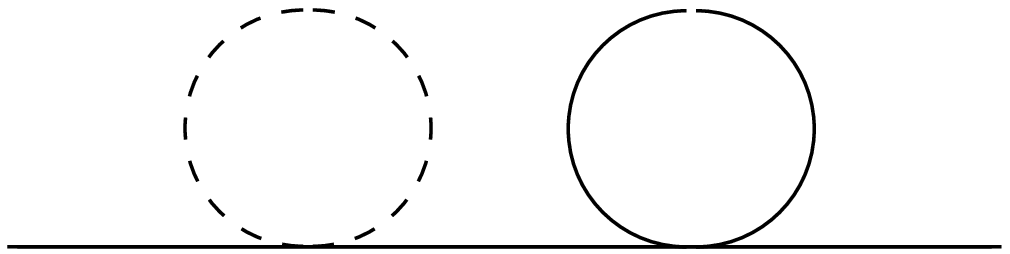} & \, -256g^2 & &
  \Graph{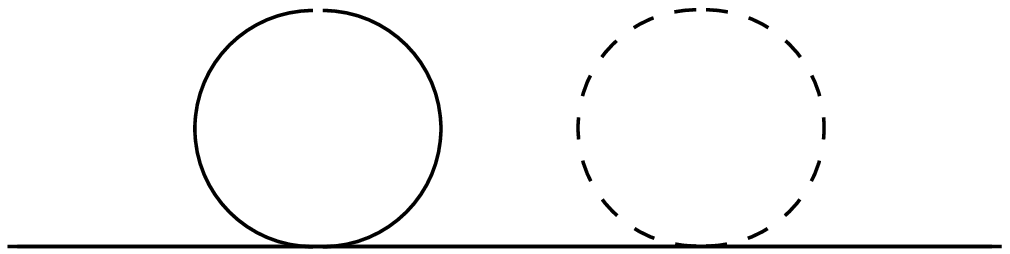} & \, -256g^2 & & & \\ \\
\Graph{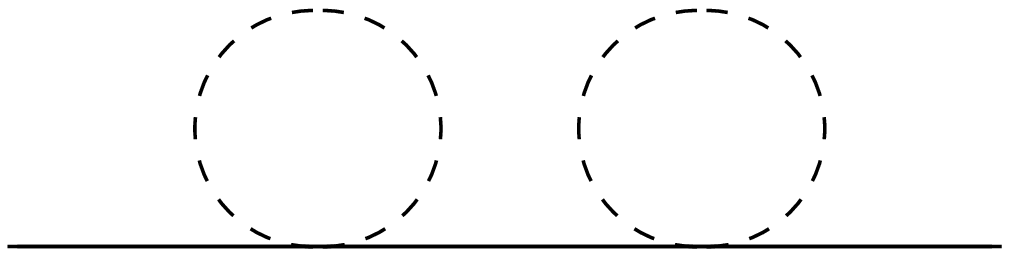} & 128g^2 & & & & & & & & & \\ 
\end{array}
\eeq[GrG01]
This constitutes an independent combinatorial check of our results.
\par
Similarly one can look at the amplitudes with 4--bosonic legs of planar
topology by expanding \Math{G^0_2}(g)
\beq
G^0_2(g)=\lim_{N\ra\infty}\,\frac{1}{N}\, \langle\, \tr \Mvf^4\,\rangle=
8\, -\, 160\, g\, + \, 4032\,g^2\, +\,\ldots,
\eeq[G02]
whose first two terms stem from the planar graphs
\beq
\begin{array}{ccccccccccr}
\Graph{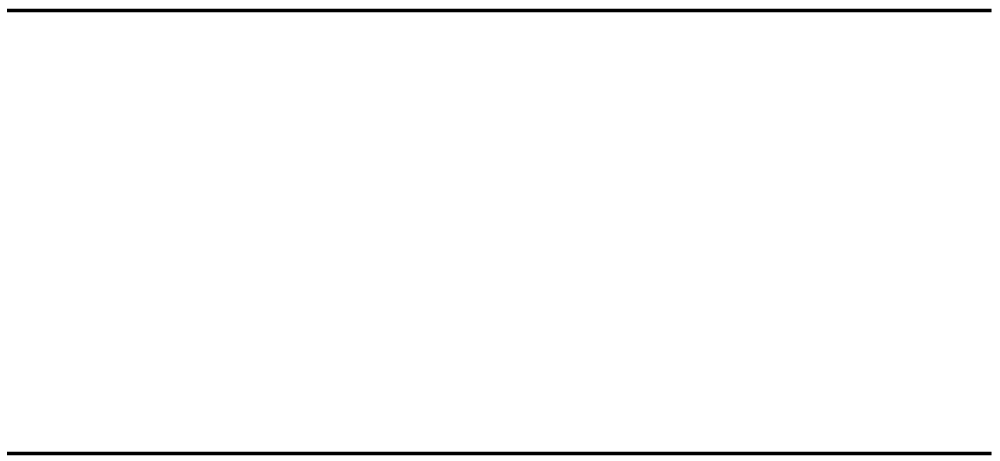} &\, 8 & & & & & & &\quad &\,{\fett \S:} & \: 8  \\ \\ \\
\Graph{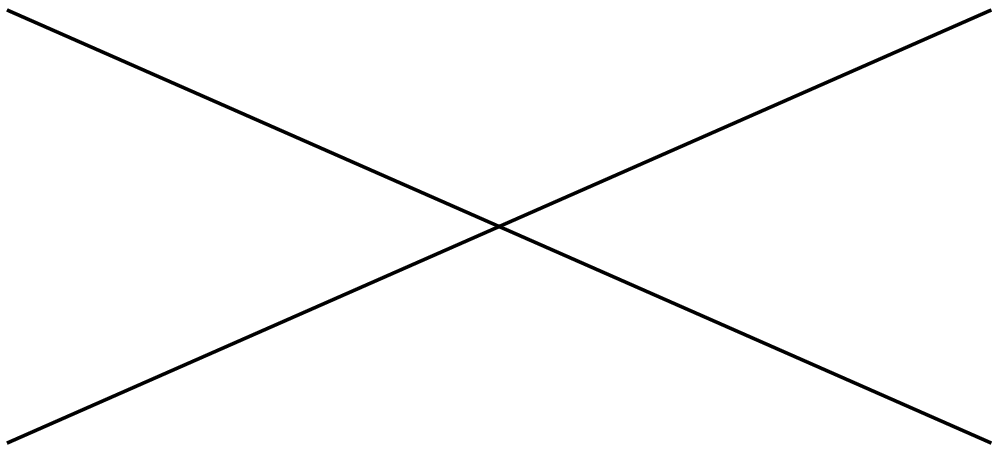} &\, -32g &\quad & \Graph{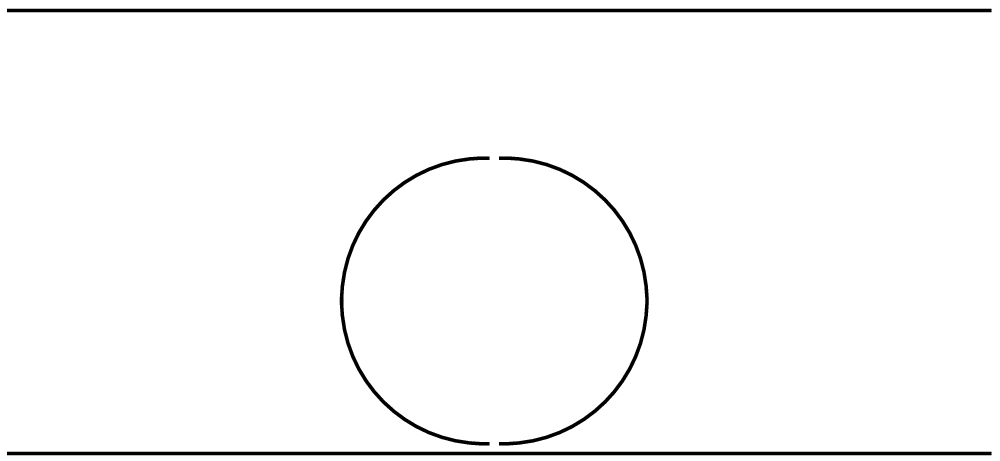} & \, -256g 
 &\quad &\Graph{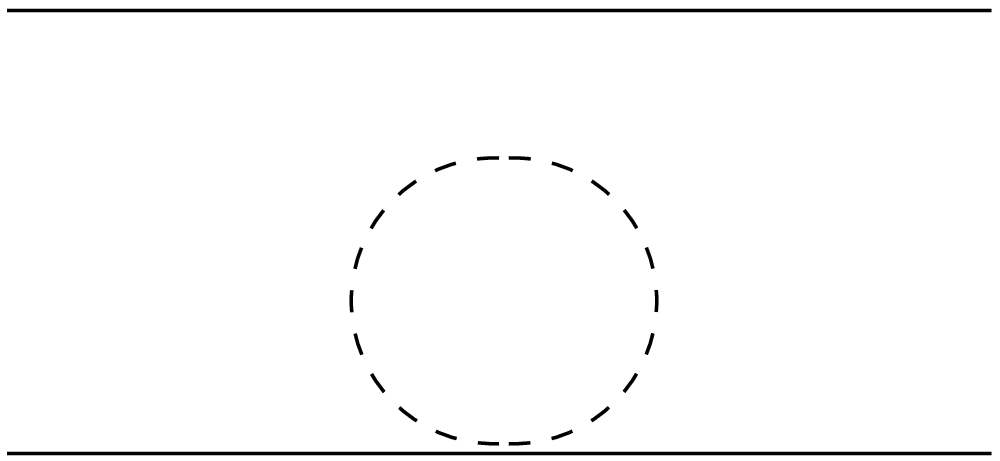}  & \, 128g  &\quad & \, {\fett \S :} &\:  -160g \\
\end{array}
\eeq[GrG02]
\par
Planar diagrams with 2--fermionic legs are counted with the help of the
observed Ward identities of eq.\ \gl{WardId}. For our
quartic potential we have
\beq
\langle\, \tr \MY\,\MYb\, \rangle = \frac{1}{2}\, \langle\, \tr \Mvf^2\,
\rangle + \frac{g}{2}\, \langle\, \tr \Mvf^4\, \rangle .
\eeq[WI0]
Hence with the results of eqs.\ \gl{G01} and \gl{G02}
the sum of the planar amplitudes containing 2--fermionic legs reads 
\beq
\lim _{N\ra\infty}\,\frac{1}{N}\, \langle\,\tr \MY\,\MYb\,\rangle =
1\, -\, 4\, g\, + \, 60\, g^2\, +\, \ldots ,
\eeq[F00]
and may be seen to originate from the diagrams
\beq
\begin{array}{ccccccccccr}
\Graph{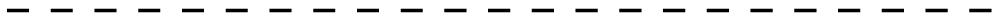} &\, 1 & & & & & & &\quad &\,{\fett \S:} & \: 1  \\ \\ \\
\Graph{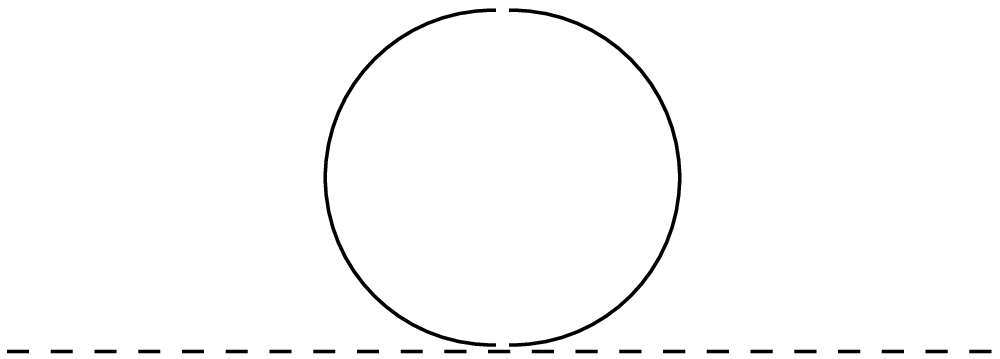} &\, -4g & & & & & & & & \, {\fett \S :} & \:  -4g \\
 \\ \\
\Graph{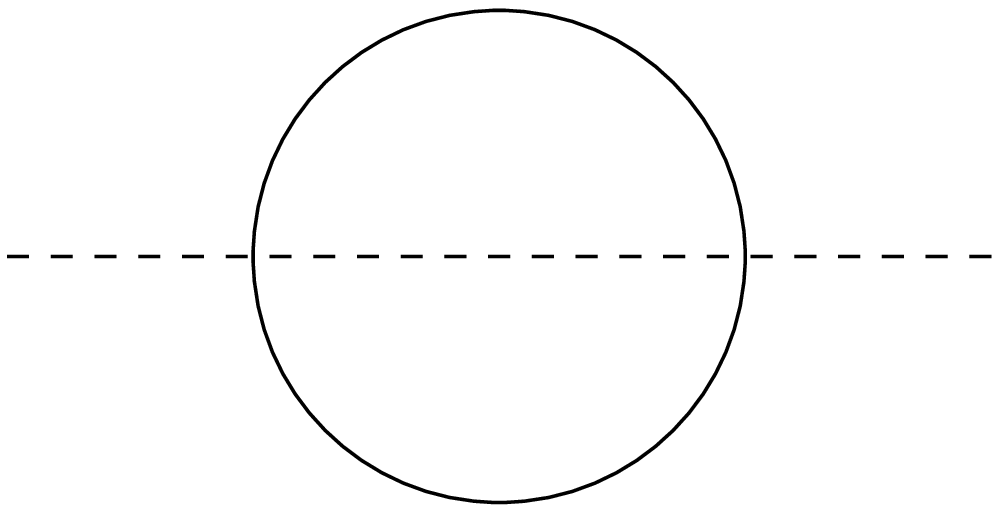} & \, 4 g^2 & \quad & \Graph{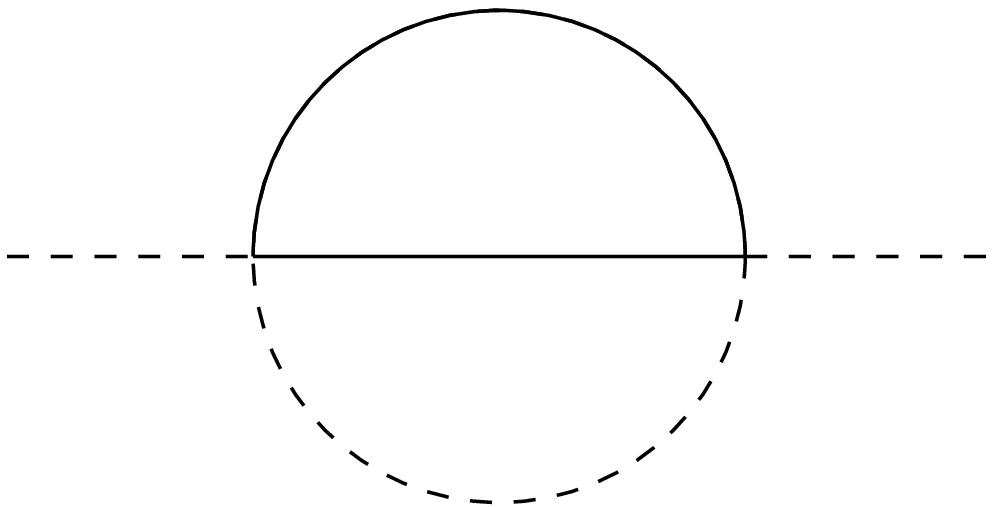} & \, 8g^2 & \quad 
   & \Graph{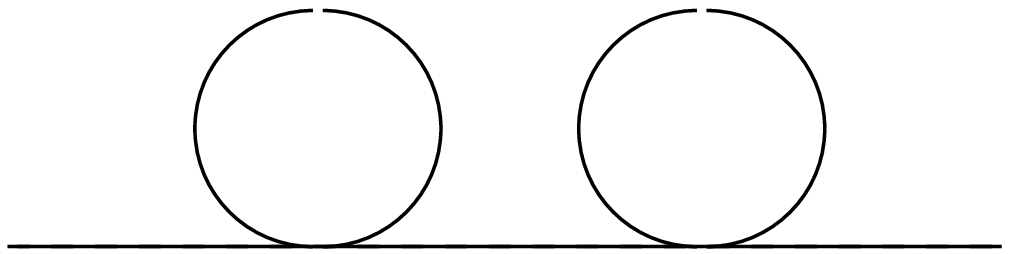}
      & \, 16g^2 &\quad  && \\ \\
\Graph{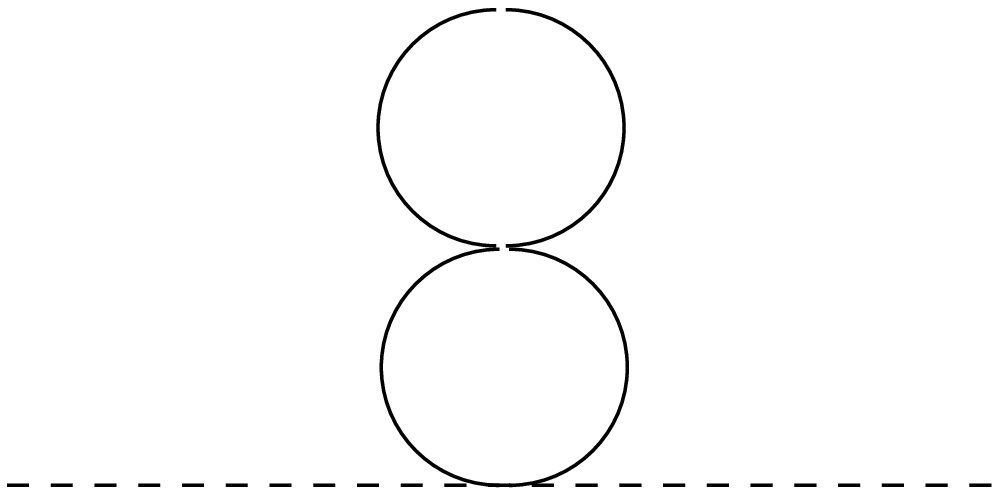} &\, 64g^2 & & \Graph{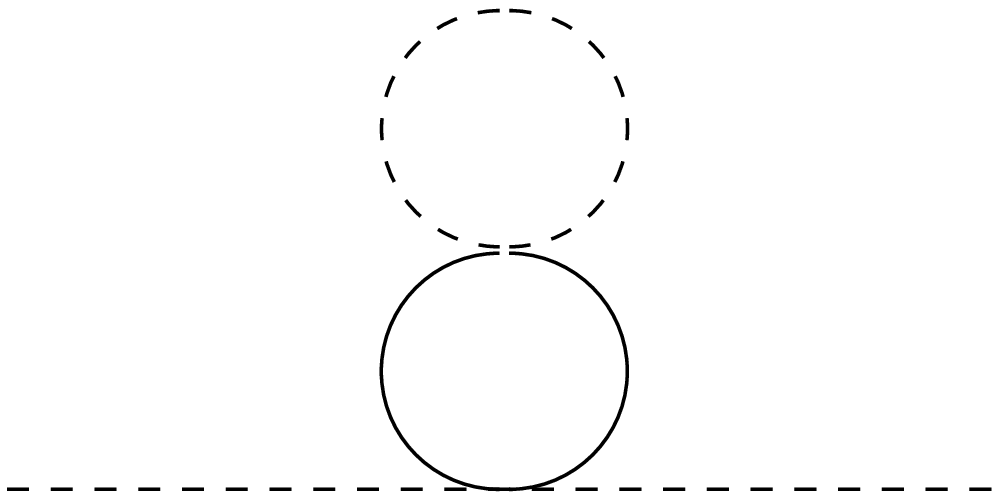} & \, -32g^2 & 
    & & & & \,{\fett \S:} & \: 60g^2 \\ 
\end{array}
\eeq[GrF01]

confirming our calculations.
\par
Of course our results allow the counting of non--planar diagrams as well.
For example the toroidal diagrams with 2--bosonic legs are contained in
\Math{G^1_1(g)}, i.e.\
\beq
G^1_1(g)=\lim_{N\ra\infty} N\, \langle\, \tr\, \Mvf^2\,\rangle =
-8\,g\, +\, 560\, g^2\, +\, \ldots ,
\eeq[G11]
whose first term stems from the toroidal diagrams
\beq
\begin{array}{ccccccccccr}
\Graph{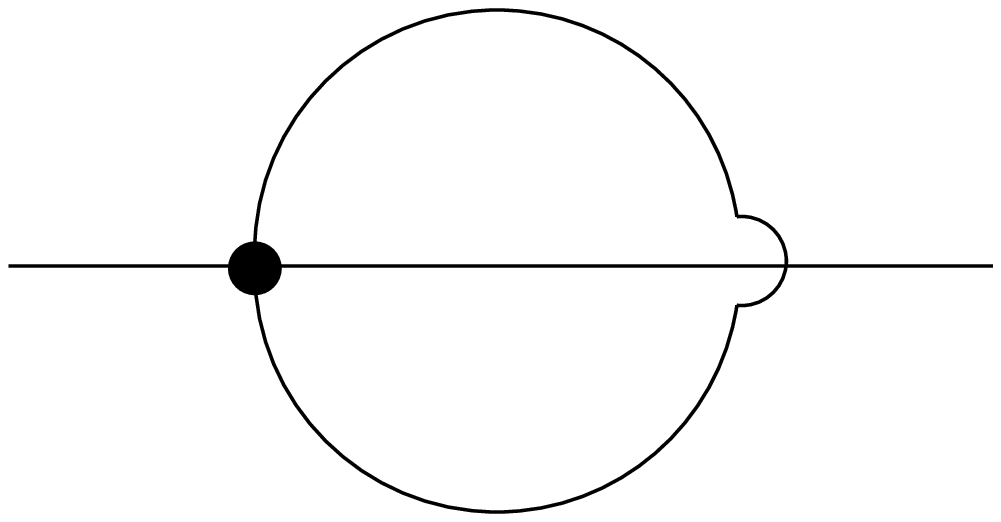} &\, -16g &\quad & \Graph{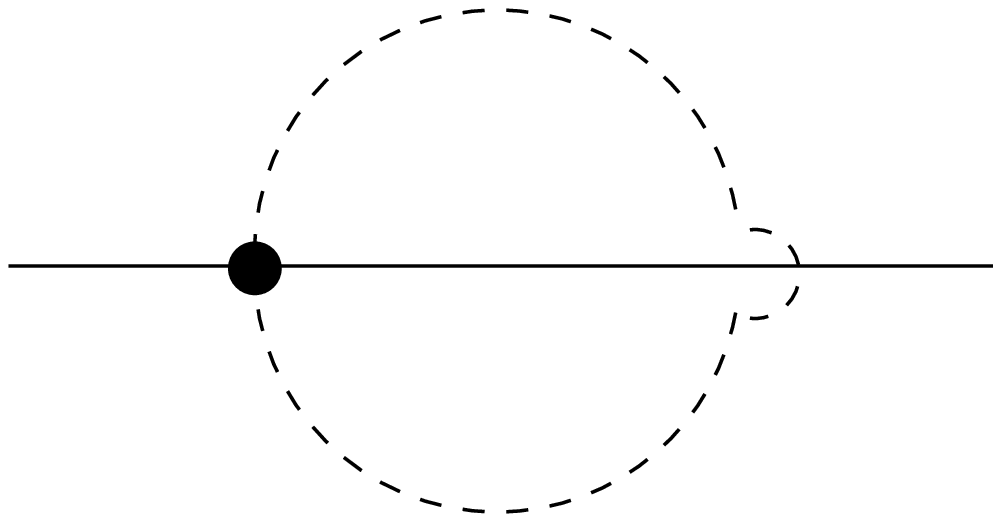} &\, 8g &\quad & 
\hspace{1.5cm} & \phantom{12g^2} &\quad & \,{\fett \S:} & - 8g  \\ 
\end{array}
\eeq[GrT01]
This concludes our analysis of the $\SF$--matrix model.
\Subkapitel{Conclusions}
In this chapter we have studied the \Math{\SF}--matrix model, defined
by a matrix valued superfield in (0,2)-dimensional superspace. After
integrating out the auxiliary and fermionic degrees of freedom its
partition function was shown to be identical to the partition 
function of a Hermitian matrix model with pure Gaussian potential. These 
two models are related to each other through a Nicolai--map. The analysis of
the Feynman rules of the  \Math{\SF}--matrix model revealed that its dual 
graphs may be interpreted as discretized random surfaces decorated with 
sign producing loops. Subsequently the one point--functions of the 
\Math{\SF}--matrix
model with quartic potential were calculated for all genera, as well as
the planar two-- and three--point functions. Despite of the triviality of
the partition function, these correlators are non--trivial and display
critical behaviour. The scaling limit of these correlators
reproduced the string susceptibility exponent of \Math{c=-2}
Liouville theory. Finally we showed how these results represent the
solution to the combinatorial problem of supersymmetric graph
counting.

%% file: supereigen
%
%
Following the successful application of Hermitian matrix models to
2d gravity and lower dimensional bosonic strings, it was natural to
ask how these methods could be generalized to the supersymmetric 
case. Unfortunately the description of discretized 2d random surfaces
by the Hermitian matrix model to date has no analogue in terms of
some supersymmetric matrix model describing a discretization of
super--Riemann surfaces. In a sense that might not be too astounding
as there exists no geometrical picture of superspace.
\par
Nevertheless the supereigenvalue model proposed by Alvarez--Gaum\'e,
Itoyama, Ma\~nes and Zadra \cite{Alv} appears to have all the virtues of
a discrete approach to 2d quantum supergravity. Guided by the prominent 
role the Virasoro constraints (cf.\ eq.\ \gl{Virasoroconstraint}) played 
for the Hermitian matrix model, the authors constructed a partition
function built out of $N$ Grassmann even and odd variables
(the ``supereigenvalues'') obeying a set of super--Virasoro constraints.
In order to formulate these constraints it is necessary to 
augment the bosonic coupling constants $g_k$ by a set of anticommuting 
couplings usually denoted by \Math{\x_{k+1/2}}.
As it is unknown whether there exists a matrix--based 
formulation of this model, we do not have a geometric interpretation
of it at hand. Nevertheless many of the well known features of the
Hermitian matrix model, such as the genus expansion, the loop equations,
the double scaling limit, the moment description and the loop insertion
operators, find their supersymmetric counterparts in the supereigenvalue
model. From this point of view the supereigenvalue model appears as
the natural supersymmetric generalization of the Hermitian one
matrix model.
\par
The supereigenvalue model is solvable nonperturbatively in coupling
constants but perturbatively in its genus expansion. Again the problem
of solving the model may be reformulated in a set of superloop equations 
obeyed by the superloop correlators.
Away from the double scaling limit these equations were
first solved for general potentials in the planar limit in ref.\ \cite{Alv2}.
We were able to develop a complete iterative solution of these equations
for all genera and general potentials in ref.\ \cite{Ple1}. 
The key point in this scheme is the change of variables from coupling 
constants to the so--called moments of the bosonic and fermionic 
potentials, thus generalizing the approach of ref.\ \cite{Amb93}
for the Hermitian matrix model. An alternative approach was pursued
by the authors of refs.\ \cite{Bec,McA} who managed to directly integrate
out the Grassmann odd variables on the level of the partition function.
No generalization of a solution based on orthogonal polynomials is 
known. The supereigenvalue model also displays a connection to 
supersymmetric integrable hierarchies \cite{FigMcA2}.
\par
In order to make contact to continuum physics the supereigenvalue 
model has to be studied in its double scaling limit. This was done in
refs.\ \cite{Alv,Ito,Alv2} for planar (and partially for higher genus)
topologies. These studies revealed that in its continuum limit the
supereigenvalue model describes the coupling of minimal superconformal
field theories (of type \Math{(2,4m)}) to 2d supergravity. Moreover
Abdalla and Zadra \cite{Zad} were able to develop a dictionary between 
\Math{N=1} super--Liouville amplitudes and supereigenvalue correlators.
It turns out that the moment description is very useful for the 
identification of the critical points in the space of coupling constants.
In ref.\ \cite{Ple2} we applied this method to the scaling limit, and
presented an improved iterative scheme to directly compute only
the higher genus terms relevant in the double scaling limit. 
\par
In the following we derive the supereigenvalue model directly from the 
super--Virasoro constraints in a superconformal field theory formulation
of the problem. After some short remarks on the general structure
of the free energy, we introduce superloop correlators and the
superloop insertion operators. The superloop equations are derived
and solved through the iterative procedure discussed above. We then turn
to the double scaling limit, identify the scaling of moments and
basis functions and present our improved iterative scheme. Finally
some remarks on the identification of the double scaled supereigenvalue
model with \Math{N=1} super--Liouville theory are made.
\Subkapitel{Super--Virasoro Generators}{SGen}
As we have discussed, we wish to regard the super--Virasoro 
constraints as the 
fundamental property of a supersymmetric generalization of the 
Hermitian one matrix model.  The Virasoro
generators of the Hermitian matrix model in eq.\ \gl{LVir} are differential
operators in the coupling constants $g_k$. In order to write down an
analogous set of super--Virasoro generators it will be necessary to 
introduce anticommuting coupling constants $\x_{k+1/2}$ as the
``superpartners'' of the bosonic couplings $g_k$. The super--Virasoro
generators are then given by \cite{Alv}:
\beql
\SG{n+1/2} &=& \sum_{k\geq 0} k\, g_k\, \del_{\x_{k+n+1/2}}
                          +\sum_{k\geq 0} \x_{k+1/2}\, \del_{g_{k+n+1}}
                          +\frac{1}{N^2}
                          \sum_{k=0}^n \del_{\x_{k+1/2}}\, 
                  \del_{g_{n-k}}, \zeile
\SL{n} &=& \sum_{k\geq0} k\, g_k\, \del_{g_{k+n}} + \frac{1}{2\,N^2}\,
                \sum_{k=0}^n \del_{g_k}\, \del_{g_{n-k}}  + 
                \sum_{k\geq 0} \Bigr ( k+\frac{n+1}{2}\Bigl )\, \x_{k+1/2}
             \, \del_{\x_{k+n+1/2}} \zeile
             && \quad + \frac{1}{2\, N^2}\, \sum_{k=0}^{n-1}
              k\,\del_{\x_{n-k-1/2}}\,\del_{\x_{k+1/2}}, \qquad 
             \qquad n\geq -1.      
\eeql[SuperVirGen] 
Note that in the above we define \Math{\sum_{k=0}^{-1}\ldots \equiv 0}.
The super--Virasoro generators satisfy
\beql
[\,\SL{n}\, , \, \SL{m}\, ] &=& (n-m)\, \SL{n+m} \zeile
[\, \SL{n}\, , \, \SG{m+1/2}\, ] &=& ( n/2-1/2-m)\, \SG{n+m+1/2} \zeile
\{\SG{n+1/2}\, , \, \SG{m+1/2}\} &=& 2\, \SL{n+m+1},
\eeql[SuperVirAlg]
which constitutes a closed subalgebra of the \Math{N=1} superconformal
algebra in the Neveu--Schwarz sector (remember that \Math{n,m\geq -1}).
\Subkapitel{Construction of the Model}{ModelConstr}
The supereigenvalue model will now be constructed from the requirement
that its partition function \Math{\SEM[\, g_k,\x_{k+1/2}]}
is annihilated by the \Math{\{ \SG{n+1/2},
\SL{n};\, n\geq -1\}}. However, equation
\gl{SuperVirAlg} shows that it is sufficient to impose 
\beq
\SG{n+1/2}\, \SEM= 0,\qquad n\geq -1,
\eeq[SuperVirConstr]
since the constraints \Math{\SL{n}\, \SEM=0}  for \Math{n\geq-1}
come out of the 
consistency conditions of eq.\ \gl{SuperVirConstr} using the algebra
\gl{SuperVirAlg}. 
\par
The method to construct \Math{\SEM[\, g_k,\x_{k+1/2}]} is based on a 
comment made in ref.\ 
\cite{Marsh} to generalize the conformal
field theory formulation of matrix models discussed in chapter I.
We seek for a representation of \Math{\SEM[\, g_k,\x_{k+1/2}]} in
the form of a correlator in a superconformal field theory. Let us consider
a free, holomorphic superfield \Math{{\schnorkel X}(z,\q)=\f(z)+\q\,\y(z)}
with the mode expansions
\beql
\f(z)&=&\widehat{q} + \widehat{p}\,\ln z - \sum_{k\neq 0}\frac{
\Jh{k}}{k}\, z^{-k} ,\zeile
\y(z) &=& \sum_{k\in\Z} \bh{k+1/2}\, z^{-k-1} ,
\eeql[Fields]
whose modes obey the (anti)--commutation relations
\beq
\begin{array}{cc}
[\, \Jh{n}\, ,\, \Jh{m}\,  ] = n\,\d_{n+m,0},\qquad  & [\, \widehat{q}\, ,
\, \widehat{p}\,]=1, \\
&\\
\{ \,\bh{n+1/2}\, ,\,\bh{m-1/2}\, \}= \d_{n+m,0}, \qquad& [\, \Jh{n}\, ,\,
\bh{m}\, ]=[\, \widehat{q}\, ,\, \bh{m}\, ]=[\, \widehat{p}\, ,\, \bh{m}\, ]=0.
\end{array}
\eeq[Scomrel]
Define the vacuum states 
\beql
\Jh{k}\, |0\rangle &=& 0 \qquad \langle 2 N|\, \Jh{-k}=0 \quad
\quad\qquad k>0 \zeile
\widehat{p}\, |0\rangle &=& 0 \zeile
\bh{n+1/2}\, |0\rangle &=& 0 \qquad \langle 2N |\, \bh{-n-1/2}=0 \qquad 
n\geq 0,
\eeql[SVac]
where \Math{|2N\rangle \equiv \exp [\, 2N\,\widehat{q}\, ]\, |0\rangle}
and \Math{\widehat{q}} is anti--Hermitian. The super--energy--momentum
tensor of this theory is given by
\beq
{\schnorkel T}(z,\q)= T^{\cal F}(z) + 2\, \q\, T^{\cal B}(z) 
              = \y\, \del \f + \q :( \del \f\,\del \f  + \y\, \del \y):\: , 
\eeq[T]
whose fermionic part has the mode expansion
\beq
T^{\cal F}(z) = \sum_{n\in\Z} T^{\cal F}_{n+1/2}\, z^{-n-2}
\eeq[TFdef]
\beqx
T^{\cal F}_{n+1/2} = \sum_{k>0} \Bigl ( \,\Jh{-k}\,\bh{k+n+1/2} +
 \bh{-k+1/2}\, \Jh{k+n}\,\Bigr ) + \sum_{a+b=n} \bh{a+1/2}\, \Jh{b},
 \qquad n\geq 0.
 \eeqx
Similar to the bosonic case we introduce a Hamiltonian built from  
Grassmann even and odd coupling constants
\beql
H(g_k, \x_{k+1/2} )&=& \sum_{k\geq 0} \Bigl ( \,g_k\, \Jh{k} + \x_{k+1/2}\,
\bh{k+1/2}\, \Bigr ) \zeile
&=& \oint_{C_0}\frac{dz}{2\p i}\, \Bigl
(\, V(z)\, \del\f(z) + \Y(z)\,\y(z)\, \Bigr ),
\eeql[SHam]
where \Math{V(z)=\sum_{k\geq 0}g_k\, z^k} and \Math{\Y(z)=\sum_{k\geq 0}
\x_{k+1/2}\, z^k}.  Using these definitions one shows that
\beqy
\SG{n+1/2}\, \langle 2N |\, \exp [ \, H(g_k, \x_{k+1/2})\, ]\, \ldots=
\eeqy
\beq
\langle 2N |\, \exp [ \, H(g_k, \x_{k+1/2})\, ]\, T^{\cal F}_{n+1/2}\,\ldots .
\eeq[SCor]
Hence any operator \Math{\schnorkel O} satisfying
\footnote{ We again extend the definition of \Math{T^{\cal F}_{n+1/2}} to
\Math{n=-1} by dropping the second term in the second eq.\ of \gl{TFdef}}
\beq
[\, T^{\cal F}_{n+1/2}\, , \, {\schnorkel O}\, ] = 0, \qquad n\geq -1,
\eeq[Ocon]
will give us a correlator obeying the super--Virasoro constraints
\beq
\SG{n+1/2}\, \langle 2N |\, \exp [ \, H(g_k, \x_{k+1/2})\, ]\, {\schnorkel O}\,
| 0\rangle =0,
\eeq[SSV]
as \Math{T^{\cal F}_{n+1/2}\, |0\rangle = 0}. Thus every nonvanishing
correlator of this form may be used for the definition of the
partition function
\beq
\SEM=\langle 2N |\, \exp [ \, H(g_k, \x_{k+1/2})\, ]\, {\schnorkel O}\,
| 0\rangle .
\eeq[SEMcordef]
The problem of finding the operators ${\schnorkel O}$ obeying eq.\ \gl{Ocon}
is solved in superconformal field theory. It turns out that
any function of the super screening charges \Math{{\schnorkel Q}_\pm} 
will do
\beq
{\schnorkel Q}_\pm  = \oint_C dz\, \int d\q\, :\exp [\, \pm\, {\schnorkel
X}(z,\q)\, ] : \: ,
\eeq[SuperSC]
where \Math{{\schnorkel X}(z,\q)} again denotes the holomorphic superfield.
We choose \Math{{\schnorkel Q}_+^{2N}} to get a nonvanishing 
correlator in eq.\ \gl{SEMcordef} \footnote{ More generally one
could take \Math{{\schnorkel Q}_+^{2N+M}\, {\schnorkel Q}_-^M}, 
supereigenvalue models constructed in this fashion have not been
analyzed so far.}. Note that we take an even \Math{2N} in order to 
have a Grassmann
 even partition function \Math{\SEM}. The model
is then defined by
\beqy
 \SEM[\, g_k, \x_{k+1/2}\, ]  = 
\eeqy
\beq
 \langle 2N|\, :\exp [ \, H(g_k, \x_{k+1/2})\, ]\,
 \, \prod_{i=1}^{2N}
 \oint_{C_i}dz_i \int d\q_i  :\exp[\, \f(z_i)+\q_i\,\y (z_i) \, ] :\, |0
 \rangle .
 \eeq[SEMcordef2]
This correlator may be evaluated by using the operator product
expansions \Math{\f(z)\,\f(\w)\sim \ln (z-\w)} and 
\Math{\y(z)\, \y(\w) \sim (z-\w)^{-1}}. The result is
\beqy
\SEM[\, g_k, \x_{k+1/2}\, ] =  
\eeqy
\beq
\prod_{i=1}^{2N} \oint_{C_i}dz_i\, \int d\q_i \, \exp \Bigl [\,
\sum_{i=1}^{2N} \Bigl ( V(z_i) + \q_i\, \Y(z_i)\Bigr ) \, \Bigr  ]\, 
\prod_{i<j}( z_i-z_j-\q_i\q_j).
\eeq[SEMdef]
Note the resemblance of this model to the eigenvalue description of the
Hermitian matrix model of eqs.\ \gl{hermeigenvaluemodel} and
\gl{ZN2}. This is the reason for calling \Math{\SEM} the ``supereigenvalue
model'' despite of the fact that the $z_i$ and $\q_i$ are no eigenvalues
of any matrix \footnote{The only attempt in literature to formulate eq.\ 
\gl{model2} in terms of matrices may be found in ref.\ \cite{Tak}, where
an explicit solution to this problem was found only for low $N$.}. 
It was first constructed in ref.\ \cite{Alv}
using different methods.
\Comment{
The derivation of eq.\ \gl{SEMdef} goes along the same lines as the
bosonic case of eq.\ \gl{ZN2}. Using \Math{:e^A:\, :e^B:\, = e^{\AB}\, :e^{A+B}:}
we have
\beql
\SEM &=& \Bigl (\prod_{i=1}^{2N} \oint_{C_i}dz_i\int d\q_i\Bigl )\,
\langle 2N|\, : \exp[\, \oint_{C_0} \{\,V(z)\, \del\f (z)+\Y (z)\, \y(z)\, \} +
        \f(z_1) + \q_1\, \y(z_1)\, ] :\zeile
        & & \quad \cdot \: \prod_{i=2} ^{2N} :\exp[ \, \f(z_i)
        +\q_i\,\y(z_i)\,]:\, |0 \rangle \, \cdot \, 
      e^{V(z_1)+\q_1\,\Y(z_1)} \zeile
      &=&  \Bigl (\prod_{i=1}^{2N} \oint_{C_i}dz_i\int d\q_i\Bigl )\,
\langle 2N|\, : \exp[\, \oint_{C_0} [\, V(z)\, \del\f (z)+\Y (z)\, \y(z) \, ]+
        \sum_{j=1}^2\f(z_j) + \q_j\, \y(z_j)\, ] :\, \zeile
        & & \quad \cdot\: \prod_{i=3} ^{2N} :  \exp[ \, \f(z_i) +\q_i\,
        \y(z_i)\,]:\, |0 \rangle \, e^{\ln (z_1-z_2) 
         - \frac{\q_1\q_2}{z_1-z_2}} 
           \, \exp [{\sum_{j=1}^2 V(z_j)+\q_j\, \Y(z_j)}]\zeile
&\vdots & \zeile
&=& \Bigl (\prod_{i=1}^{2N} \oint_{C_i}dz_i\int d\q_i\Bigl )
\langle 2N| : \exp[\, \oint_{C_0}[  V(z)\, \del\f (z)+\Y (z)\, \y(z) ] +
        \sum_{j=1}^{2N}\f(z_j) + \q_j\, \y(z_j)\, ] : |0 \rangle \zeile
        & & \quad \cdot \:\prod_{i<j} (z_i -z_j-\q_i\q_j)\, 
         \exp [{\sum_{j=1}^{2N} V(z_j)+\q_j\, \Y(z_j)}]
            \zeile
&=&\Bigl (\prod_{i=1}^{2N} \oint_{C_i} \int d\q_i\Bigl )\,
 \prod_{i<j} (z_i -z_j-\q_i\q_j) 
 \,\exp [{\sum_{j=1}^{2N} V(z_j)+\q_j\, \Y(z_j)}]\, 
   \underbrace{\langle 2N|\, : \exp[\, 2N \hat{q}\, ] :\, |0 \rangle }_{=1}
\nonumber
\eeql      
and we recover eq.\ \gl{SEMdef}.}
\Subkapitel{Structure of the Free Energy}{StructFree}
After deforming the contours of integration in eq.\ \gl{SEMdef} to the real
line we recover the supereigenvalue model proposed by 
Alvarez--Gaum\'e, Itoyama, Ma\~nes and Zadra \cite{Alv}. After a slight
change of notation the partition function is given by
\beqy
\SEM[\,g_k, \x_{k+1/2}; N]=
\eeqy
\beq
 \prod_{i=1}^{2N} \Bigl ( \int_{-\infty}^\infty 
d\l_i\int d\q_i\Bigr )
\, \prod_{i<j}(\, \l_i-\l_j -\q_i\q_j\, )
 \exp \Bigl [ -2N\,\sum_{i=1}^{2N}
\Bigl(\, V(\l_i) -\q_i\, \Y(\l_i) \, \Bigr ) \Bigr ],
\eeq[model2]
built from a set of $2N$ bosonic and fermionic variables $\l_i$ and
$\q_i$ respectively. The bosonic and fermionic
potentials read
\beq
V(\l_i)=\sum_{k=0}^\infty g_k\, {\l_i}^k, \qquad \mbox{and}\qquad
\Y(\l_i)=\sum_{k=0}^\infty \x_{k+1/2}\, {\l_i}^k,
\eeq[Pots]
with the Grassmann even and odd coupling constants $g_k$ and $\x_{k+1/2}$
respectively.
The free energy of the supereigenvalue model is naturally
defined by
\beq
\SEM(g_k, \x_{k+1/2}; \, 2N) = e^{\, (2N)^2\, F(g_k, \x_{k+1/2}; \, 2N)}.
\eeq[Fdefinition]
Before developing an iterative procedure to solve the
supereigenvalue model, let us discuss the fermionic structure
of the free energy. Consider the supereigenvalue model of eq.\
\gl{model2} in the absence of fermionic couplings, i.e.\ \Math{\x_{k+1/2}
=0}. The integration over the fermionic variables $\q_i$ may then
be performed \footnote{Simply write \Math{\prod_{i<j}
(\l_i-\l_j-\q_i\q_j)= \prod_{i<j}(\l_i-\l_j)\, \exp[\,{\sum_{i<j} 
\frac{\q_i\q_j}{\l_i-\l_j}}]}.} giving rise to the effective bosonic
eigenvalue model
\beqy
\SEM [\, g_k, 0; 2N]=
\eeqy
\beq
 \prod_{i=1}^{2N} \Bigl ( \int_{-\infty}^\infty d\l_i\,
\Bigr )\, \prod_{i<j}(\, \l_i-\l_j \, )\, \Pfaff (\l_{ij}^{-1})\,
\exp \Bigl [ -2N\,\sum_{i=1}^{2N}  V(\l_i) \Bigr ],
\eeq[bosmodel]
where \Math{\Pfaff (\l_{ij}^{-1})} is the Pfaffian of the antisymmetric
matrix
\beql
\l_{ij}^{-1} &=& \frac{1}{\l_i -\l_j}, \qquad i\neq j  \zeile
\Pfaff( A) &=& \sqrt{\, \det A\,} = \frac{1}{2^N\, N!}\, \e^{i_1 i_2 \ldots
i_{2N}}\, A_{i_1 i_2} A_{i_3i_4} \ldots A_{i_{2N-1}i_{2N}}.
\eeql[PfaffDef]
There exists an interesting identity between the effective bosonic model of
eq.\ \gl{bosmodel} and the Hermitian matrix model
\beq
\SEM (g_k, 0; 2N)= c(N)\, \Bigl [ \,
{\cal Z}_B(2g_k; N)\, \Bigr ]^2,
\eeq[SEMZB]
where \Math{{\cal Z}_B(2g_k; N)} is the partition function of the 
purely bosonic \Math{N\times N}
Hermitian matrix model of eq.\ \gl{hermmmodel}. We prove this identity in
the following.
\Comment{
The statement we wish to prove is
\beqy
\int\Bigl ( \prod_{i=1}^{2N} d\m(\l_i)\Bigr ) \, \Pfaff (\l_{ij}^{-1}) \,
\D(\l_1,\ldots,\l_{2N}) = 
\nonumber\eeqy
\beq
c(N)\,\int\Bigl ( \prod_{i=1}^{2N} d\m(\l_i)\Bigr )
\, \D^2( \l_1,\ldots,\l_N)\, \D^2(\l_{N+1},\ldots, \l_{2N}),
\eeq[statement]
where \Math{\m(\l_i)} is a measure factor depending only on the eigenvalue
\Math{\l_i}, \Math{\D(\l_1,\ldots,\l_M)} denotes the van der Monde determinant
and \Math{c(N)} some irrelevant $N$ dependent factor.
\par
We prove eq.\ \gl{statement} by induction in $N$. First note that due
to the antisymmetry of \Math{\D(\l_1,\ldots,\l_{2N})} under the exchange
of two eigenvalues we have
\beqy
\int\Bigl ( \prod_{i=1}^{2N} d\m(\l_i)\Bigr ) \, \Pfaff (\l_{ij}^{-1}) \,
\D(\l_1,\ldots, \l_{2N}) =
\eeqy
\beq
\int\Bigl ( \prod_{i=1}^{2N} d\m(\l_i)\Bigr ) \, 
\D(\l_1,\ldots,\l_{2N})\, \frac{1}{\l_1-\l_2}\,\frac{1}{\l_3-\l_4}\,\cdots
\frac{1}{\l_{2N-1}-\l_{2N}},
\eeq[Rem1]
up to irrelevant factors depending on $N$.
Moreover from \Math{\D(\l)= (-)^{N(N-1)/2} \det({\l_i}^{j-1})} we have the
identity
\beq
\int\Bigl ( \prod_{i=1}^{N} d\m(\l_i)\Bigr ) \, 
\D^2(\l_1,\ldots, \l_{N})= \int\Bigl ( \prod_{i=1}^{N} d\m(\l_i)\Bigr ) \, 
\D(\l_1,\ldots, \l_{N})\, \prod_{i=1}^N {\l_i}^{i-1}.
\eeq[Rem2]
Let us now show that eq.\ \gl{statement} is true for \Math{N=2}
\beql
Z_4&=&\int\Bigl ( \prod_{i=1}^{4} d\m(\l_i)\Bigr ) \, 
(\l_1-\l_3)\, (\l_2-\l_3)\, (\l_1-\l_4)\, (\l_2-\l_4) \zeile
&=&\int\Bigl ( \prod_{i=1}^{4} d\m(\l_i)\Bigr ) \, 
\det \left ( \matrix{ 1& \l_1 \cr 1 & \l_3}\right ) \,
\det \left ( \matrix{ 1& \l_2 \cr 1 & \l_4}\right ) \,
(\l_2-\l_3)\, (\l_1-\l_4) \zeile
&=&2  \int\Bigl ( \prod_{i=1}^{4} d\m(\l_i)\Bigr )\, \Bigr [
\det \left ( \matrix{ 1& \l_1 \cr \l_3 & {\l_3}^2}\right ) 
\,\det \left ( \matrix{ 1& \l_2 \cr \l_4 & {\l_4}^2}\right ) \,-
\det \left ( \matrix{ \l_1& {\l_1}^2 \cr \l_3 & {\l_3}^2}\right ) \,
\det \left ( \matrix{ 1& \l_2 \cr 1 & \l_4}\right )\Bigr ] ,
\nonumber
\eeql[Calc0]
the second term in the last expression vanishes.
Applying  eq.\ \gl{Rem2} to the first term then proves the assumption 
\gl{statement} for \Math{N=2}.
\par
Now by using the induction hypothesis the left hand side of eq.\ \gl{statement}
for \Math{(N+1)} becomes
\beql
Z_{2N+2}&=& \int\Bigl ( \prod_{i=1}^{2N+2} d\m(\l_i)\Bigr )\,
\D^2(\l_1,\ldots ,\l_N)\, \D^2(\l_{N+1},\ldots , \l_{2N})\, \prod_{i=1}
^{2N}(\l_i-\l_{2N+1})\, (\l_i-\l_{2N+2}) \zeile
&=&\int\Bigl ( \prod_{i=1}^{2N+2} d\m(\l_i)\Bigr )\,
\D(\l_1,\ldots ,\l_N)\: \prod_{i=1}^{N}{\l_i}^{i-1} \:\:
\D(\l_{N+1},\ldots , \l_{2N})\, \prod_{i=N+1}^{2N}{\l_i}^{i-N-1}\zeile
&& \quad \cdot\: \prod_{i=1}^{2N} (\l_i-\l_{2N+1})\, (\l_i-\l_{2N+2})\zeile
&=& \int\Bigl ( \prod_{i=1}^{2N+2} d\m(\l_i)\Bigr )\,
\D(\l_1,\ldots ,\l_N,\l_{2N+1})\, \prod_{i=1}^{N}{\l_i}^{i-1} \:
\D(\l_{N+1},\ldots , \l_{2N},\l_{2N+2})
\zeile
&& \quad \cdot\:  \prod_{i=N+1}^{2N}{\l_i}^{i-N-1}
\, \prod_{i=1}^{N} (\l_{i+N}-\l_{2N+1})\, (\l_i-\l_{2N+2}).
\eeql[Calc2]
Note that
\beqx
\D(\l_1,\ldots ,\l_N,\l_{2N+1})\, \prod_{i=1}^{N}{\l_i}^{i-1} =
\det \left ( \matrix{ 1& \l_1 & {\l_1}^2 & \ldots & {\l_1}^{N-1} \cr
\l_2 & {\l_2}^2 & {\l_2}^3 & \ldots & {\l_2}^N \cr \vdots & &&& \cr
{\l_N}^{N-1}& {\l_N}^{N}& {\l_N}^{N+1}& \ldots &{\l_N}^{2N-2}\cr
1 & {\l_{2N+1}}& {\l_{2N+1}}^2& \ldots & {\l_{2N+1}}^{N-1}} \right )
\eeqx
as well as an analogue expression for the second van der Monde 
determinant in eq.\ \gl{Calc2}. One may then convince oneself
that in the product 
\beqy
\prod_{i=1}^{N} (\l_{i+N}-\l_{2N+1})\, (\l_i-\l_{2N+2}) =
\eeqy
\beqx
\sum_{l,k=0}^N \sum_{i_1\neq\ldots\neq i_l\atop \in (1,\ldots, N)}
\sum_{j_1\neq \ldots \neq j_k\atop \in (N+1,\ldots , 2N)} (-)^{k+l}\,
{\l_{2N+1}}^{N-k}\, \l_{i_1}\ldots \l_{i_l}\, {\l_{2N+2}}^{N-l} \,
\l_{j_1}\ldots \l_{j_k},
\eeqx
only the following terms contribute to \Math{Z_{2N+2}} of eq.\ 
\gl{Calc2} 
\beqx
{\l_{2N+1}}^{N}\, {\l_{2N+2}}^{N} + \sum_{r=1}^N {\l_{2N+1}}^{r-1} \,
{\l_{2N+2}}^{r-1} \, \l_r\ \l_{r+1}\ldots \l_N\: \l_{N+r}\,\l_{N+r+1}\ldots
\l_{2N}
\eeqx
Hence after some relabeling of indices and  via eq.\ \gl{Rem2} we arrive at
\beql
Z_{2N+2}&=&  \int\Bigl ( \prod_{i=1}^{2N+2} d\m(\l_i)\Bigr )\,
\D(\l_1,\ldots ,\l_N,\l_{2N+1})\, \prod_{i=1}^{N}{\l_i}^{i-1}\, {\l_{2N+1}}^N
\zeile && \quad
\D(\l_{N+1},\ldots , \l_{2N},\l_{2N+2})\, \prod_{i=N+1}^{2N}{\l_i}^{i-N-1}
\, {\l_{2N+2}}^N\zeile
&=& 
\int\Bigl ( \prod_{i=1}^{2N+2} d\m(\l_i)\Bigr )\,
\D^2(\l_1,\ldots ,\l_N,\l_{2N+1})\:
\D^2(\l_{N+1},\ldots , \l_{2N},\l_{2N+2}),
\nonumber
\eeql
and we have proven eq.\ \gl{statement}.}
Eq.\ \gl{SEMZB} was first proposed in ref.\ \cite{Bec}. It implies
that up to irrelevant additive constants the part of the free energy
independent of the fermionic couplings obeys
\beq
F^{(0)}_S[\, g_k, \x_{k+1/2}=0; 2N]= 2\, F_B[\, g_k;N] ,
\eeq[Frel]
a relation which we will recover in the explicit solution
of the supereigenvalue model.
\par
Including the fermionic couplings $\x_{k+1/2}$ in eq.\ \gl{model2} and 
performing the fermionic integrations to obtain the complete
effective bosonic eigenvalue model is a more complicated
task. Nevertheless McArthur \cite{McA} succeeded in doing so.
The outcome important to us is the remarkable 
result that the free energy contains
contributions only up to second order in the fermionic couplings, i.e.
\beq
F_S[\, g_k,\x_{k+1/2}] = F^{(0)}_S[\, g_k] + \sum_{k,l}\x_{k+1/2}\, 
\x_{l+1/2}\, F^{(2)}_{k,l}[\, g_k],
\eeq[McAForm]
confirming a conjecture made in ref.\ \cite{Bec}. This observation will
be the key to the solution of the superloop equation which we will
describe in the following section.
\Subkapitel{Superloop Equations}{SuperloopEq}
Our solution of the supereigenvalue model
is based on an integral form of the superloop equations, which are the 
analogue of the loop equations of the Hermitian matrix model discussed
in chapter I.
Generalizing the approach of Ambj\o rn, Chekhov,
Kristjansen and Makeenko \cite{Amb93}
for the Hermitian one matrix model we develop an iterative procedure which
allows us to calculate the genus $g$ contribution to the 
\Math{(n|m)}-superloop
correlators for (in principle) any $g$ and any \Math{(n|m)} and 
(in practice) for
any potential. The possibility of going to arbitrarily high genus is
provided by
the superloop equations, whereas the possibility of obtaining arbitrary
\Math{(n|m)}-superloop results is due to the superloop insertion operators
introduced below. A change of variables from the coupling constants to 
moments allows us to explicitly present results for arbitrary potentials.
The remainder of this chapter is essentially an enlarged version of the
authors papers \cite{Ple1} on the general and \cite{Ple2} on the
double scaled solution of the supereigenvalue model.
\Subsubkapitel{Superloop Insertion Operators}
For notational simplicity let us from now on write the supereigenvalue
model with $N$ taken to be even as
\beqy
{\cal Z}_S(g_k, \x_{k+1/2}; N)= e^{N^2 \, F} =\eeqy
\beq
\int (\prod_{i=1}^{N} d\l_i \, d\q_i ) \, \prod_{i<j}\, (\l_i -
\l_j - \q_i\q_j )
\, \exp \Bigl ( - N
\sum_{i=1}^N \, [ V(\l_i) - \q_i \Y (\l_i)] \, \Bigr ),
\eeq[model]
with the potentials
\Math{V(\l_i)= \sum_{k=0}^\infty g_k {\l_i}^k} and
\Math{\Y(\l_i) = \sum_{k=0}^\infty \x_{k+1/2}  {\l_i}^k} of eq.\ \gl{Pots}.
Expectation values are defined in the usual way by
\beqy
\Bigl \langle {\cal O}(\l_j,\q_j) \Bigr \rangle =\eeqy\beq
 \frac{1}{\cal Z}_S \int (\prod_{i=1}^{N}
d\l_i \, d\q_i )\, \,\D (\l_i, \q_i ) \,\, {\cal O} (\l_j ,\q_j) \, \exp
\Bigl ( - N
\sum_{i=1}^N [ \, V(\l_i) - \q_i \Y (\l_i)\, ] \Bigr ) ,
\eeq[expval]
where we write \Math{\D (\l_i, \q_i)= \prod_{i<j} (\l_i -\l_j -
\q_i\q_j)} for the measure.
We introduce the one--superloop correlators
\beq
\widehat{W}(p \mid\, ) = N \,\Blangle \,\sum_i \frac{\q_i}{p-\l_i} \,
\Brangle
\qquad
\mbox{and}
\qquad
\widehat{W}(\, \mid p) = N \, \Blangle \,\sum_i \frac{1}{p-\l_i}\, \Brangle ,
\eeq[Wwidehat]
which act as generating functionals  for the one--point correlators
\Math{\langle \,\sum_i {\l_i}^k \,\rangle} and 
\Math{\langle \,\sum_i \q_i {\l_i}^k\,\rangle}
upon expansion in $p$.  This easily generalizes to higher--point correlators
with the \Math{(n|m)}--superloop correlator
\beqy
\widehat{W}(p_1,\ldots ,p_n \mid q_1,\ldots ,q_m) =
\eeqy
\beq
 N^{n+m} \,\Blangle \,
\sum_{i_1} \frac{\q_{i_1}}{p_1-\l_{i_1}} \, \ldots \, \sum_{i_n}
\frac{\q_{i_n}}{p_n-\l_{i_n}} \, \sum_{j_1}\frac{1}{q_1-\l_{j_1}} \,
\ldots\, \sum_{j_m} \frac{1}{q_m-\l_{j_m}} \,\Brangle . 
\eeq[Wwidehatnm]
Quite analogously to the bosonic case 
these correlators may be obtained from
the partition function ${\cal Z}_S$ by application of the superloop insertion
operators \Math{\d/ \d V(p)} and  \Math{\d/ \d \Y (p)}:
\beqy
\widehat{W}(p_1,\ldots ,p_n \mid q_1,\ldots ,q_m)= 
\eeqy
\beq
\frac{1}{\cal Z}_S
\dY{p_1} \,\ldots\,
\dY{p_n}\, \dV{q_1}\,\ldots\, \dV{q_m}
\,\, {\cal Z}_S,
\eeq[Wwidehatnm2]
where
\beq
\dV{p}= -\sum_{k=0}^\infty\frac{1}{p^{\, k+1}} \frac{\pa}{\pa g_k}
 \quad\mbox{and}\quad
\dY{p}= -\sum_{k=0}^\infty\frac{1}{p^{\, k+1}} \frac{\pa}{\pa \x_{k+1/2}}.
\eeq[superloopinsertionops]
In particular equation \gl{Wwidehat} can now be written as
\Math{\widehat{W}(p\mid\,)=
\d\, \ln {\cal Z}_S/ \d \Y (p)} and \Math{\widehat{W}(\, \mid p)= \d\, \ln 
{\cal Z}_S /\d V
(p)}.
\par
However, it is convenient to work with the connected part of the
\Math{(n|m)}--superloop correlators, denoted by $W$. They may
be obtained from the free energy \mbox{$F=N^{-2}\,\ln\,{\cal Z}_S$} through
\beqy
W(p_1,\ldots ,p_n \mid q_1,\ldots ,q_m)=
\eeqy
\beq
\dY{p_1} \,\ldots\,
\dY{p_n}\, \dV{q_1}\,\ldots\, \dV{q_m}
\,\,  F=\phantom{\qquad \qquad \qquad \qquad gggg}
\eeq[Wnm]
\beqx
N^{n+m-2} \, \Blangle \,
\sum_{i_1} \frac{\q_{i_1}}{p_1-\l_{i_1}} \, \ldots \, \sum_{i_n}
\frac{\q_{i_n}}{p_n-\l_{i_n}} \, \sum_{j_1}\frac{1}{q_1-\l_{j_1}} \,
\ldots\, \sum_{j_m} \frac{1}{q_m-\l_{j_m}} \,\Brangle _{
\mbox{\scriptsize conn}} ,
\eeqx
in complete analogy to the bosonic case of eq.\ \gl{hermnloopcorrelator3}.
Note that  \mat{n\leq 2} due to the maximally quadratic dependence
of $F$ on the $\x_{k+1/2}$ mentioned in section \Sub{StructFree}.
\par
With the normalizations chosen above, one assumes that these correlators
enjoy the genus expansion 
\beq
W(p_1,\ldots ,p_n \mid q_1,\ldots ,q_m)= \sum_{g=0}^\infty \, 
\frac{1}{N^{2g}}
\, W_g (p_1,\ldots ,p_n \mid q_1,\ldots ,q_m).
\eeq[genusexpWnm]
Similarly one has the genus expansion
\beq
F= \sum_{g=0}^\infty \, \frac{1}{N^{2g}} \, F_g
\eeq[genusexpF]
for the free energy. In contrast to the Hermitian matrix model there
is no geometrical argument based on Feynman diagrams and the 
Euler relation available to justify the genus expansion. This is of course
due to a missing formulation of the supereigenvalue model in terms
of some generalized matrix model. However, the genus expansion is
motivated from the structure of the superloop equations, which we
derive in the following subsection.
\Subsubkapitel{Superloop Equations}
The superloop equations of our model are two Schwinger--Dyson equations, 
which are derived through a shift in integration variables $\l_i$ and
$\q_i$. They were
first stated in \cite{Alv,Alv2}, and we present them in an integral form for
the loop correlators $W(p\mid \, )$ and $W(\, \mid p )$. 
\par
The Grassmann--odd superloop equation reads
\beqy
\ccint{C}{\w} \, \frac{V^\prime (\w)}{p-\w}\, W (\w \mid\, ) \, + \,
\ccint{C}{\w} \, \frac{\Y (\w)}{p-\w} \, W (\, \mid \w) =
\eeqy
\beq
W(p \mid \, ) \, W(\,\mid p) \, + \, \frac{1}{N^2} \, W(p \mid p)
\eeq[superloop1]
and its counterpart, the Grassmann--even superloop equation, takes the form
\beqx
\ccint{C}{\w} \, \frac{ V^\prime (\w )}{p-\w} \, W(\, \mid \w ) \, + \,
\cint{\w} \frac{\Y ^\prime (\w )}{p-\w} \,
W(\w \mid \, )\, -\, \frac{1}{2} \, \frac{d}{dp}\, \cint{\w} \,
\frac{\Y (\w )\, W( \w \mid )\, }{p-\w} = \phantom{W(p \mid \, )\,}
\eeqx
\beq
 \frac{1}{2} \, \Bigl [ \, W(\, \mid p)^2 \, - \, W(p \mid \, )\, W^\prime (p
 \mid \, ) \, + \, \frac{1}{N^2}\, \Bigl ( \, W(\, \mid p,p) \, - \,
\frac{d}{dq}\,
 W( p,q \mid \, ) \,  \Bigr | _{p=q} \, \Bigr ) \Bigr ] .
\eeq[superloop2]
In the derivation we have assumed that the loop correlators have one--cut
structure,
i.e.\ in the limit $N\ra \infty$ we assume that the eigenvalues \Math{\l_i}
are contained
in a finite interval \mbox{$[x.y]$}.
Moreover $C$ is a curve around the cut.
\Comment{
In order to derive the first of the two superloop equations for the model
of eq.\ \gl{model} consider the shift in integration variables
\beqx
\l_i \ra \l_i + \q_i \frac{\e}{p-\l_i} \quad \mbox{and} \quad \,\,\,
\q_i \ra \q_i + \frac{\e}{p-\l_i}
\eeqx
where $\e$ is an odd constant. Under these we find that
\beqx
\prod_i d\l_i\, d\q_i \ra (1-\e\sum_i \frac{\q_i}{(p-\l_i)^2})\,
\prod_i d\l_i\,
\q_i
\eeqx
and the measure transforms as
\beqx
\D (\l_i,\q_i) \ra (1-\e \sum_{i\neq j} \frac{\q_i}{(p-\l_i)(p-\l_j)})\,
\D (\l_i,\q_i).
\eeqx
The vanishing of the terms proportional to $\e$ then gives us
the Schwinger--Dyson equation
\beq \Bigl
\langle \, N \Bigl\{ \sum_i \frac{1}{p-\l_i} \, \Bigr ( V^\prime (\l_i) \q_i +
\Y (\l_i) \Bigr ) \Bigr \}  \, - \,\sum_i \frac{\q_i}{p-\l_i} \, \sum_j
\frac{1}{p-\l_j} \, \Bigr \rangle = 0.
\eeq[Schwinger-Dyson-1]
Note that \Math{\langle \,\sum_i \q_i\,(p-\l_i)^{-1} \,
\sum_j (p-\l_j)^{-1} \, \rangle = N^{-2}\, \widehat{W} (p\mid p)}
with the definitions of the previous subsection.  In order to
transform eq.\ \gl{Schwinger-Dyson-1} into an
integral equation we define the
bosonic and fermionic density operators
\beqx
\r (\l ) = \frac{1}{N} \, \sum_i \langle \, \d (\l-\l_i ) \, \rangle
\quad \mbox{and} \quad \,\,\,
r (\l)= \frac{1}{N} \, \sum_i \langle \, \q_i \, \d (\l-\l_i ) \, \rangle .
\eeqx
With these the first sum in eq.\ \gl{Schwinger-Dyson-1} may be written as
\beqx
N^2\, \int d\l \, \Bigl ( \, r(\l)\, \frac{V^\prime (\l)}{p-\l}\,+\,
\r (\l)\, \frac{\Y (\l)}{p-\l}\, \Bigr ) =\phantom{N^2\, \int d \l
\,\, r(\l )\, \Bigl
[ \cint{\w} \, \frac{1}{\w -\l}\,  \frac{V^\prime (\w)}{p-\w}
\, \Bigr ] + }
\eeqx
\beq
 N^2\, \int d \l \,\, r(\l )\, \Bigl [ \cint{\w}
 \, \frac{1}{\w -\l}\,  \frac{V^\prime (\w)}{p-\w} \, \Bigr ] +
   N^2\, \int d \l \, \,\r (\l )\, \Bigl [ \cint{\w} \, \frac{1}{\w -\l}\,
 \frac{\Y (\w)}{p-\w} \, \Bigr ]
 \eeq[calc2.2]
where we assume that the real eigenvalues $\l_i$ are contained 
within a finite interval \Math{x \, <\l_i \, < y \, ,\, \forall i} in the limit
\Math{N \ra \infty}.  
Moreover $C$ is a curve around the cut \Math{[x,y]} and $p$ lies outside
this curve. Performing the $\l$
integrals in eq.\ \gl{calc2.2} gives us the one--superloop
 correlators. The full Schwinger--Dyson equation \gl{Schwinger-Dyson-1}
 may then be expressed in the integral form
\beqx
\cint{\w} \, \frac{V^\prime (\w)}{p-\w}\, \widehat{W} (\w \mid\, ) \, + \,
\cint{\w} \, \frac{\Y (\w)}{p-\w} \, \widehat{W} (\, \mid \w) = N^{-2}\,
\widehat{W} (p \mid p) .
\eeqx
Rewriting this in terms of the connected superloop correlators $W$ 
of eq.\ \gl{Wnm} yields eq. \gl{superloop1}.
\par
The derivation of the second superloop equation goes along the same lines by
performing the shift
\beqx
\l_i \ra \l_i + \frac{\e}{p-\l_i} \quad \mbox{and} \quad \,\,\,
\q_i \ra \q_i + \frac{1}{2}\, \frac{\e \, \q_i}{(p-\l_i)^2 }
\eeqx
with $\e$ even and infinitesimal. Similar steps as the ones discussed above
then lead us to the second superloop equation
\beqx
\cint{\w} \, \frac{V^\prime (\w )\, \widehat{W} (\, \mid \w ) \, +\,
\Y^\prime (\w) \,
\widehat{W}(\w \mid \, )}{p-\w} \, -\, \frac{1}{2} \,
\frac{d}{dp}\, \cint{\w}\,
\frac{\Y (\w )\, \widehat{W}(\w \mid \, )}{p-\w}=
\eeqx
\beqx
\frac{1}{2}\, \frac{1}{N^2}\, \widehat{W}(\, \mid p,p) \, - \, \frac{1}{2}\,
\frac{1}{N^2}\, \frac{d}{dq}\, \widehat{W}(p,q \mid \, ) \Bigr | _{p=q}
\eeqx
which, after rephrasing in connected quantities, gives eq.\ \gl{superloop2}.
}
Note the similarity to the loop equation for the hermitian matrix model
\gl{LoopEquation}. 
The eqs.\ \gl{superloop1} and \gl{superloop2} are equivalent to the
super--Virasoro constraints. The constraints follow
from the superloop equations by expanding them in $p$ and evaluating the 
resulting contour integrals. The resulting equations at each power of
$p$ then correspond to \Math{\SG{n+1/2}\, {\cal Z}_S=0} and \Math{\SL{n}
\, {\cal Z}_S=0} for \Math{n\geq-1}.
\par
The key to the solution of these complicated equations order by order
in $N^{-2}$ is the observation stated in eq.\ \gl{McAForm} that
the free energy $F$ depends at most quadratically on fermionic coupling
constants \cite{Bec,McA}. 
Via eq.\ \gl{Wnm} this directly translates to the one-loop
correlators, which we from now on write as
\beql
W(p \mid \, ) &=& v(p) \zeile
W(\, \mid p)  &=& u(p) \, + \, \widehat{u}(p).
\eeql
Here \Math{v(p)} is of order one in fermionic couplings, whereas
\Math{u(p)} is taken
to be of order zero and \Math{\widehat{u} (p)} of order two in the fermionic
coupling constants \Math{\x_{k+1/2}}. This
observation allows us to split up the two superloop equations
\gl{superloop1} and \gl{superloop2} into a set of four equations, 
sorted by their order in the \Math{\x_{k+1/2}}'s. Doing this we obtain
\par
\medskip
\leftline{Order 0:}
\beqy
\cint{\w} \, \frac{V^\prime (\w )}{p-\w} \, u(\w ) = 
\eeqy
\beq
\frac{1}{2}\, u(p)^2
\, + \, \frac{1}{2}\, \frac{1}{N^2}\, \dV{p}\, u(p) \, - \, \frac{1}{2}\,
\frac{1}{N^2}
\, \frac{d}{dq}\, \dY{p} \, v(q) \Bigr | _{p=q}
\eeq[order0]
\leftline{Order 1:}
\beqy
\cint{\w}\, \frac{V^\prime (\w )}{p-\w }\, v(\w )\, + \, \cint{\w} \,
\frac{\Y (\w )}{p-\w}\, u(\w ) = 
\eeqy
\beq
v(p)\, u(p) \, + \, \frac{1}{N^2}\,
\dV{p}\, v(p)
\eeq[order1]
\leftline{Order 2:}
\beqy
\cint{\w} \, \frac{V^\prime (\w )}{p-\w }\, \widehat{u}(\w ) \, + \,
\cint{\w} \, \frac{\Y^\prime (\w )}{p-\w}\, v(\w ) \, - \, \frac{1}{2}\,
\frac{d}{dp} \, \cint{\w}\,  \frac{\Y (\w )}{p-\w }\, v(\w ) =
\phantom{\frac{d}{dp}\, \cint{\w}}
\eeqy
\beq
 u(p)\, \widehat{u}(p)\, -\, \frac{1}{2}\, v(p)\, \frac{d}{dp}\, v(p) \, + \,
\frac{1}{2}\, \frac{1}{N^2}\, \dV{p}\, \widehat{u}(p) 
\eeq[order2]
\leftline{Order 3:}
\beq
\cint{\w}\, \frac{\Y (\w )}{p-\w }\, \widehat{u} (\w ) = v(p)\, \widehat{u}
(p).
\eeq[order3]
It is the remarkable form of these four equations which allows us to develop
an iterative procedure to determine \Math{u_g(p), v_g(p), \widehat{u}_g(p)}
and $F_g$
genus by genus. Plugging the genus expansions into these equations lets them
decouple partially, in the sense that the equation of order 0 at genus $g$
only involves
$u_g$ and lower genera contributions. The order 1 equation then only contains
$v_g$, $u_g$ and lower genera results and so on. The first thing to do,
however, is to find the solution for $g=0$.
\Subkapitel{The Planar Solution}{PlanSol}
In the following the planar solution for the superloop correlators is given for
a general potential. It was first obtained in \cite{Alv2}. We present
it in a very compact integral form augmented by the use of new
variables to characterize the potentials, the moments.
\Subsubkapitel{Solution for $u_0(p)$ and $v_0(p)$}
In the limit \Math{N\ra \infty} the order 0 equation \gl{order0} becomes
\beq
\cint{\w}\, \frac{V^\prime (\w )}{p-\w }\, u_0(\w )\,
= \, \frac{1}{2}\, u_0(p)^2.
\eeq[order0genus0]
This equation is well known, as up to a factor of $1/2$ it is
nothing but the planar loop equation of the 
hermitian matrix model \gl{LoopEquation0}. 
With the above assumptions on the one--cut structure
and by demanding that $u(p)$ behaves as $1/p$ for $p\ra \infty$ one
finds \cite{Mig83}
\beq
u_0 (p)\, = \, \cint{\w} \, \frac{V^\prime (\w )}{p-\w}\, \biggl [ \,
\frac{(p-x)(p-y)}{(\w -x)(\w -y)}\, \biggr ]^{1/2} ,
\eeq[u0]
as we verified in chapter I section \Sub{hermsolution}.
The endpoints  $x$ and $y$ of the cut on the real axis
are determined by the following requirements:
\beq
0=\cint{\w}\, \frac {V^\prime(\w )}{\sqrt{(\w -x)(\w -y)}} , \quad \quad
1=\cint{\w}\, \frac{\w\, V^\prime(\w )}{\sqrt{(\w -x)(\w -y)}},
\eeq[determinexy]
deduced from our knowledge that \Math{W(\, \mid p)= 1/p + {\cal O}(p^{-2})}.
\par
The order 1 equation \gl{order1} in the \Math{N\ra \infty} limit determining
the odd loop correlator $v_0(p)$ reads
\beq
\cint{\w}\, \frac{V^\prime (\w )}{p-\w} \, v_0 (\w ) \, + \, \cint{\w}\,
\frac{\Y (\w )}{p-\w}\, u_0(\w ) \, =\, v_0(p)\, u_0 (p).
\eeq[order1genus0]
It is solved by
\beq
v_0(p) \, =\, \cint{\w}\,
\frac{\Y (\w )}{p-\w}\, \biggl [ \, \frac{(\w -x)(\w -y)}
{(p-x)(p-y)} \, \biggr ]^{1/2} \,\, + \,\, \frac{\c}{\sqrt{(p-x)(p-y)}} .
\eeq[v0]
Here $\c$ is a Grassmann odd constant not determined by eq.\ 
\gl{order1genus0}, in fact
\Math{\c=N^{-1}\, \langle \, \sum_i \q_i \, \rangle} in the planar limit.
It will be
determined in the analysis of the two remaining equations \gl{order2}
and \gl{order3}. 
\Comment{
One verifies the above solution by direct computation.
Using eqs.\ \gl{u0} and \gl{v0} the right hand side of
eq. \gl{order1genus0} reads
\beql
u_0(p)\, v_0(p) &=& \ccint{C_1}{\w}\,\ccint{C_2}{z}\, 
\frac{V^\prime(\w)\,\Y (z)}
{(p-\w )(p-z)}\, \biggl [ \frac{(z-x)(z-y)}{(\w -x)(\w -y)}\, \biggr ]^{1/2}
\zeile
& &
+ \ccint{C_1}{\w}\,
\frac{V^\prime(\w )}{p-\w}\, \frac{\c}{[(\w -x)(\w -y)]^{1/2}},
\eeql[B1]
and the left hand side becomes
\beqx
\ccint{C_1}{\w}\, \ccint{C_2}{z}\, \frac{V^\prime(\w )\, 
\Y (z)}{(p-\w)(\w -z)}\,
\biggl [ \frac{(z-x)(z-y)}{(\w -x)(\w -y)}\biggr ]^{1/2}  \,
+ \ccint{C_1}{\w}\,\frac{V^\prime (\w )}{p-\w}\, 
\frac{\c}{[(\w -x)(\w -y)]^{1/2}}
\eeqx
\beqx
\phantom{\ccint{C_1}{\w}\, \ccint{C_2}{z}}
+\, \ccint{C_1}{\w}\,\ccint{C_2}{z}\, \frac{\Y (\w )\, V^\prime (z)}{(p-\w )
(\w -z)}\, \biggl [ \frac{(\w -x)(\w -y)}{(z-x)(z-y)} \biggr]^{1/2}.
\eeqx
Now in the last term pull the contour integral $C_2$  over the 
curve $C_1$. One can show that the contribution from the extra pole 
vanishes. After renaming $\w \leftrightarrow z$ and combining the 
first and third terms one gets eq.\ \gl{B1}. Thus eq.\ \gl{order1genus0} 
is verified.
}
\Subsubkapitel{Moments and Basis Functions}
Let us now define new variables characterizing the
potentials \Math{V(p)} and \Math{\Y (p)}. Instead of the couplings $g_k$
we introduce the bosonic moments $M_k$ and $J_k$ defined by \cite{Amb93}
\beql
M_k &=& \cint{\w}\, \frac{V^\prime (\w )}{ (\w - x)^k} \,
\frac{1}{[\, (\w -x)\, (\w-y)\, ]^{1/2}}
, \quad k\geq 1 \zeile
J_k &=& \cint{\w}\, \frac{V^\prime (\w )}{ (\w - y)^k} \,
\frac{1}{[\, (\w -x)\, (\w-y)\, ]^{1/2}}
, \quad k\geq 1 ,
\eeql[MJDef]
and the couplings $\x_{k+1/2}$ are replaced by the fermionic moments
\beql
\X_k &=& \cint{\w}
\, \frac{\Y (\w )}{(\w -x)^k}\, [\,(\w -x)(\w -y)\, ]^{1/2}
,\quad k\geq 1 \zeile
\L_k &=& \cint{\w}\,
\frac{\Y (\w )}{(\w -y)^k}\, [\,(\w -x)(\w -y)\, ]^{1/2}
,\quad k\geq 1 .
\eeql[XLDef]
These moments depend on the coupling constants, both explicitly
and through $x$ and $y$:
\beql
M_k &=& (k+1)\, g_{k+1} + \Bigl [\, (k+\frac{1}{2}) x + \frac{1}{2}y \, \Bigr ] 
\, (k+2)\, g_{k+2} + \ldots
\zeile & & \zeile
J_k &=& (k+1)\, g_k + \Bigl [\, \frac{1}{2} x + (k+\frac{1}{2}) y \, \Bigr ] \, 
(k+2)\, g_{k+2} + \ldots
\zeile & & \zeile
\X_k &=& (\, 1 -\d_{1,k}\, )\, \x_{k-3/2} +  \Bigl [\, (k-\frac{1}{2}) x -
\frac{1}{2}y \, \Bigr ] \, \x_{k-1/2} + \ldots
\zeile & & \zeile
\L_k &=& (\, 1 -\d_{1,k}\, )\, \x_{k-3/2} + \Bigl [\, (k-\frac{1}{2}) y
 -\frac{1}{2}x\,\Bigr ] \, \x_{k-1/2} + \ldots
 \eeql[momentsexplicit]
The main motivation for introducing these new variables is that, for each 
term in the genus expansion of the free energy and the correlators, 
the dependence on an
infinite number of coupling constants arranges itself nicely into a function
of a {\it finite} number of moments.
\par
We further introduce the basis functions $\c ^{(n)}(p)$
and $\Y^{(n)}(p)$ recursively
\beql
\c ^{(n)}(p) &= &
\frac{1}{M_1}\, \Bigl ( \, \f^{(n)}_x(p)\, - \,\sum_{k=1}^{n-1} \c^{(k)}
(p)\, M_{n-k+1}\, \Bigr ) , \zeile
\Y^{(n)}(p) &= &\frac{1}{J_1}\,
\Bigl ( \, \f^{(n)}_y(p)\, - \,\sum_{k=1}^{n-1} \Y^{(k)}
(p)\, J_{n-k+1} \Bigr ) ,
\eeql
where
\beql
\f^{(n)}_x  (p) & = & (p-x)^{-n}\, [\, (p-x)(p-y)\, ]^{-1/2}, \zeile
\f^{(n)}_y  (p) & = & (p-y)^{-n}\, [\, (p-x)(p-y)\, ]^{-1/2},
\eeql
following ref.\ \cite{Amb93}.
\par
It is easy to show that for the linear operator 
\Math{\widehat{\schnorkel V}^\prime} defined by
\beq
\Vop f(p) = \cint{\w}\, \frac{V^\prime (\w )}{p-\w}\,
f(\w ) - u_0(p)\, f(p)
\eeq[opVprime]
and appearing in the superloop equations we have
\beql
\Vop \c ^{(n)}(p) &=& \frac{1}{(p-x)^n}, \quad n\geq 1, 
\zeile
\Vop  \Y^{(n)}(p) &=& \frac{1}{(p-y)^n}, \quad n\geq 1. 
\eeql[VprimePsi]
Moreover, \Math{\f^{(0)}_x(p)=\f^{(0)}_y(p)\equiv \f^{(0)}(p)} 
lies in the kernel of \Math{\widehat{\schnorkel V}^\prime}.
\Subsubkapitel{Solution for $\widehat{u}_0$ and $\c$}
Next consider the order 2 equation \gl{order2} at genus 0
\beql
\Vop \widehat{u}_0 &=& \frac{1}{2}\,
\frac{d}{dp}\,\cint{\w}\, \frac{\Y (\w )}{p-\w}\, v_0(\w ) \, -\,
\cint{\w}\, \frac{\Y^\prime (\w )}{p-\w }\, v_0(\w ) \zeile && \qquad
 - \frac{1}{2}\, v_0(p)\, \frac{d}{dp}\, v_0(p) .
\eeql[order2genus0]
Plugging  eq.\ \gl{v0} into the right hand side of this equation 
yields after a somewhat lengthy calculation
\beq
\Vop \widehat{u}_0 \, =\,
\frac{1}{2}\, \frac{\X_2\, (\X_1 -\c )}{(x-y)}\, \frac{1}{p-x}\, - \,
\frac{1}{2}\, \frac{\L_2\, (\L_1 -\c )}{(x-y)}\, \frac{1}{p-y}.
\eeq[page20g]
With eq.\ \gl{VprimePsi} this immediately tells us that
\beq
\widehat{u}_0(p) = \frac{1}{2}\, \frac{\X_2\, (\X_1 -\c )}{(x-y)}
\,  \c ^{(1)}(p) \,- \,
\frac{1}{2}\, \frac{\L_2\, (\L_1 -\c )}{(x-y)}\,\Y^{(1)}(p).
\eeq[uhat0first]
There can be no contributions proportional to the zero mode \Math{
\f^{(0)}(p)}, as we know that \Math{\widehat{u}(p)} behaves as
\Math{{\cal O}(p^{-2})} for \Math{p \ra \infty}.
\par
Finally we determine the odd constant $\c$. This is done by employing the
\hbox{order 3} equation \gl{order3} for $g=0$, i.e.
\beq
\cint{\w}\, \frac{\Y (\w )}{p-\w}\,\widehat{u}_0(\w ) \, -\, v_0 (p)\,
\widehat{u}_0(p)=0.
\eeq[order3genus0]
After insertion of eqs.\ \gl{v0} and \gl{uhat0first} one can show that
\beq
0=\Yop \widehat{u}_0(p) =\frac{1}{2}\,
\frac{ \X_2\, (\X_1 -\c )\, (\L_1 -\c )}{M_1\, (x-y)^{3}\, (p-y)}\,
-\, \frac{1}{2}\,
\frac{ \L_2\, (\L_1 -\c )\, (\X_1 -\c )}{J_1\, (x-y)^{3}\, (p-x)},
\eeq[cresult]
where we have defined the linear operator $\widehat{\eckig \Y}$ by
\beq
\Yop f(p) = \cint{\w}\, \frac{\Y (\w )}{p-\w}\, f(\w) - v_0(p)\, f(p),
\eeq[Psiop]
in accordance to \Math{\widehat{\schnorkel V}^\prime}.
The result \gl{cresult} lets us finally read off the coefficient $\c$ as
\beq
\c \, = \, \frac{1}{2}\, (\, \X_1 \, + \, \L_1\, ).
\eeq[c]
Putting it all
together, we may now write down the complete genus 0 solution for
the one--superloop correlators \Math{W(\, \mid p)} and 
\Math{W(p\mid \,)}:
\beql
W_0(\,\mid p) &=&\cint{\w}\, \frac{V^\prime (\w )}{p-\w}\,
\biggl [ \, \frac{(p-x)(p-y)}{(\w -x)(\w -y)}\, \biggr ]^{1/2}\,
\phantom{+\,
\frac{1}{4}\, \frac{\X_2\, \XL}{M_1\, (x-y)}\, \f^{(1)}_x(p)} \label{W0p_0}
\zeile & & \: +\,
\frac{1}{4}\, \frac{\X_2\, \XL}{M_1\, (x-y)}\, \f^{(1)}_x(p)
+\, \frac{1}{4}\, \frac{\L_2\, \XL}{J_1\, (x-y)}\, \f^{(1)}_y(p)\zeile
&& \label{Wp0_0} \\
W_0(p\mid \, ) &=& \cint{\w}\, \frac{\Y (\w )}{p-\w}\, \biggl [\,
\frac{(\w -x)(\w -y)}{(p-x)(p-y)}\, \biggr ] ^{1/2} \, +\, \frac{1}{2}\,
\frac{\X_1\, +\, \L_1}{[\, (p-x)(p-y)\, ]^{1/2}}. \nonumber
\eeql
One can show that this solution is equivalent to the less compact one 
obtained in ref.\ \cite{Alv2}.
\par
We shall make use of the following rewriting of the purely bosonic part of
$W_0(\,\mid p)$ 
\beq
u_0(p) = V^\prime (p) \, -\, \frac{1}{2}\, [(p-x)(p-y)]^{1/2}\,
\sum_{q=1}^\infty\,
\Bigl\{ (p-x)^{q-1}\, M_q\, +\, (p-y)^{q-1}\, J_q \Bigr\},
\eeq[u0alternative]
derived by deforming the contour integral in eq.\ \gl{W0p_0} into one
surrounding the point $p$ and the other encircling infinity.
To take the residue at infinity one rewrites \Math{(p-\w)^{-1}} as
\beq
\frac{1}{p-\w} = \frac{1}{2}  \frac{1}{(p-x)-(\w-x)}  + \frac{1}{2}
\frac{1}{(p-y)-(\w-y)},
\eeq[trick]
and expands in \Math{\Bigl( \frac{p-x}{\w-x}\Bigr )}  and
\Math{\Bigl( \frac{p-y}{\w-y}\Bigr )} respectively. Doing this for
the fermionic \Math{W_0(p\mid \, )} yields
\beq
v_0(p)=\Y (p) -
\, \frac{1}{2}\, [(p-x)(p-y)]^{-1/2}\, \sum_{q=2}^\infty\,
\Bigl\{ (p-x)^{q-1}\, \X_q\, +\, (p-y)^{q-1}\, \L_q \Bigr\}.
\eeq[v0alternative]
It is important to realize that the bracketed terms in eq.\
\gl{u0alternative}
as well as in eq.\ \gl{v0alternative} are actually identical.
Here we see that the planar solution is special in the sense that it depends on
the full set of moments. Interestingly enough this is not the case for higher
genera.
\Subkapitel{The Iterative Procedure}{IteraticeProc}
Our iterative solution of the superloop equations results in a certain
representation of the free energy and the loop correlators in terms of the
moments and basis functions defined in section \Sub{PlanSol}.
 We will show that it
suffices to know $u_g(p)$ and $v_g(p)$ only up to a zero mode in order to
calculate $F_g$. We give explicit results for genus one.
\Subsubkapitel{The Iteration for $u_g$ and $v_g$}
The correlators $u_g(p)$ and $v_g(p)$ are determined by the order 0 
and order 1 equations \gl{order0} and \gl{order1} after insertion of 
the genus expansions \gl{genusexpWnm} of these operators. We find
\beqx
\Vop u_g(p) =
\frac{1}{2}\, \sum_{g^\prime=1}^{g-1}\, u_{g^\prime}(p)\, u_{g-g^\prime}(p)
\, \eeqx
\beq +\, \frac{1}{2}\, \dV{p}\, u_{g-1}(p)\, 
-\, \frac{1}{2}\, \frac{d}{dq}\,\dY{p}\,
v_{g-1}(q)\Bigr | _{p=q}
\eeq[order0genusg]
and
\beq
\Vop  v_g(p)  = -\Yop u_g(p)\, +\,  \sum_{g^\prime=1}^{g-1}\, v_{g^\prime}(p)
\, u_{g-g^\prime}(p) \, +\, \dV{p}\, v_{g-1}(p)
\eeq[order1genusg]
at genus \Math{g\geq 1}.
From the structure of these equations we directly deduce that
\Math{u_g(p)} and \Math{v_g(p)} will be linear combinations of the 
basis functions \Math{\c^{(n)}(p)}
and \Math{\Y^{(n)}(p)}. By eq.\  \gl{VprimePsi}
the coefficients of this linear
combination may be read off the poles \Math{(p-x)^{-k}} and 
\Math{(p-y)^{-k}} of
the right hand sides of eqs. \gl{order0genusg}
and \gl{order1genusg} after a partial fraction decomposition.
\par
Let us demonstrate how this works for $g=1$.
According to eq.\ \gl{order0genusg} for $u_1(p)$
we first calculate \Math{\d\, u_0/\d V(p)}. 
We then need to know the derivatives
\Math{\d\, x/\d V(p)} and \Math{\d\, y/ \d V(p)}. 
They can be obtained from eq.\ \gl{determinexy} and read
\beq
\frac{\d \, x}{\d V(p)}= \frac{1}{M_1}\, \f^{(1)}_x (p), \quad
\frac{\d \, y}{\d V(p)}= \frac{1}{J_1}\, \f^{(1)}_y (p).
\eeq[dVxy]
Using the relation
\beq
\dV{p}\, V^\prime (\w ) = \frac{d}{dp}\, \frac{1}{p-\w}
\eeq[dVVprime]
one finds \footnote{Of course this is nothing but \Math{2\, W_0(p,p)} of
the Hermitian matrix model of eq.\ \gl{W0PP}.}
\beq
\dV{p}\, u_0(p) = \frac{1}{8}\, \px{2} + \frac{1}{8}\,\py{2}
- \frac{1}{4\, d}\, \px{} + \frac{1}{4\, d}\, \py{},
\eeq[dVu0]
where $d= x-y$. 
\par
Next we determine \Math{\d\, v_0(q)/\d \Y (p)}.  Using the relation
\beq
\dY{p}\, \Y (q) = -\, \frac{1}{p-q}
\eeq[dYY]
and the result
\beql
\frac{\d\,
\X_k}{\d \Y (p)} &=& \d_{k1}\, -\,  \frac{[\, (p-x)(p-y)\, ]^{1/2}}{ (p-x)^k}
\label{dYX} \zeile
\frac{\d\, \L_k}{\d\Y (p)}
&=& \d_{k1}\, - \, \frac{[\, (p-x)(p-y)\, ]^{1/2}}{ (p-y)^k}
\eeql[dYL]
for \Math{k \geq 1}, one finds
\beq
\frac{d}{dq}\, \dY{p}\, v_0(q)\Bigr |_{p=q} = - \dV{p} \, u_0 (p).
\eeq[dqdYv0]
This enables us to write down \Math{u_1(p)},
\beq
u_1 (p) = \frac{1}{8}\, \c^{(2)} (p) + \frac{1}{8}\, \Y^{(2)}(p)
-\frac{1}{4\, d}\, \c^{(1)} +\frac{1}{4\, d}\, \Y^{(1)}(p).
\eeq[u1]
Note that up to the overall factor of two this is identical to the one--loop
correlator of
the hermitian matrix model \cite{Amb93}, as it has to be due to eq.\ 
\gl{Frel}.
\par
Now we solve eq.\ \gl{order1genusg} at \Math{g=1} for \Math{v_1(p)}. 
It is important to
realize that generally eq.\ \gl{order1genusg} fixes \Math{v_g(p)} only up to a
zero mode
contribution \Math{\k_g\, \f^{(0)}(p)}. This comes from the fact that, unlike
for the bosonic $u(p)$, we do not know the coefficient of the 
\Math{p^{-1}} term for \Math{v(p)}. The zero mode coefficient 
\Math{\k_g} will be fixed later on by  requiring
\Math{v_g(p)} to be a total derivative of the free energy $F_g$.
\par
In order to calculate \Math{\d v_0/ \d V(p)} we make use of the relation
\beql
\frac{\d\, \X_k}{\d V(p)} &=& (k-\frac{1}{2})\, \X_{k+1}\,
\frac{1}{M_1}\, \f^{(1)}_x (p)\zeile && \quad
 +\frac{1}{2}\, \Bigl [ \, \sum_{r=2}^k \frac{\X_r}{(-d)^{1+k-r}} +
\frac{\X_1 -\L_1}{(-d)^k} \, \Bigr ]\, \frac{1}{J_1}\, \f^{(1)}_y (p),
\eeql[dVXi]
as well as \Math{\d\, \L_k/\d V(p)} obtained from the above by 
the replacements \Math{x \lra y}, \Math{M_k \lra J_k}, 
\Math{\X_k \lra \L_k} and \Math{d \ra -d}.
The derivatives \Math{\d\, M_k/\d V(p)} and \Math{\d\, J_k/\d V(p)} 
were calculated in ref.\ \cite{Amb93}
\beql
\frac{\d \, M_k}{\d V(p)} &=& -\frac{1}{2}\, (p-x)^{-k-1/2}(p-y)^{-3/2} -
                (k+1/2)\, \f^{(k+1)}_x (p) \zeile
&& + \frac{1}{2}\,
\Bigl [ \, \frac{1}{(-d)^k} - \sum_{i=1}^k \frac{1}{(-d)^{k-i+1}} \,
\frac{M_i}{J_1}\, \Bigr ] \, \f^{(1)}_y (p)
\zeile
& & + (k+1/2)\, \frac{M_{k+1}}{M_1}\, \f^{(1)}_x (p), 
\eeql[dMdV]
and $\d\, J_k/\d V(p)$ is obtained by the usual replacements. Using these
and the earlier results one has
\beql
\frac{\d\, v_0}{\d V(p)}
 = W_0(p\mid p) &=& \bra{ - \frac{\XL}{4\, d\, M_1}}\, \px{3} \, +\,
  \bra{- \frac{\XL}{4\, d\, J_1} }\, \py{3} \zeile  & &+
  \bra{\frac{\X_2}{4\, d\, M_1}}\, \px{2}\,
  +\,\bra{- \frac{\L_2}{4\, d\, J_1}}\, \py{2}  \zeile  &  & +
  \bra{\frac{\L_2}{4\, d^2\, J_1}- \frac{\X_2}{4\, d^2\, M_1}}\, \px{}
  \zeile & & +
  \bra{\frac{\X_2}{4\, d^2\, M_1} -
  \frac{\L_2}{4\, d^2\, J_1}}\, \py{}. 
\eeql[Wpp]
For the evaluation of the right hand side of eq.\ \gl{order1genusg} at
genus $g$ we also need to know how the operator 
\Math{\widehat{\eckig \Y}} acts on the functions \Math{\f^{(n)}_x(p)} 
and \Math{\f^{(n)}_y(p)}, in terms of which \Math{u_g(p)} is given. 
A straightforward calculation yields
\beql
\Yop \f^{(n)}_x (p) &=&
\sum_{k=1}^{n+1}\, \px{k}\, \biggl [ \, -\frac{\XL}{2\,
(-d)^{n+2-k}} \, -\, \sum_{l=2}^{n+2-k} \frac{\X_l}{(-d)^{n+3-k-l}} \, \biggr ]
\zeile && \quad +\, \py{}\, \biggl[ \, -\frac{\XL}{2\, (-d)^{n+1}}\, \biggr ] 
\eeql[Yopfxn]
as well as the
analogous expression for \Math{\Yop \f^{(n)}_y(p)}
obtained from eq.\ \gl{Yopfxn} by the replacements
\Math{x \lra y}, \Math{M_k \lra J_k}, \Math{\X_k \lra \L_k} and 
\Math{d \ra -d}.
\par
We now have collected
all the ingredients needed to evaluate the right hand side
of eq.\ \gl{order1genusg}. After a partial fraction decomposition we
may read off the poles at $x$ and $y$, and therefore obtain the coefficients
of the linear combination in the basis functions. We arrived at
the result for $g=1$ with the aid of {\it Maple}, namely
\beq
v_1(p) =\sum_{i=1}^3\Bigl ( \, B^{(i)}_1\, \c^{(i)}(p) \, +\, E^{(i)}_1\,
\Y^{(i)}(p)\, \Bigr ) \, + \, \k_1\, \f^{(0)}(p),
\eeq[v1]
where the coefficients \Math{B^{(i)}_1} and \Math{E^{(i)}_1} are given by
\beql
B^{(1)}_1 & = & -\frac{1}{8}\, \frac{\X_3}{d\, M_1} +
\frac{1}{8}\, \frac{\X_2}{d^2\, M_1} + \frac{1}{4}\,\frac{\L_2}{d^2\, J_1}
\zeile
& & +\frac{1}{8}\,\frac{M_2\, \X_2}{d\, {M_1}^2}
-\frac{1}{16}\,\frac{M_2\,\XL}{d^2\, {M_1}^2} +\frac{1}{16}\,\frac{J_2\,\XL}
{d^2\, {J_1}^2} \zeile
& & -\frac{3}{16}\,
\frac{\XL}{d^3\, M_1} - \frac{3}{16}\, \frac{\XL}{d^3\, J_1},
\zeile  &&\zeile
B^{(2)}_1 &=&
\frac{1}{8}\,\frac{\X_2}{d\, M_1} + \frac{1}{16}\,\frac{M_2\, \XL}
{d\, {M_1}^2} + \frac{3}{16}\,\frac{\XL}{d^2\, M_1}, \zeile
&& \zeile
B^{(3)}_1 &=& -\frac{5}{16}\,\frac{\XL}{d\, M_1}, 
\eeql[v1coeffs]
and \Math{E^{(i)}_1 = B^{(i)}_1 ( M\lra J, \X \lra \L, d\ra -d)}.
\par
Yet $\k_1$ is still undetermined. To compute it and the remaining doubly
fermionic part $\widehat{u}_g(p)$ of the loop correlator $W_g(\,\mid p)$ one
can employ the order 2 and order 3 eqs.\ \gl{order2} and \gl{order3}
at genus $g$. It is, however, much easier to construct the free energy 
$F_g$ at this stage from our knowledge of $u_g(p)$ and $v_g(p)$.
\Subsubkapitel{The Computation of $F_g$ and $\k_g$}
As mentioned earlier, the free energy of the supereigenvalue model depends
at most quadratically on the fermionic coupling constants.
In this subsection we present an algorithm which
allows us to determine $F_g$ and $\k_g$ as soon as the results for 
$u_g(p)$ and $v_g(p)$ (up to the zero mode coefficient $\k_g$) are known.
\par
The purely bosonic part of the free energy $F_g$ is just twice the 
free energy of the hermitian matrix model. 
By using the results of Ambj\o rn et al.\ \cite{Amb93} one may 
then compute the bosonic part of $F_g$ from $u_g(p)$. Here one
rewrites $\c^{(n)}(p)$ and $\Y^{(n)}(p)$ as derivatives with respect
to $V(p)$. One easily verifies the following relations for
two lowest basis functions:
\beql
\c^{(1)}(p) &=& \frac{\d\, x}{\d V(p)}, \zeile && \zeile
\Y^{(1)}(p) &=& \frac{\d\, y}{\d V(p)}, \zeile && \zeile
\c^{(2)}(p) &=& \frac{\d}{\d V(p)}( -\frac{2}{3}\ln M_1- \frac{1}{3}\ln d\, ),
\zeile && \zeile
\Y^{(2)}(p) &=& \frac{\d}{\d V(p)}( -\frac{2}{3}\ln J_1- \frac{1}{3}\ln d\, ).
\eeql[c1c2]
Combining this with the result for $u_1(p)$ of eq.\ \gl{u1} yields the
purely bosonic piece of $F_1$
\beq
F^{\mbox{\scriptsize bos}}_1= -\frac{1}{12}\, \ln M_1 - \frac{1}{12}\,
\ln J_1 - \frac{1}{3}\, \ln d,
\eeq[F1bos]
which as expected is just twice the result for the Hermitian matrix model
(cf.\ \cite{Amb93}).
\par 
For the higher basis functions the situation is not as simple. However, a
rewriting of the basis functions allows one to identify $u_g(p)$ 
as a total derivative. For $\c^{(n)}(p)$ one uses 
the recursive form \cite{Amb93}
\beql
\c^{(n)}(p) &=& \frac{1}{M_1}\Bigl [ \, \frac{1}{2n-1}\, \sum_{i=1}^{n-1}
(-d)^{i-n}\, \Bigl ( \f^{(i)}_x (p)- M_i\, \frac{\d \, y}{\d V(p)}\, \Bigr )\zeile
&& - \frac{2}{2n-1}\, \frac{\d \, M_{n-1}}{\d V(p)} - \sum_{k=2}^{n-1}
\c^{(k)}(p)\, M_{n-k+1}\, \Bigr ] \qquad n\geq 2,
\eeql[cnrew]
where $\f^{(n)}_x$ should be rewritten as
\beql
\f^{(n)}_x(p) &=& \frac{1}{2n-1}\sum_{i=1}^{n-1}(-d)^{i-n}\, 
\Bigl ( \f^{(i)}_x(p)
-M_i\, \frac{\d\, y}{\d V(p)}\, \Bigr ) \zeile
&& + M_n\, \frac{\d\, x}{\d V(p)} - \frac{2}{2n-1}\, \frac{\d\, M_{n-1}}
{\d V(p)} \qquad n\geq 2 \zeile && \zeile
\f^{(1)}_x(p) &=& M_1\, \frac{\d \, x}{\d V(p)}.
\eeql[fxrew]
The rewriting of $\Y^{(n)}(p)$ follows from the above by performing the usual
replacements \Math{J \lra M} and \Math{d \ra -d}.
\par
The strategy for the part of $F_g$ quadratic in fermionic couplings
consists in rewriting
$v_g(p)$ as a total derivative in the fermionic potential
$\Y (p)$. We know that eq.\  \gl{dYL} implies
\beq
\frac{\d\,\XL}{\d\Y (p)} = - d\, \f^{(0)}(p)
\eeq[dYXL]
and
\beq
\frac{\d\, \X_k}{\d\Y (p)} = -\,\vq^{(k)}_x(p), \quad
\frac{\d\,\L_k}{\d\Y (p)}=-\vq^{(k)}_y(p),\quad
k\geq 2
\eeq[dYXkLk]
where
\beql
\vq^{(k)}_x(p)&=&
(p-x)^{-k}\, [\, (p-x)(p-y)\, ]^{1/2} \quad k \geq 1, \zeile
\vq^{(k)}_y(p)&=&
(p-y)^{-k}\, [\, (p-x)(p-y)\, ]^{1/2} \quad k \geq 1. 
\eeql[vfdef]
Let us again specialize to \Math{g=1}.
Using the above we can reexpress eq.\ \gl{v1} as
\beqx
\dY{p}\, F_1 \, +\, \k_1 \, \frac{1}{d}\, \dY{p}\,\XL = \sum_{i=1}^3
\Bigl ( \, B^{(i)}_1\, \c^{(i)}(p) \, +\, E^{(i)}_1\, \Y^{(i)}(p)\, \Bigr )
\eeqx
\beq
= \sum_{r=2}^4 \Bigl ( \, \b^{(r)}_1\, \vq^{(r)}_x(p) \, +\, \e^{(r)}_1\,
\vq^{(r)}_x(p)\, \Bigr )\, +\, \g_1\, [\, \vq^{(1)}_x(p)-\vq^{(1)}_y(p)\, ],
\eeq[p49]
with the new coefficients $\b^{(r)}_1$, $\e^{(r)}_1$ and $\g_1$ completely
determined by the known coefficients $B^{(i)}_1$
and $E^{(i)}_1$. As the new functions
\Math{\vq^{(r)}_x(p)} and \Math{\vq^{(r)}_y(p)} are total derivatives 
in $\Y (p)$, this equation allows us to calculate $\k_1$ and $F_1$.
It is obvious that this method works for higher $g$ as well.
\par
With the help of {\it Maple} the zero mode coefficient $\k_1$
of $v_1(p)$ becomes
\beql
\k_1 &=&
{\frac {11\,\X_{{2}}}{16\,{d}^{3}{M_{{1}}}^{2}}}
-{\frac {11\,\L_{{2}}}{16\,{d}^{3}{J_{{1}}}^{2}}}
+{\frac {5\,\L_{{2}}J_{{2}}}{8\,{d}^{2}{J_{{1}}}^{3}}}
+{\frac {5\,\X_{{2}}M_{{2}}}{8\,{d}^{2}{M_{{1}}}^{3}}}
-{\frac {5\,\X_{{2}}M_{{3}}}{16\,d{M_{{1}}}^{3}}} \zeile &&
+{\frac {5\,\L_{{2}}J_{{3}}}{16\,d{J_{{1}}}^{3}}}
-{\frac {\X_{{2}}}{16\,{d}^{3}J_{{1}}M_{{1}}}}
+{\frac {\L_{{2}}}{16\,{d}^{3}J_{{1}}M_{{1}}}}
-{\frac {\X_{{2}}J_{{2}}}{16\,{d}^{2}{J_{{1}}}^{2}M_{{1}}}}
-{\frac {\L_{{2}}M_{{2}}}{16\,{d}^{2}{M_{{1}}}^{2}J_{{1}}}} \zeile &&
+{\frac {3\,\X_{{2}}{M_{{2}}}^{2}}{8\, d{M_{{1}}}^{4}}}
-{\frac {3\,\L_{{2}}{J_{{2}}}^{2}}{8\,d{J_{{1}}}^{4}}}
+{\frac {3\,\L_{{3}}J_{{2}}}{8\,d{J_{{1}}}^{3}}}
-{\frac {3\,\X_{{3}}M_{{2}}}{8\,d{M_{{1}}}^{3}}}
+{\frac {5\,\X_{{4}}}{16\,d{M_{{1}}}^{2}}}\zeile &&
-{\frac {5\,\L_{{4}}}{16\,d{J_{{1}}}^{2}}}
-{\frac {5\,\X_{{3}}}{8\,{d}^{2}{M_{{1}}}^{2}}}
-{\frac {5\,\L_{{3}}}{8\,{d}^{2}{J_{{1}}}^{2}}}
+{\XL}\,
\Bigl \{ {\frac {3\,{J_{{2}}}^{2}}{8\,{d}^{2}{J_{{1}}}^{4}}} \zeile &&
     -{\frac {3\,{M_{{2}}}^{2}}{8\,{d}^{2}{M_{{1}}}^{4}}}
     -{\frac {5\,M_{{2}}}{8\,{d}^{3}{M_{{1}}}^{3}}}
     -{\frac {5\,J_{{2}}}{8\,{d}^{3}{J_{{1}}}^{3}}}
     +{\frac {5\,M_{{3}}}{16\,{d}^{2}{M_{{1}}}^{3}}}
     -{\frac {5\,J_{{3}}}{16\,{d}^{2}{J_{{1}}}^{3}}} \zeile &&
     +{\frac {11}{16\,{d}^{4}{J_{{1}}}^{2}}}
     -{\frac {11}{16\,{d}^{4}{M_{{1}}}^{2}}}
     +{\frac {M_{{2}}}{16\,{d}^{3}J_{{1}}{M_{{1}}}^{2}}}
     +{\frac {J_{{2}}}{16\,{d}^{3}{J_{{1}}}^{2}M_{{1}}}} \Bigr \}
\eeql[kappa]
and the doubly fermionic part of $F_1$ is constructed as well.
The result for the full free energy at genus 1 then reads
\beql
F_1 &=& -\frac{1}{12}\,
\ln\, M_1 \,-\,\frac{1}{12}\,\ln\, J_1\, -\,\frac{1}{3}\,
\ln\, d \zeile &&
-\XL\, \Bigl \{ \,
  {\frac {11\,\X_{{2}}}{16\,{d}^{4}\,{M_{{1}}}^{2}}}
-{\frac {11\,\L_{{2}}}{16\,{d}^{4}\,{J_{{1}}}^{2}}}
+{\frac {5\,\L_{{2}}\, J_{{2}}}{8\,{d}^{3}\,{J_{{1}}}^{3}}}
+{\frac {5\,\X_{{2}}\, M_{{2}}}{8\,{d}^{3}\,{M_{{1}}}^{3}}} \zeile &&
-{\frac {5\,\X_{{2}}\, M_{{3}}}{16\,d^2\,{M_{{1}}}^{3}}}
+{\frac {5\,\L_{{2}}\, J_{{3}}}{16\,d^2\,{J_{{1}}}^{3}}}
-{\frac {\X_{{2}}\, J_{{2}}}{16\,{d}^{3}\,{J_{{1}}}^{2}M_{{1}}}}
-{\frac {\L_{{2}}\, M_{{2}}}{16\,{d}^{3}\, {M_{{1}}}^{2}J_{{1}}}}
\zeile &&
+{\frac {3\,\X_{{2}}\,{M_{{2}}}^{2}}{8\,d^2\,{M_{{1}}}^{4}}}
-{\frac {3\,\L_{{2}}\,{J_{{2}}}^{2}}{8\,d^2\,{J_{{1}}}^{4}}}
+{\frac {3\,\L_{{3}}\, J_{{2}}}{8\,d^2\,{J_{{1}}}^{3}}}
-{\frac {3\,\X_{{3}}\, M_{{2}}}{8\,d^2\,{M_{{1}}}^{3}}}
+{\frac {5\,\X_{{4}}}{16\,d^2\,{M_{{1}}}^{2}}}\zeile &&
-{\frac {5\,\L_{{4}}}{16\,d^2\,{J_{{1}}}^{2}}}
-{\frac {\X_{{3}}}{2\,{d}^{3}\,{M_{{1}}}^{2}}}
-{\frac {\L_{{3}}}{2\,{d}^{3}\,{J_{{1}}}^{2}}}
+{\frac {3\,\X_{{2}}}{16\,{d}^{4}\, J_{{1}}M_{{1}}}}
-{\frac {3\,\L_{{2}}}{16\,{d}^{4}\, J_{{1}}M_{{1}}}}\Bigr \}\zeile &&
+{\frac {\X_{2}\, \X_{{3}}}{8\,{d}^{2}\, {M_{{1}}}^{2}}}
+{\frac {\L_2\, \L_{{3}}}{8\,{d}^{2}\,{J_{{1}}}^{2}}}
-{\frac {\X_2\,\L_2}{4\,{d}^{3}\, J_{{1}}M_{{1}}}}.
\eeql[F1]
The above results hold true for generic bosonic and fermionic potentials.
Note that for symmetric bosonic potentials, i.e.\ \Math{x=-y}, and generic
fermionic potentials one has
\beq
M_k= (-)^{k+1}\, J_k,
\eeq[MJSymm]
easily seen from the definitions of eq.\ \gl{MJDef}. If one  takes
the fermionic potential to be symmetric as well one finds
\beq
\X_k = (-)^k\, \L_k \qquad k\geq 2,
\eeq[XLSymm]
from the defining equation \gl{XLDef}. Then one checks
\Math{F^{\mbox{\scriptsize ferm}}_1=0} and this continues to hold true
for higher genera as well, seen from an analysis of the odd superloop
equation in the totally symmetric case.
\Subsubkapitel{The Iteration for $\widehat{u}_g(p)$}
The remaining part of the loop correlator \Math{W(\,\mid p)} is now 
easily derived from $F_g$ by applying the loop insertion operator 
\Math{\d/\d V(p)} to its doubly fermionic part.
For genus 1 the result is
\beq
\widehat{u}_1(p)= \sum_{i=1}^4
\Bigl ( \, \widehat{A}^{(i)}_1\, \c^{(i)}(p) \, +\,
\widehat{D}^{(i)}_1\, \Y^{(i)}(p)\, \Bigr ) ,
\eeq[uhat1]
where
\beql
\widehat{A}^{(4)}_1 &=& -\frac{35\, \XL\, \X_2}{32\, d^2\, {M_1}^2}
\zeile && \zeile
\widehat{A}^{(3)}_1 &=& \XL\, \Bigl \{ \,
                -\frac{15\,  \X_3}{16\, d^2\,{M_1}^2} +
                \frac{45\,\X_2}{32\, d^3\, {M_1}^2} +
                \frac{25\,\X_2\, M_2}{32\, d^2\, {M_1}^3} -
                \frac{5\,\L_2}{32\, d^3\, J_1\, M_1} \, \Bigr \}
\zeile & & \zeile
\widehat{A}^{(2)}_1 &=&
{\frac {21\,\XL\,\X_{{3}}}{16\,{d}^{3}{M_{{1}}}^{2}}}
-{\frac {3\,\XL\,\X_{{2}}{M_{{2}}}^{2}}{8\,{d}^{2}{M_{{1}}}^{4}}}
+{\frac {3\,\XL\,\X_{{2}}J_{{2}}}{32\,{d}^{3}{J_{{1}}}^{2}M_{{1}}}}
+{\frac {\X_{{2}}\,\X_{{3}}}{8\,{d}^{2}{M_{{1}}}^{2}}} \zeile &&
+{\frac {3\,\L_{{2}}\,\X_{{2}}}{8\,J_{{1}}M_{{1}}{d}^{3}}}
+{\frac {3\,\XL\,\X_{{3}}M_{{2}}}{4\,{d}^{2}{M_{{1}}}^{3}}}
-{\frac {15\,\XL\,\X_{{4}}}{16\,{d}^{2}{M_{{1}}}^{2}}}
-{\frac {\X_3\,\X_{{2}}}{4\,{d}^{2}{M_{{1}}}^{2}}} \zeile &&
-{\frac {51\,\XL\,\X_{{2}}}{32\,{d}^{4}{M_{{1}}}^{2}}}
-{\frac {\L_{{2}}\,\XL\, M_{{2}}}{32\,J_{{1}}{d}^{3}{M_{{1}}}^{2}}}
-{\frac {\L_2\,\XL}{4\,J_{{1}}{d}^{4}M_{{1}}}}\zeile &&
-{\frac {33\,\XL\,\X_{{2}}M_{{2}}}{32\,{d}^{3}{M_{{1}}}^{3}}}
+{\frac {5\,\XL\,\X_{{2}}M_{{3}}}{16\,{d}^{2}{M_{{1}}}^{3}}}
-{\frac {9\,\XL\,\X_{{2}}}{32\,J_{{1}}{d}^{4}M_{{1}}}}
\zeile & & \zeile
\widehat{A}^{(1)}_1 &=&
{\frac {33\,\XL\,\X_{{2}}M_{{2}}}{32\,{d}^{4}{M_{{1}}}^{3}}}
-{\frac {\XL\,\X_{{2}}M_{{2}}}{32\,{d}^{4}J_{{1}}{M_{{1}}}^{2}}}
-{\frac {35\,\XL\,\X_{{5}}}{32\,{d}^{2}{M_{{1}}}^{2}}} \zeile &&
-{\frac {3\,\XL\,\X_{{2}}{J_{{2}}}^{2}}{16\,{d}^{3}{J_{{1}}}^{4}}}
+{\frac {15\,\XL\,\X_{{4}}M_{{2}}}{16\,{d}^{2}{M_{{1}}}^{3}}}
+{\frac {45\,\XL\,\X_{{4}}}{32\,{d}^{3}{M_{{1}}}^{2}}}\zeile &&
+{\frac {5\,\XL\,\X_{{2}}}{16\,{d}^{5}J_{{1}}M_{{1}}}}
-{\frac {7\,\XL\,\L_{{3}}}{32\,{d}^{4}{J_{{1}}}^{2}}}
-{\frac {\X_2\,\L_{{3}}}{16\,{d}^{3}{J_{{1}}}^{2}}}
+{\frac {\X_3\,\X_{{2}}}{16\,{d}^{3}{M_{{1}}}^{2}}} \zeile &&
+{\frac {51\,\XL\,\X_{{2}}}{32\,{d}^{5}{M_{{1}}}^{2}}}
-{\frac {51\,\XL\,\X_{{3}}}{32\,{d}^{4}{M_{{1}}}^{2}}}
-{\frac {\X_2\,\X_{{3}}}{8\,{d}^{3}{M_{{1}}}^{2}}}
+{\frac {5\,\X_2\,\X_{{4}}}{16\,{d}^{2}{M_{{1}}}^{2}}}\zeile &&
-{\frac {\L_2\,\X_{{2}}}{2\,J_{{1}}M_{{1}}{d}^{4}}}
-{\frac {7\,\L_2\,\XL\, J_{{2}}}{32\,{J_{{1}}}^{3}{d}^{4}}}
+{\frac {3\,\L_2\,\X_{{3}}}{8\,{d}^{3}J_{{1}}M_{{1}}}}
+{\frac {7\,\L_2\,\XL}{16\,{J_{{1}}}^{2}{d}^{5}}}\zeile &&
+{\frac {5\,\XL\,\X_{{2}}\, J_{{2}}}{16\,{J_{{1}}}^{3}{d}^{4}}}
-{\frac {11\,\XL\,\X_{{2}}}{32\,{J_{{1}}}^{2}{d}^{5}}}
-{\frac {5\,\XL\,\X_{{2}}M_{{3}}}{16\,{d}^{3}{M_{{1}}}^{3}}}  \zeile &&
-{\frac {9\,\XL\,\X_{{3}}{M_{{2}}}^{2}}{16\,{d}^{2}{M_{{1}}}^{4}}}
+{\frac {3\,\XL\,\X_{{3}}J_{{2}}}{32\,{d}^{3}{J_{{1}}}^{2}M_{{1}}}}
-{\frac {9\,\XL\,\X_{{3}}}{32\,J_{{1}}M_{{1}}{d}^{4}}} \zeile &&
+{\frac {15\,\XL\,\X_{{3}}M_{{3}}}{32\,{d}^{2}{M_{{1}}}^{3}}}
-{\frac {9\,\XL\,\X_{{3}}M_{{2}}}{8\,{d}^{3}{M_{{1}}}^{3}}}
+{\frac {5\,\XL\,\X_{{2}}J_{{3}}}{32\,{d}^{3}{J_{{1}}}^{3}}} \zeile &&
-\frac{\X_2}{2\, d}\,\biggl ( \,
 -{\frac {11\,\L_{{2}}}{16\,{d}^{3}{J_{{1}}}^{2}}}
+{\frac {5\,\L_{{2}}J_{{2}}}{8\,{d}^{2}{J_{{1}}}^{3}}}
+{\frac {5\,\L_{{2}}J_{{3}}}{16\,d{J_{{1}}}^{3}}}
+{\frac {\L_{{2}}}{16\,{d}^{3}J_{{1}}M_{{1}}}} \zeile &&
-{\frac {\L_{{2}}M_{{2}}}{16\,{d}^{2}{M_{{1}}}^{2}J_{{1}}}}
-{\frac {3\,\L_{{2}}{J_{{2}}}^{2}}{8\,d{J_{{1}}}^{4}}}
+{\frac {3\,\L_{{3}}J_{{2}}}{8\,d{J_{{1}}}^{3}}}
-{\frac {3\,\X_{{3}}M_{{2}}}{8\,d{M_{{1}}}^{3}}}
+{\frac {5\,\X_{{4}}}{16\,d{M_{{1}}}^{2}}} \zeile &&
-{\frac {5\,\L_{{4}}}{16\,d{J_{{1}}}^{2}}}
-{\frac {5\,\X_{{3}}}{8\,{d}^{2}{M_{{1}}}^{2}}}
-{\frac {5\,\L_{{3}}}{8\,{d}^{2}{J_{{1}}}^{2}}}\,\biggr )
+{\frac {3\,\XL\,\X_{{2}}{M_{{2}}}^{2}}{8\,{d}^{3}{M_{{1}}}^{4}}}
\zeile &&
-{\frac {\XL\,\X_{{2}}J_{{2}}}{8\,{d}^{4}{J_{{1}}}^{2}M_{{1}}}}
+{\frac {\L_2\,\XL\, M_{{2}}}{16\,{d}^{4}J_{{1}}{M_1}^{2}}}
+{\frac {11\,\L_2\,\XL}{32\,J_{{1}}{d}^{5}M_{{1}}}} \zeile &&
+\XL\, \Bigl \{\,
     {\frac {3\,{J_{{2}}}^{2}}{8\,{d}^{2}{J_{{1}}}^{4}}}
-{\frac {3\,{M_{{2}}}^{2}}{8\,{d}^{2}{M_{{1}}}^{4}}}
-{\frac {5\,M_{{2}}}{8\,{d}^{3}{M_{{1}}}^{3}}}
-{\frac {5\,J_{{2}}}{8\,{d}^{3}{J_{{1}}}^{3}}} \zeile &&
+{\frac {5\,M_{{3}}}{16\,{d}^{2}{M_{{1}}}^{3}}}
-{\frac {5\,J_{{3}}}{16\,{d}^{2}{J_{{1}}}^{3}}}
+{\frac {11}{16\,{d}^{4}{J_{{1}}}^{2}}}
-{\frac {11}{16\,{d}^{4}{M_{{1}}}^{2}}} \zeile &&
+{\frac {M_{{2}}}{16\,{d}^{3}J_{{1}}{M_{{1}}}^{2}}}
+{\frac {J_{{2}}}{16\,{d}^{3}{J_{{1}}}^{2}M_{{1}}}}
  \, \Bigr \} ,
\eeql[uhat1coeff]
and the analogous expressions for the \Math{\widehat{D}^{(i)}_1} 
obtained from the above by replacing \Math{M\lra J}, 
\Math{\X\lra\L} and \Math{d\ra -d}.
\Subsubkapitel{General Structure of $u_g$, $v_g$, $\widehat{u}_g$ 
and $F_g$}
In the following subsection we deduce
the number of moments and basis functions the
free energy and the superloop correlators at genus $g$ depend on.
\par
Ambj\o rn  et al.\ \cite{Amb93, Amb2} have shown that the free 
energy of the hermitian matrix model depends on $2(3g-2)$ moments. 
This directly translates to \Math{F_g^{\mbox{\scriptsize bos}}}.
Similarly as \Math{u_g=\d\, F^{\mbox{\scriptsize bos}}_g/\d  V(p)} and
with eq.\ \gl{dMdV} we see that $u_g$
contains \mat{2(3g-1)} bosonic moments
and basis functions up to order \mat{(3g-1)}, i.e.
\beq
u_g(p)= \sum_{k=1}^{3g-1}\, A^{(k)}_g\, \c^{(k)}(p) \, +\,
D^{(k)}_g\, \Y^{(k)}(p).
\eeq[structureug]
For the structure of $v_g$ consider the leading--order poles on the right
hand side of
eq.\ \gl{order1genusg}. Label this order by $n_g$, then
with eqs.\ \gl{dVXi}, \gl{dMdV}
and \gl{Yopfxn} the three terms on the right hand side
of eq.\ \gl{order1genusg} give rise to the following poles of leading
order
\beql
\Yop  u_g(p) & : & (3g-1) + 1 \zeile
v_{g^\prime}\, u_{g-g^\prime} & : & n_{g^\prime} + \Bigl (\, 3 \,
(g-g^\prime) -1\,
\Bigr ) \, +\, 1 \zeile
\dV{p}\, v_{g-1} & : & n_{g-1} \, +\, 3 .
\eeql[ng]
From the above we deduce that \Math{n_g=3\, g}, and therefore
\beq
v_g(p)= \sum_{k=1}^{3g}\, B^{(k)}_g\, \c^{(k)}(p) \, +\, E^{(k)}_g\,
\Y^{(k)}(p) \, + \,\kappa_g\, \phi^{(0)}(p).
\eeq[structurevg]
As the highest bosonic moments in $v_g$ come from the highest--order
basis functions, we see that $v_g$ depends on $2(3\, g)$ bosonic moments.
To find the dependence on
the number of fermionic moments recall eq. \gl{dYL}. In
order for $v_g$ to have a leading contribution of \mat{\c^{(3g)}(p)} the
fermionic part of the free energy \Ffg must contain
$\X_{3g+1}$. We are thus led to the conclusion that \Ffg and $v_g$
both depend on \mat{2(3\, g+1)} fermionic moments
\footnote{In some sense this is counterintuitive: In ref.\ \cite{Amb93}
it was shown that there is a relation of the dependence on the number
of moments to the dimension of the moduli space of a Riemann surface
of genus $g$, which is \Math{3g-3}. So one could have speculated 
that something similar holds true
for the fermionic moments and the number of fermionic moduli of
a super--Riemann surface, which is \Math{2g-2}. However, the discovered 
slope of $3$ in $g$ seems to contradict such an interpretation.}.
As the application of the
loop insertion operator \mat{\d /\d\Y (p)} does not change the number of
bosonic moments, \Ffg must contain \mat{2(3\, g)} bosonic moments.
\par
Knowing the structure of \Ffg then tells us with eqs. \gl{dVXi} and
\gl{dMdV}
that $\widehat{u}_g$ depends on \mat{2(3\, g +1)} bosonic and 
\mat{2(3\, g+2)}
fermionic moments. For genus $g$ it reads
\beq
\widehat{u}_g (p) = \sum_{k=1}^{3g+1} \, \widehat{A}^{(k)}_g \, \c^{(k)}(p)
\, + \, \widehat{D}^{(k)}_g\, \Y^{(k)}(p).
\eeq[strustureuhatg]
The genus $g$ contribution to the \mat{(\, |s)}--superloop correlator
\mat{W_g(\, |p,\ldots ,p)} will then depend on \mat{2(3\, g +s)} bosonic
and \mat{2(3\, g+s+1)} fermionic moments. Similarly the genus $g$
\mat{(1|s)}--superloop correlator \mat{W_g(p|p,\ldots ,p)} is a function
of \mat{2(3\, g + s)} bosonic and \mat{2(3\, g+s+1)} fermionic moments.
\par
This concludes our analysis of the iterative process away from the 
double scaling limit.
\Subkapitel{The Double Scaling Limit}{DScal}
In view of our lack of understanding the supereigenvalue model on a
geometrical basis as a discretization of super--Riemann surfaces it
appears quite ambitious to speak of a continuum limit. However, due to 
the striking similarities to the Hermitian matrix model with its well
understood continuum limit (discussed in chapter I 
section \Sub{discret}) it is natural to export these techniques to the
supersymmetric case. We shall see that again one finds critical values
for the coupling constants $g_k$ and $\x_{k+1/2}$ allowing one to
perform a double scaling limit. 
\par
Similar to the situation in the Hermitian matrix model the ``naive''
\mbox{$N\ra\infty$} limit of the supereigenvalue model is
unsatisfactory as it leaves us only with the planar contributions, easily seen
from eqs.\
\gl{genusexpWnm} and \gl{genusexpF}. Crucial for the double scaling
limit is the observation that there exists a subspace in the space of
couplings \mbox{$\{ g_k,\x_{k+1/2}\}$} where all higher genus contributions
to the free energy \mbox{$F_{g}$} diverge. This enables us to take the
double scaling limit, where one simultaneously approaches the critical
subspace of couplings {\it and} takes \mbox{$N\ra \infty$}, giving
contributions
to the free energy from all genera. Just as in the Hermitian matrix
model Kazakov \cite{Kaz} multicritical points appear, related to extra
zeros of the eigenvalue densities accumulating at one endpoint of their 
support.
\par
For the formulation of this limit the moment technique developed in the
above turns out to be extremely useful. One can then determine
which terms in the explicit solutions for $F_g$, \Math{W(p|\mid \, )}
and \Math{W(\,\mid p)} calculated in the previous section contribute
to the double scaling limit. However, the calculation of these quantities
away from the double scaling limit is rather time consuming,
especially as a large number of terms turn out to be irrelevant in this
limit. In this section we will therefore develop a procedure which
directly produces only the terms relevant in the double scaling limit.
\par
The first thing to do is to study the scaling of the moments and
basis functions by approaching the critical point. 
\Subsubkapitel{Multicritical Points}
The analysis of the scaling behaviour for the bosonic quantities was carried
out by Ambj\o rn et al.\ \cite{Amb93,Amb3} in the framework
of the hermitian matrix model.
Consider the case of generic, i.e.\ non--symmetric, potentials $V(p)$ and 
$\Y(p)$.
The $m$'th multicritical point is reached when the eigenvalue density
\mbox{$(u_0(p)-V^\prime (p))$} of eq.\ \gl{u0alternative}, which under
normal circumstances vanishes as a square root on both ends of its support,
acquires \Math{(m-1)} additional zeros at one end of the cut, say $x$. 
Alternatively one would define the critical point by a diverging free
energy. A glance at our genus one result of eq.\ \gl{F1} tells us that
$F_1$ diverges for vanishing $M_1$ (or $J_1$). The only moments 
appearing in the denominators of higher genus $F_g$'s will be $M_1$ 
and $J_1$ as well. Hence the condition for being at an $m$'th multicritical 
point is taken to be
\beq
M_1^c=M_2^c=\cdots M_{m-1}^c=0, \quad M_k^c\neq 0,\,\,\,\,  k\geq m,
\quad\quad J_l^c\neq 0, \,\,\,\, l\geq 1,
\eeq[MJcritical]
defining a critical subspace in the space of bosonic couplings $g_k$.
Denote by \Math{g_k^c} a particular point in this subspace for which the
eigenvalue
distribution is confined to the interval \Math{[x_c,y_c]}. If we now move 
away from this point the conditions of eq.\
\gl{MJcritical} will no longer be fulfilled and the cut will move to
the interval \Math{[x,y]}. Assume we control this movement by the parameter 
$a$ and set \cite{Amb93}
\beql
x &=& x_c - a\, \L^{1/m} \zeile
p &=& x_c + a\, \p 
\eeql[xscaling]
The scaling of $p$ and the introduction of its scaling variable $\p$ is
necessary in order to speak of the double scaling limit of
the superloop correlators. $\L$ plays the role of the cosmological constant.
We will now deduce the scaling of $y$ by
further assuming that the critical subspace $\{g_k^c\}$ is reached as
\beq
g_k = g\, \cdot\, g_k^c,
\eeq[gscaling]
where \Math{g} is a function of $a$ to be determined. Imposing the boundary
conditions \gl{determinexy} yields
\beq
y-y_c \sim a^m, \quad \mbox{and} \quad  g-1 \sim a^m.
\eeq[ygscale]
\Comment{
This may be seen as follows. Consider the first boundary condition
\gl{determinexy} using \Math{y=y_c+\D y}, eqs.\ \gl{xscaling} and
\gl{gscaling} and the fact that the \Math{M_i^c} for \Math{i\in [1,m-1]}
vanish lead to
\beql
0&=& \cint{\w}\frac{g\,V^\prime_c(\w)}{(\w-x_c+a\, \L^{1/m})^{1/2}\,
(\w-y_c-\D y)^{1/2}} \zeile
&=& g \cint{\w}\frac{g\, V^\prime_c(\w)}{(\w-x_c)^{1/2}\, (\w-y_c)^{1/2}}
[\sum_{i=0}^\infty c_i\, \frac{(a\L^{1/m})^i}{(\w-x_c)^i}\, ]\,
[\sum_{i=0}^\infty c_i\, \frac{(\D y)^i}{(\w-y_c)^i}\, ] \zeile
&=&g \sum_{i=m}^\infty c_i\, M^c_i\, (a\L^{1/m})^i + g\, c_1\, J_1^c\,\D y 
\zeile && + {\cal O}(a\cdot \D y) ,\nonumber
\eeql
from which one deduces \Math{\D y\sim a^m}. The computation of
\Math{g-1\sim a^m} goes along the same lines starting from the
second boundary condition \gl{determinexy}.}
Knowing this one easily computes the $m$'th multicritical scaling behaviour
of the bosonic moments

\beq
M_k \sim a^{m-k}, \quad k\in [1,m-1],
\eeq[Mscale]
while the higher $M$--moments and the $J$--moments do not scale.
\Comment{
See this by direct computation:
\beql
M_k &=& \cint{\w} \frac{g\,V^\prime_c(\w)}{(\w-x_c+a\, \L^{1/m})^{k+1/2}
\, (\w-y_c-a^m\, \tilde{\L})^{1/2}} \zeile 
&=& g\, \cint{\w} \frac{V^\prime_c(\w)}{(\w-x_c)^{k+1/2}
\, (\w-y_c)^{1/2}}\, [\sum_{i=0}^\infty c^{(k)}_i\, \frac{(a\,\L^{1/m})^i}
{(\w-x_c)^i} ]\, [\sum_{j=0}^\infty c^{(0)}_j\, \frac{(a^m\,\tilde{\L})^j}
{(\w-y_c)^j} ],
\nonumber\eeql
where for \Math{k\in [1,m-1]} the leading nonvanishing term
is \Math{a^{m-k}} (from \Math{i=m-k} and \Math{j=0} in the sums).}
Moreover the functions $\f^{(n)}_x(p)$  and $\f^{(n)}_y(p)$ are found to behave
like
\beq
\f^{(n)}_x(p) \sim a^{-n-1/2}, \quad\quad \f^{(n)}_y(p) \sim a^{-1/2},
\eeq[phiscaling]
from which one proves the scaling behaviour of the basis functions
\beq
\c^{(n)}(p) \sim a^{-m-n+1/2},\quad\quad \Y^{(n)}(p) \sim a^{-1/2},
\eeq[basisscaling]
by induction.
\par
Let us now turn to the scaling of the fermionic moments $\X_k$ and $\L_k$.
Similar to the bosonic case the function \Math{(v_0(p)-\Y (p))}  of eq.\
\gl{v0alternative} usually vanishes at the endpoints of the cut like a
square root.
We will fine tune the coupling constants \Math{\x_{k+1/2}} in such a manner
that \Math{(n-1)} extra zeros accumulate at $x$, i.e.\
\beq
\X_2^c=\cdots \X_{n-1}^c=0,\quad \quad \X_k^c\neq 0,\,\,\,\,  k\geq n,
\quad \quad \L_l\neq 0, \,\,\,\, l\geq 2,
\eeq[XLscale]
where \Math{\X_k^c \equiv \X_k [ x_c,y_c,\x^c_{k+1/2}]}. 
In addition the analysis of the solution away from the scaling limit 
tells us that the moments $\X_1$ and $\L_1$ will always
appear in the combination \Math{\XL}
\footnote{The only exception is the planar zero mode $\c$ of eq.\
\gl{c}, however, as seen from eq.\ \gl{v0alternative} \Math{v_0(p)} 
effectively does not depend on this combination.}. This suggests 
to impose the
constraint \Math{\X_1^c-\L_1^c=0} on these moments. As there is 
no analogue to
the boundary conditions \gl{determinexy} for the fermionic quantities we
are free to choose the scaling of the coupling constants $\x_{k+1/2}$. We
set
\beq
\x_{k+1/2} = a^{1/2}\, \, \x^c_{k+1/2},
\eeq[xiscale]
and will comment on this choice later on.  From this one derives
\Math{(\X_1-\L_1) \sim a^{n-1/2}} and \Math{\X_k \sim a^{n-k+1/2}} for
\Math{k\in [2,n-1]}.
All other fermionic moments scale uniformly with $a^{1/2}$. So far the
fermionic
scaling is completely independent of the bosonic scaling, governed
by the integer $n$. We shall, however, introduce the requirement that the
scaling part of the bosonic one--superloop correlator 
\Math{W_0(\,\mid p)} of eq.\ \gl{W0p_0} scales
uniformly, i.e.\ we require \Math{(u_0(p)-V^\prime(p))} and
\Math{\widehat{u}_0(p)}
to scale in the same way \cite{Alv2}. As \Math{(u_0(p)-V^\prime(p))\sim
a^{m-1/2}}
we arrive at the following condition on $n$:
\beq
n = m+1.
\eeq[nfix]
And therefore
\beql
\X_1-\L_1 \sim a^{m+1/2},&& \quad\quad \X_k \sim a^{m-k+3/2}, \quad
k\in [2,m] \zeile && \zeile
\X_k \sim a^{1/2}, \quad k>m &&\quad\quad \L_l\sim a^{1/2}, \quad l>
1.
\eeql[Xscale]
The double scaling limit is now defined by letting \Math{N\ra\infty}
and \Math{a\ra 0} but keeping the string coupling constant
\Math{\a=a^{-2m-1}\, N^{-2}} fixed.
\footnote{Let us now comment on the choice of eq.\
\gl{xiscale}. At first sight one might have expected a scaling like
\Math{\x_{k+1/2}=[1+o(a)]\,\x_{k+1/2}^c}. This turns out to be inconsistent
because then the condition \gl{nfix} demands $n$ to be half--integer which
it can not be.
If one takes the more general ansatz \Math{\x_{k+1/2}=
a^l\, \x_{k+1/2}^c} the condition of uniform scaling of \Math{W(\,\mid p)}
yields the allowed sequence
\Math{\{n,l\}=\{m+1,1/2\},\{m,3/2\},\{m-1,5/2\},\ldots}. The scaling of the
lowest moments in eq.\ \gl{Xscale} remains unchanged, the uniform scaling
however already starts with \Math{\X_{n}} scaling like $a^l$ and thus simply
reduces the number of double scaling relevant terms.}
\Subsubkapitel{The Iteration for $u_g(p)$ and $v_g(p)$}
By making use of the above scaling properties of the moments and basis
functions
we may now develop the iterative procedure which allows us to calculate
directly the double scaling relevant versions of $u_g(p)$ and
$v_g(p)$. Recall the iterative scheme of the previous sections:
By eq.\ \gl{VprimePsi} every $u_g$ and $v_g$ may be
written as a linear combination of basis functions $\c^{(n)}$ and $\Y^{(n)}$,
where the coefficients of this expansion are simply read off the poles at
$x$ and $y$ of the right hand sides of the superloop equations
\gl{order0genusg} and \gl{order1genusg}
after a partial fraction decomposition. To optimize the procedure to only
produce terms which are relevant in the double scaling limit we have to
analyze the operators appearing on the right hand sides of the superloop
equations, i.e.\ the superloop insertion
operators $\d/\d V(p)$ and $\d/\d\Y(p)$ as well as the operator
\Math{\widehat{\eckig \Y}}.
\par
From the point of view of the $a\ra 0$ limit the effect of a given operator
in
$\d/\d V(p)$ acting on an expression which scales with $a$ to some
power is to lower this power by a certain amount. Carefully examining 
each term in $\d/\d V(p)$ shows that a is maximally lowered by a 
power of \Math{(m+3/2)}.
All operators which do not lower $a$ by this amount are subdominant
in the scaling limit and may be neglected. The outcome of this analysis for
$\d/\d V(p)$ is
\beql
\frac{\d}{\d V(p)_x} &=& \sum_{k=1}^\infty \frac{\d M_k}{\d V(p)_x}\,
\frac{\d}{\d M_k} \, + \, \frac{\d x}{\d V(p)}\, \frac{\d}{\d x} \zeile &&
\quad \, +\, \frac{\d\XL }{\d V(p)_x}\, \frac{\d}{\d \XL }
\, + \,  \sum_{k=2}^\infty \frac{\d\X_k}{\d V(p)_x}\,
\frac{\d}{\d\X_k} 
\eeql[dVx]
where \footnote{Here the subscript $x$ indicates that the critical
behaviour is associated with the endpoint $x$.}
\beq
\frac{\d M_k}{\d V(p)_x} \phantom{\X_1}
 = - (k+\frac{1}{2} )\, \f^{(k+1)}_x(p) \, +\, (k+\frac{1}{2})
\, \frac{M_{k+1}}{M_1}\,\f^{(1)}_x(p) 
\eeq[dM]
\beql
\frac{\d x}{\d V(p)} \phantom{\X_1}
&=& \frac{1}{M_1}\,\f^{(1)}_x(p) \label{dx} \zeile && \zeile
\frac{\d \XL }{\d V(p)_x} &=& \frac{1}{2}\, \frac{\X_2}{M_1}\, \f^{(1)}_x(p)
\label{dXL1} \zeile && \zeile
\frac{\d\X_k}{\d V(p)_x}\phantom{\X_1}
 &=& (k-\frac{1}{2})\, \frac{\X_{k+1}}{M_1}\, \f^{(1)}_x(p)
\eeql[dXk]
and
\beq
\f^{(n)}_x(p) = (p-x)^{-n-1/2}\, {d_c}^{-1/2},
\eeq[PhiXscale]
with \Math{d_c=x_c-y_c}. Repeating this analysis for the fermionic
superloop insertion operator $\d/\d\Y (p)$ shows that here $a$ is maximally
lowered by a power of \Math{(m+1)} and the relevant contributions are
\beq
\frac{\d}{\d\Y (p)_x} = \frac{\d \XL }{\d\Y (p)_x}\, \frac{\d}{\d \XL } \, +\,
\sum_{k=2}^\infty \frac{\d\X_k}{\d\Y (p)_x}\, \frac{\d}{\d\X_k}
\eeq[dYx]
with
\beql
\frac{\d \XL }{\d\Y (p)_x} &=& - d_c\, \f^{(0)}_x(p) \zeile
&&\zeile
\frac{\d\X_k}{\d\Y (p)_x} \phantom{\,\X_1}
&=& - d_c \, \f^{(k-1)}_x(p). 
\eeql[dYXk]
Finally we state the double scaling version of the operator
\Math{\widehat{\eckig \Y}} acting on the function $\f^{(n)}_x(p)$
\beq
\Yopx
\f^{(n)}_x(p) = \sum_{k=1}^n \frac{\X_{n+2-k}}{d_c}\,
\frac{1}{(p-x)^k} \, + \, \frac{\XL }{2\, d_c}\, \frac{1}{(p-x)^{n+1}}.
\eeq[YopPhi]
Here the operator \Math{\widehat{\eckig \Y}_x} is seen to increase the 
power of $a$ of the expression it acts on by a power of $m$.
\par
We are now in a position to calculate the double scaling limit of the right
hand
sides of the loop equations \gl{order0genusg} and \gl{order1genusg} for
$u_g(p)$ and $v_g(p)$ provided we know the double scaled versions of
\Math{u_1(p),\ldots, u_{g-1}(p)} and \Math{v_1(p),\ldots, v_{g-1}(p)}. As all
the $y$ dependence has disappeared we do not have to perform a 
partial fraction decomposition
of the result. Moreover no $J_k$ and $\L_k$ dependent terms will 
contribute if we do not start out with any and we do not.
\par
The starting point of the iterative procedure are of course the genus 0
correlators. Keeping only the double scaling relevant
parts of eqs.\ \gl{dVu0} and \gl{dqdYv0} one has
\beq
-\frac{d}{dq}\,\frac{\d v_0(q)}{\d\Y (p)_x}\, |_{p=q} =
\frac{\d u_0(p)}{\d V(p)_x}
= \frac{1}{8}\, \frac{1}{(p-x)^2}
\eeq[u(p,p)]
Similarly eq.\ \gl{Wpp} turns into
\beq
\frac{\d v_0(p)}{\d V(p)_x} = \Bigl [ - \frac{\XL}{4 \, d_c\, M_1}\Bigr ]\,
\frac{1}
{(p-x)^3}\, + \, \Bigl [ \frac{\X_2}{4\, d_c\, M_1}\Bigr ]\,\frac{1}{(p-x)^2}.
\eeq[W0(p|p)]
\par
The higher genus correlators $u_g(p)$ and $v_g(p)$ are expressed as linear
combinations of the basis function $\c^{(n)}(p)$ and take the general form
\beql
u_g(p)&=& \sum_{k=1}^{3g-1}\, A^{(k)}_g \, \c^{(k)}(p) \zeile && \zeile
v_g(p)&=& \sum_{k=1}^{3g}\, B^{(k)}_g \, \c^{(k)}(p)\, + \, \k_g\, \f^{(0)}(p),
\eeql[uvgeneral]
where $\k_g$ is the zero mode coefficient not determined by the
first two superloop equations \gl{order0genusg} and \gl{order1genusg}. Note
that the \Math{A^{(k)}_g}
coefficients should up to a factor of two be identical to those of the
double scaled Hermitian matrix model obtained in ref.\ \cite{Amb93}.
We have calculated the $A^{(k)}_g$ and $B^{(k)}_g$
coefficients in the double--scaling limit for \Math{g=1,2} and $3$ with the 
aid of {\it Maple}. The results of Ambj\o rn et al.\ \cite{Amb93}
for the $A^{(k)}_g$ coefficients could be reproduced.
\par
Before we state our explicit results let us turn to the general
scaling behaviour of the one--superloop correlators
\beq
W_g(\,\mid p) \sim a^{(1-2g)\, (m+1/2)\, -\, 1}, \quad
\mbox{and}\quad
W_g(p\mid \, ) \sim a^{(1-2g)\, (m+1/2)\, -\, 1/2},
\eeq[Wscaling]
which one proves by induction. As shown in ref.\ \cite{Amb93} the
coefficients \Math{A^{(k)}_g} of eq.\ \gl{uvgeneral} take the form
\beq
A^{(k)}_g = \sum_{\a_j} \langle\, \a_1,\ldots , \a_s \mid \a\, \rangle_{g,k}
\, \frac{M_{\a_1}\ldots M_{\a_s}}{{M_1}^\a\, {d_c}^{g-1}},
\eeq[Agkdef]
where the brackets stand for rational numbers and where \Math{\a_j,\a} and
\Math{s} are subject to the constraints
\beq
\a=2g+s-2 \qquad\mbox{and}\qquad \sum_{j=1}^s(\a_j-1) = 3g-k-1
\eeq[AgkRels]
with \Math{\a_j\in [2,3g-1]}. Explicitly up to genus two one finds
\beql
A^{(1)}_1&=&0 \qquad A^{(2)}_1=\frac{1}{8} \zeile && \zeile
A^{(1)}_2&=&0 \qquad A^{(2)}_2=0 \qquad A^{(3)}_2=\frac{49\, {M_2}^2}
{128\, d_c\, {M_1}^4}- \frac{5\, M_3}{16\, d_c\ {M_1}^3} \zeile
A^{(4)}_2&=& -\frac{49\, M_2}{64\, d_c\, {M_1}^3} \qquad \frac{105}{128\,
d_c\, {M_1}^2}.
\eeql[Agkg2res]
The terms above are only potentially relevant, as for the $m$'th multicritical
point all terms containing \Math{M_k} with \Math{k>m} are subleading in
the double scaling limit.
\par 
Similarly the structure of the coefficients $B_g^{(k)}$ may be determined
from the iterative procedure and is seen to be
\beq
B_g^{(k)}= \sum_{\a_j, \b}\, \langle \a_1,\ldots , \a_s, \b \mid \a
\rangle_{g,k}
\,\frac{M_{\a_1} \ldots M_{\a_s}\, \X_\b}{{M_1}^\a\, {d_c}^g},
\eeq[Bgkgeneral]
where the brackets denote rational numbers and where we write $\X_1$ for
\Math{\XL}. One shows that the
$\a$, $\a_j$, $\b$ and $s$ obey the conditions
\beq
\a=2g+s-1, \quad\mbox{and} \quad \sum_{j=1}^s (\a_j -1) = 3g+1-\b-k
\eeq[Bgkcond]
with \Math{\a_j\in [2,3g]} and \Math{\b\in [1,3g]}.
For the zero mode coefficient $\k_g$ the general structure is given by 
the same expansion as eq.\ \gl{Bgkgeneral} with \Math{k=0}.
The conditions on $\a$, $\a_j$, $\b$ and $s$ then read
\beq
\a=2g+s, \quad\mbox{and} \quad \sum_{j=1}^s (\a_j -1) = 3g+1-\b
\eeq[kappacond]
where \Math{\a_j\in [2,3g]} and \Math{\b\in [1,3g+1]}.
\par
The explicit results for the $B^{(k)}_g$
coefficients for \Math{g=1} and \Math{g=2} are given by
\beql
B_1^{(1)} &=&-{\frac {\X_{{3}}}{8\,M_{{1}}\,d_c}}+{\frac {M_{{2}}\X_{{2
}}}{8\,{M_{{1}}}^{2}\, d_c}},
\zeile
B_1^{(2)}&=&{\frac {M_{{2}}\,\XL}{16\,{
M_{{1}}}^{2}d_c}}+{\frac {\X_{{2}}}{8\,M_{{1}}\, d_c}},
\qquad
B_1^{(3)}=-{\frac {5\,{\XL}}{16\,M_{{1}}\, d_c}}.
\zeile && \zeile
B_2^{(1)}&=&{\frac {203\,M_{{2}}\X_{{5}}}{128\,{d_c}^{2}{M_{{1}}}^{
4}}}-{\frac {145\,{M_{{3}}}^{2}\X_{{2}}}{128\,{M_{{1}}}^{5}{d_c}^{2}}}-{
\frac {105\,\X_{{6}}}{128\,{d_c}^{2}{M_{{1}}}^{3}}}+{\frac {63\,{M_{{2}}}
^{3}\X_{{3}}}{32\,{M_{{1}}}^{6}{d_c}^{2}}} \zeile
 &&+{\frac {105\,M_{{4}}\X_{{3}}}{
128\,{d_c}^{2}{M_{{1}}}^{4}}}+{\frac {145\,M_{{3}}\X_{{4}}}{128\,{d_c}^{2}{
M_{{1}}}^{4}}}+{\frac {105\,M_{{5}}\X_{{2}}}{128\,{d_c}^{2}{M_{{1}}}^{4}}
}-{\frac {77\,M_{{4}}M_{{2}}\X_{{2}}}{32\,{M_{{1}}}^{5}{d_c}^{2}}}\zeile
&&
-{\frac {87\,M_{{3}}M_{{2}}\X_{{3}}}{32\,{M_{{1}}}^{5}{d_c}^{2}}}
+{\frac {
75\,M_{{3}}{M_{{2}}}^{2}\X_{{2}}}{16\,{M_{{1}}}^{6}{d_c}^{2}}}-{\frac {63
\,{M_{{2}}}^{4}\X_{{2}}}{32\,{M_{{1}}}^{7}{d_c}^{2}}}-{\frac {63\,{M_{{2}
}}^{2}\X_{{4}}}{32\,{M_{{1}}}^{5}{d_c}^{2}}},\zeile
B_2^{(2)}&=&-{\frac {21\,{M_{{
2}}}^{3}\X_{{2}}}{64\,{M_{{1}}}^{6}{d_c}^{2}}}-{\frac {21\,M_{{2}}\X_{{4}}
}{64\,{d_c}^{2}{M_{{1}}}^{4}}}+{\frac {77\,M_{{3}}M_{{2}}\X_{{2}}}{128\,{
M_{{1}}}^{5}{d_c}^{2}}}-{\frac {35\,M_{{4}}\X_{{2}}}{128\,{d_c}^{2}{M_{{1}}
}^{4}}} \zeile
&&+{\frac {75\,M_{{3}}{M_{{2}}}^{2}{\XL}}{32\,{M_{{1}}}^
{6}{d_c}^{2}}}+{\frac {105\,M_{{5}}{\XL}}{256\,{d_c}^{2}{M_{{1}}}
^{4}}}-{\frac {63\,{M_{{2}}}^{4}{\XL}}{64\,{M_{{1}}}^{7}{d_c}^{
2}}} \zeile
&&+{\frac {35\,\X_{{5}}}{128\,{M_{{1}}}^{3}{d_c}^{2}}}-{\frac {145\,{M_{
{3}}}^{2}{\XL}}{256\,{M_{{1}}}^{5}{d_c}^{2}}}+{\frac {21\,{M_{{
2}}}^{2}\X_{{3}}}{64\,{M_{{1}}}^{5}{d_c}^{2}}}-{\frac {35\,M_{{3}}\X_{{3}}
}{128\,{d_c}^{2}{M_{{1}}}^{4}}}, \zeile
&& -{\frac {77\,M_{{4}}M_{{2}}{\XL}}{64\,{M_{{1}}}^{5}{d_c}^{2
}}} \zeile
B_2^{(3)}&=&-{\frac {599\,M_{{3}}M_{{2}}
{\XL}}{128\,{M_{{1}}}^{5}{d_c}^{2}}}+{\frac {105\,{M_{{2}}}^{3}
{\XL}}{32\,{M_{{1}}}^{6}{d_c}^{2}}}-{\frac {5\,M_{{3}}\X_{{2}}}{
16\,{d_c}^{2}{M_{{1}}}^{4}}} \zeile
&& +{\frac {385\,M_{{4}}{\XL}}{256\,{d_c
}^{2}{M_{{1}}}^{4}}} +{\frac {21\,{M_{{2}}}^{2}\X_{{2}}}{64\,{M_{{1}}}^{
5}{d_c}^{2}}}+{\frac {7\,M_{{2}}\X_{{3}}}{128\,{d_c}^{2}{M_{{1}}}^{4}}},
\zeile
B_2^{(4)}&=&-{\frac {357\,{M_{{2}}}^{2}{\XL}}{64\,{M_{{1}}}^{5}{
d_c}^{2}}}+{\frac {875\,M_{{3}}{\XL}}{256\,{d_c}^{2}{M_{{1}}}^{4}
}}-{\frac {35\,\X_{{3}}}{128\,{M_{{1}}}^{3}{d_c}^{2}}}-{\frac {63\,M_{{2}
}\X_{{2}}}{128\,{d_c}^{2}{M_{{1}}}^{4}}},\zeile
B_2^{(5)}&=&{\frac {105\,\X_{{2}}}{
128\,{M_{{1}}}^{3}{d_c}^{2}}}+{\frac {1617\,M_{{2}}{\XL}}{256\,
{d_c}^{2}{M_{{1}}}^{4}}},
\zeile
B_2^{(6)}&=&-{\frac {1155\,{\XL}}{256\,{M_{{1}}}^{3}{d_c}^{2}}}. 
\eeql[Bees]
Note that the terms listed above are only potentially relevant, depending on
which multi--critical model one wishes to consider. For an $m$'th
multi--critical
model all terms containing $M_k$, \Math{k>m}, or $\X_l$, \Math{l>m+1}, 
vanish in the double--scaling limit. We remind the reader that we assumed 
to have a non--symmetric potential and that the critical behaviour was 
associated with the endpoint $x$. In the case where the critical behaviour 
is associated with the
endpoint $y$ all formulas in this section still hold provided $d_c$ is replaced
by $-d_c$, $M_k$ by $J_k$, $\X_k$ by $\L_k$ and $x$ by $y$.
\Subsubkapitel{The Iteration for $F_g$, $\k_g$ and $\widehat{u}_g(p)$}
Having computed $u_g(p)$ and $v_g(p)$ (up to the zero mode) we may now
proceed to calculate the free energy $F_g$ and the zero mode coefficient
$\k_g$.
This is done by rewriting $u_g$ and $v_g$ as total derivatives in the 
superloop insertion operators \Math{\d/\d V(p)} and \Math{\d/\d\Y (p)} 
respectively, yielding
the bosonic and doubly fermionic parts of $F_g$ as well as $\k_g$. The
procedure
to compute the bosonic part of the free energy \Math{\Fbg}
works just as in the Hermitian
matrix model described in ref.\ \cite{Amb93}. We have \Math{\Fbg=2\,
F^{\mbox{\scriptsize herm}}_g.}
\par
From eq.\ \gl{Wscaling} we see that \Math{F_g} scales as
\beq
F_g= \Fbg \, +\, \Ffg\, \sim a^{(2-2g)(m+1/2)},
\eeq[Fscaling]
just as the hermitian matrix model at its $m$'th multicritical point.
\par
To find the bosonic piece of the free energy $F_g$ one uses the
following recursive rewriting of \Math{\c^{(n)}(p)} as a total
derivative
\beq
\c^{(n)}(p)=- \frac{1}{M_1}\Bigl ( \, \frac{2}{2n-1} \, \frac{\d \, M_{n-1}}
{\d V(p)}\, - \sum_{k=2}^{n-1}\c^{(k)}(p)\, M_{n-k+1}\, \Bigl )
\qquad n\geq 2,
\eeq[cscalrew]
One does not need any expression for \Math{\c^{(1)}(p)} as \Math{A^{(1)}_g=0}
in eq.\ \gl{uvgeneral} for all $g\geq 2$ \cite{Amb93}.
\par
To obtain the doubly fermionic part \Math{\Ffg} simply
rewrite the basis functions $\c^{(n)}(p)$ of $v_g(p)$ appearing in
\gl{uvgeneral}
in terms of the functions \Math{\f^{(n)}_x(p)} which by eq.\ 
\gl{dYXk} are nothing but total derivatives in \Math{\d/\d\Y (p)}. Doing this
for $v_g(p)$ allows one to directly deduce the form of \Math{\Ffg} and 
$\k_g$.
\footnote{Actually the presentation here is somewhat misleading. In order to
compute the $u_g(p)$ and $v_g(p)$ by iteration in the way outlined in the
previous subsection one needs to know the {\it full} \Math{v_1(p), \ldots,
v_{g-1}(p)}
including the zero mode coefficients \Math{\k_1, \ldots, \k_{g-1}}. In
practice one hence computes all quantities at genus $g$, i.e.\ $u_g$, $v_g$,
$\k_g$
and $F_g$, before proceeding to genus \Math{(g+1)}.}
\par
The explicit results for genus one are
\beql
F^{\mbox{\scriptsize bos}}_1 &=& -\frac{1}{12}\ln M_1, \zeile && \zeile
F^{\mbox{\scriptsize ferm}}_1&=&
\XL\Bigl (-{\frac {
5\,{\X}_{{4}}}{16\,{d_c}^{2}{M_{{1}}}^{2}}}+{\frac {3\,M_{{2}}\, {\X}_{{3}}}
{8\,{d_c}^{2}{M_{{1}}}^{3}}}+{\frac {5\,M_{{3}}\, {\X}_{{2}}}
{16\,{d_c}^{2}{M_{{1}}}^{3}}}-{\frac {3\,{M_{{2}}}^{2}\, {\X}_{{2}}}{8
\,{d_c}^{2}{M_{{1}}}^{4}}}\Bigr )\zeile && \quad
+{\frac {\X_{{2}}\, {\X}_{{3}}}{8\,{d_c}^{2}{M_{{1}}}^{2}}}
\eeql[F1scal]
and
\beq
\k_1={\frac {5\,\X_{{4}}}{16\,{M_{{1}}}^{2}
d_c}}-{\frac {3\,M_{{2}}\,\X_{{3}}}{8\,{M_{{1}}}^{3}d_c}}-{\frac
{5\,M_{{3}}\,\X
_{{2}}}{16\,{M_{{1}}}^{3}d_c}}+{\frac {3\,{M_{{2}}}^{2}\,\X_{{2}}}{8\,{M_{{
1}}}^{4}d_c}}.
\eeq[k1scal]
These expressions may of course be alternatively obtained by taking the 
double scaling limit of the genus one results computed
away from the scaling limit, except for the scaling violating term 
\Math{F^{\mbox{\scriptsize bos}}_1} at genus one. This term in eq.\ 
\gl{F1scal} was selected by producing the double scaling version of 
\Math{u_1(p)}.
\par
Before presenting explicit results for genus two, let us describe the
general structure of the free energy at genus $g$.
For the bosonic piece we have \cite{Amb93}
\beq
F^{\mbox{\scriptsize bos}}_g =\sum_{\a_j > 1}\,  \langle\, \a_1,\ldots ,\a_s
\mid \a \, \rangle _g \, \frac{M_{\a_1}\ldots M_{\a_s}}{M_1^\a\, {d_c}^{g-1}}
\qquad g\geq 2,
\eeq[Fgdsl]
where the brackets denote rational numbers. The $\a_j$, $\a$ and $s$ 
obey
\beq
\a=2g-2+s \qquad \mbox{and}\qquad \sum_{j=1}^s (\a_j -1)=3g-3,
\eeq[]
with \Math{\a_j\in [2, 3g-2]}.
\par
Similarly the general form of \Math{\Ffg} for \Math{g\geq 1} reads
\beq
\Ffg= \sum _{\a_j, \b_i}\, \langle \a_1,\ldots , \a_s, \b_1,\b_2\mid 
\a\rangle_g
\, \frac{\X_{\b_1}\, \X_{\b_2}\, M_{\a_1} \ldots M_{\a_s}}{{M_1}^\a\,
{d_c}^{g+1}},
\eeq[Ffggeneral]
where we write \Math{\X_1} for \Math{\XL} and where the 
brackets denote
rational numbers as before. The \Math{\a_j}, $\b_i$, $\a$ and $s$ are
subject to the constraints
\beq
\a=2g+s, \quad \mbox{and}\quad \sum_{j=1}^s (\a_j -1)= 3g+2-\b_1-\b_2,
\eeq[alphaconstraints]
where \Math{\a_j\in [2,3g]} and \Math{\b_i\in [1,3g+1]}.
\par
The results for genus two read
\beql
F_2^{\mbox{\scriptsize bos}}&=& \frac{29\, M_2\, M_3}{64\, {M_1}^5\,d_c}
-\frac{21\, {M_2}^3}{80\, {M_1}^5\,d_c} -\frac{35\, M_4}{192\, {M_1}^3\,
d_c} \zeile && \zeile
F_2^{\mbox{\scriptsize ferm}}&=&
\XL \, \biggl[ \,
 {\frac {1015\,\X_{{5}}\, M_{{3}}}{128\,{M_{{1}}}^{5}{d_c}^{3}}}
-{\frac {375\,{\X}_{{4}}\, M_{{2}}\, M_{{3}}}
   {16\,{M_{{1}}}^{6}{d_c}^{3}}}                                \zeile &&
+{\frac {315\,\X_{{4}}\, {M_{{2}}}^{3}}{16\,{M_{{1}}}^{7}
   {d_c}^{3}}}
+{\frac {385\,\X_{{4}}\, M_{{4}}}{64\,{M_{{1}}}^{5}{d_c}^{3}}}
+{\frac {1323\,\X_{{2}}\, {M_{{2}}}^{5}}
   {64\,{M_{{1}}}^{9}{d_c}^{3}}}
+{\frac {693\, \X_{{6}}\, M_{{2}}}{64\,{M_{{1}}}^{5}{d_c}^{3}}} \zeile &&
-{\frac {1155\,\X_{{7}}}{256\,{M_{{1}}}^{4}{d_c}^{3}}}
-{\frac {63\,\X_{{2}}\, M_{{2}}\, M_{{5}}}
   {4\,{M_{{1}}}^{6}{d_c}^{3}}}
-{\frac {525\,\X_{{5}}\, {M_{{2}}}^{2}}
   {32\,{M_{{1}}}^{6}{d_c}^{3}}}
-{\frac {1155\,\X_{{3}}\, M_{{2}}\, M_{{4}}}
  {64\,{M_{{1}}}^{6}{d_c}^{3}}}                \zeile &&
+{\frac {315\,\X_{{3}}\, M_{{5}}}{64\,{M_{{1}}}^{5}{d_c}^{3}}}
-{\frac {2175\,\X_{{3}}\,
  {M_{{3}}}^{2}}{256\,{M_{{1}}}^{6}{d_c}^{3}}}
+{\frac {675\,\X_{{3}}\, {M_{{2}}}^{2}\, M_{{3}}}
  {16\,{M_{{1}}}^{7}{d_c}^{3}}}
-{\frac {1785\,\X_{{2}}\, M_{{3}}\, M_{{4}}}
  {128\,{M_{{1}}}^{6}{d_c}^{3}}}                         \zeile &&
-{\frac {495\,\X_{{2}}\, {M_{{2}}}^{3}\, M_{{3}}}
  {8\,{M_{{1}}}^{8}{d_c}^{3}}}
+{\frac {2205\,\X_{{2}}\, {M_{{2}}}^{2}\, M_{{4}}}
  {64\,{M_{{1}}}^{7}{d_c}^{3}}}
-{\frac {1323\,\X_{{3}}\, {M_{{2}}}^{4}}
  {64\,{M_{{1}}}^{8}{d_c}^{3}}}
+{\frac {8175\,\X_{{2}}\, {M_{{3}}}^{2}M_{{2}}}
  {256\,{M_{{1}}}^{7}{d_c}^{3}}}     \zeile & &
+{\frac {1155\,\X_{{2}}\, M_{{6}}}
  {256\,{M_{{1}}}^{5}{d_c}^{3}}}
\, \biggr]
-{\frac {21\,\X_{{2}}\, {\X}_{{5}}\, M_{{2}}}{16\,{M_{{1}}}^{5}{d_c}^{3}}}
+{\frac {105\,\X_{{2}}\, {\X}_{{6}}}{128\, M_{{1}}^{4}{d_c}^{3}}}
-{\frac {145\,\X_{{2}}\, {\X}_{{4}}\, M_{{3}}}{128\,{M_{{1}}}^{5}{d_c}^{3}}}
\zeile &&
+{\frac {105\,\X_{{2}}\, {\X}_{{4}}{M_{{2}}}^{2}}{64\,{M_{{1}}}^{6}{d_c}^{3}}}
+{\frac {21\,{\X}_{{3}}\, \X_{{4}}\, M_{{2}}}{64\,{M_{{1}}}^{5}{d_c}^{3}}}
-{\frac {35\,\X_{{3}}\, {\X}_{{5}}}{128\,{M_{{1}}}^{4}{d_c}^{3}}}
-{\frac {63\,\X_{{2}}\, {\X}_{{3}}\,
{M_{{2}}}^{3}}{32\,{M_{{1}}}^{7}{d_c}^{3}}}
\zeile &&
-{\frac {35\,\X_{{2}}\, {\X}_{{3}}\, M_{{4}}}{32\,{M_{{1}}}^{5}{d_c}^{3}}}
+{\frac {195\,\X_{{2}}\, {\X}_{{3}}\, M_{{2}}\, M_{{3}}}
  {64\,{M_{{1}}}^{6}{d_c}^{3}}}.  
\eeql[Ff2]
Of course so far we have determined only the coefficients $F_g$ of the genus
expansion of the free energy (cf.\ eq.\ \gl{genusexpF}).
 For an $m$'th multicritical model the relevant expansion parameter in the
 double scaling limit is the string coupling constant \Math{\a=a^{-2m-1}\,
N^{-2}}.
 If we introduce the bosonic and fermionic scaling moments 
 $\m_{\, k}$ and $\t_{\, k}$
 by (cf.\ eqs.\ \gl{Mscale} and \gl{Xscale})
\beqx
 M_k=a^{m-k}\, \m_{\, k}, \quad k\in [1,m],
 \eeqx
 \beq
 \XL= a^{m+1/2}\, \t_{\, 1}, \quad\quad \X_l=a^{m-l+3/2}\,\t_{\, l},\quad
 l\in [2,m+1],
 \eeq[scalingmoments]
 we get by replacing $M_k$ by $\m_{\, k}$ and setting $M_k$ equal to zero for
\Math{k>m}
 as well as replacing $\X_l$ by $\t_{\, l}$ and setting $\X_l$ to zero for
\Math{l>m+1}
 in the formulas above exactly the coefficients of the expansion in the string
 coupling constant.
Needless to say that these results apply for non--symmetric potentials as
well, where the critical behaviour is associated with the endpoint $y$ by
performing the usual replacements.
\par
For the zero mode coefficient at genus two we find
\beql
{\k_2}&=&{\frac {495\,\X_{{2}}M_{{3}}{M_{{2}}}
^{3}}{8\,{d_c}^{2}{M_{{1}}}^{8}}}+{\frac {63\,\X_{{2}}M_{{5}}M_{{2}}}{4\,
{d_c}^{2}{M_{{1}}}^{6}}}-{\frac {315\,\X_{{3}}M_{{5}}}{64\,{d_c}^{2}{M_{{1}
}}^{5}}}+{\frac {1323\,\X_{{3}}{M_{{2}}}^{4}}{64\,{d_c}^{2}{M_{{1}}}^{8}}
}
\zeile &&
-{\frac {675\,\X_{{3}}M_{{3}}{M_{{2}}}^{2}}{16\,{d_c}^{2}{M_{{1}}}^{7}}}
+{\frac {2175\,\X_{{3}}{M_{{3}}}^{2}}{256\,{d_c}^{2}{M_{{1}}}^{6}}}-{
\frac {8175\,\X_{{2}}{M_{{3}}}^{2}M_{{2}}}{256\,{d_c}^{2}{M_{{1}}}^{7}}}-
{\frac {385\,\X_{{4}}M_{{4}}}{64\,{d_c}^{2}{M_{{1}}}^{5}}}
\zeile &&
-{\frac {2205\,
\X_{{2}}M_{{4}}{M_{{2}}}^{2}}{64\,{d_c}^{2}{M_{{1}}}^{7}}}+{\frac {375\,\X
_{{4}}M_{{3}}M_{{2}}}{16\,{d_c}^{2}{M_{{1}}}^{6}}}-{\frac {1155\,\X_{{2}}
M_{{6}}}{256\,{d_c}^{2}{M_{{1}}}^{5}}}+{\frac {1785\,\X_{{2}}M_{{4}}M_{{3
}}}{128\,{d_c}^{2}{M_{{1}}}^{6}}}
\zeile &&
-{\frac {693\,\X_{{6}}\, M_{{2}}}{64\,{d_c}^{
2}{M_{{1}}}^{5}}}-{\frac {315\,\X_{{4}}{M_{{2}}}^{3}}{16\,{d_c}^{2}{M_{{1
}}}^{7}}}+{\frac {525\,\X_{{5}}{M_{{2}}}^{2}}{32\,{d_c}^{2}{M_{{1}}}^{6}}
}-{\frac {1015\,\X_{{5}}M_{{3}}}{128\,{d_c}^{2}{M_{{1}}}^{5}}}
\zeile &&
-{\frac {
1323\,\X_{{2}}{M_{{2}}}^{5}}{64\,{d_c}^{2}{M_{{1}}}^{9}}}+{\frac {1155\,\X
_{{3}}M_{{4}}M_{{2}}}{64\,{d_c}^{2}{M_{{1}}}^{6}}}+{\frac {1155\,\X_{{7}}
}{256\,{d_c}^{2}{M_{{1}}}^{4}}}. 
\eeql[kappa2]
We obtained these results with the aid of a {\it Maple} program which 
performs the iteration up to arbitrary genus.
In practice the expressions become quite
lengthy, e.g.\ \Math{F^{\mbox{\scriptsize ferm}}_3} consists of 114 terms.
\par
For the knowledge of the full \Math{W_g(\,\mid p)} it remains to compute
$\widehat{u}_g(p)$. This is of course done by applying \Math{\d/\d V(p)} to
\Math{\Ffg}. The general structure of $\widehat{u}_g(p)$ turns out to be
\beq
\widehat{u}_g(p)= \sum_{k=1}^{3g+1} \widehat{A}_g^{(k)}\, \c^{(k)}(p),
\eeq[uhatgeneral]
where
\beq
\widehat{A}_g^{(k)}= \sum_{\a_j, \b_1,\b_2}\,
\langle \a_1,\ldots ,\a_s,\b_1,\b_2\mid \a\rangle_{g,k}\,
\frac{M_{\a_1}\ldots M_{\a_s}\, \X_{\b_1}\,\X_{\b_2}}{{M_1}^\a\, {d_c}^{g+1}}
\eeq[Ahatgkgeneral]
underlying the conditions
\beq
\a= s+3g, \quad\mbox{and}\quad \sum_{j=1}^s (\a_j-1)= 4+3g-k-\b_1-\b_2,
\eeq[Ahatconds]
with \Math{\a_j\in [2,3g+1]} and \Math{\b_i\in [1,3g+2]}. Due to space let us
only state the genus one results
\beql
\widehat{A}_1^{(1)}&=&{\frac {5\,\X_{{2}}\,
\X_{{4}}}{32\,{d_c}^{2}{M_{{1}}}^{2}
}}+{\frac {3\,M_{{2}}\X_{{2}}{\X}_{{3}}}{16\,{d_c}^{2}{M_{{1}}}^{3}}}
-{\frac {35\,{\XL}{\X}_{{5}}}{32\,{d_c}^{2}{M_{{1}}}^{2}}
} \zeile &&
+{\frac {15\, M_{{3}}{\XL}\,{\X}_{{3}}}{32\,{d_c}^{2}{M_{{1}}
}^{3}}}
+{\frac {15\, M_{{2}}{\XL}\,{\X}_{{4}}}{16\,{d_c}^{2}{M
_{{1}}}^{3}}}-{\frac {9\,{M_{{2}}}^{2}{\XL}\,{\X}_{{3}}}{16
\,{d_c}^{2}{M_{{1}}}^{4}}},
\zeile
\widehat{A}_1^{(2)}&=&-{\frac {15\,{\XL}\,{\X}_{{4}}}
{16\,{d_c}^{2}{M_{{1}}}^{2}}}+{\frac {3\,\X_{{2}}\, {\X}_{{3}}
}{8\,{d_c}^{2}{M_{{1}}}^{2}}}+{\frac {3\,M_{{2}}{\XL}\, {\X}_{
{3}}}{4\,{d_c}^{2}{M_{{1}}}^{3}}}\zeile &&
+{\frac {5\,M_{{3}}{\XL}{\X}_{{2}}}
{16\,{d_c}^{2}{M_{{1}}}^{3}}}-{\frac {3\,{M_{{2}
}}^{2}{\XL}\, {\X}_{{2}}}{8\,{d_c}^{2}{M_{{1}}}^{4}}}, \zeile
\widehat{A}_1^{(3)}&=&-{\frac {
15\,{\XL}{\X}_{{3}}}{16\,{d_c}^{2}{M_{{1}}}^{2}}}+{\frac {
25\, M_{{2}}\, {\XL}\, {\X}_{{2}}}{32\,{d_c}^{2}{M_{{1}}}^{3}}},
\zeile && \zeile
\widehat{A}_1^{(4)}&=&-{\frac {35\,{\XL}\, {\X}_{{2}}}{32\,{d_c}^{2}{M_{
{1}}}^{2}}}. 
\eeql[uhat1dscsal]
This concludes our analysis of the double--scaling limit for generic
potentials.
\par
The case of symmetric bosonic and generic fermionic potentials was 
considered in ref.\ \cite{Alv} where a doubling of degrees of freedom 
was observed for genus zero. We then have the independent set of moments
\Math{\{M_k, \X_k , \L_k\}}. With the methods presented in this paper 
one can see that this holds for higher genera as well, i.e. the free energy 
here takes the form
\beq
F_g^{\mbox{\scriptsize symm}}= 2\, \Fbg \, + \, 
F_g^{\mbox{\scriptsize ferm,x}}\, +\, F_g^{\mbox{\scriptsize ferm,y=-x}},
\eeq[Fgsymm]
where $\Fbg$ denotes the bosonic part of the free energy for generic
potentials,
\Math{F_g^{\mbox{\scriptsize ferm,x}}} and
\Math{F_g^{\mbox{\scriptsize ferm,y=-x}}} denote the doubly fermionic parts
in the generic case
where the critical behaviour is associated with the endpoints $x$ and
 \Math{y=-x}
respectively (cf.\ eq.\ \gl{Ffggeneral}). If one chooses to take
the fermionic potential
to be symmetric as well the doubly fermionic part of the free energy will
vanish, as we have already seen on the discrete level in section 
\Sub{IteraticeProc}.
\Subkapitel{Identification of the Model}{IdentModel}
In order to identify the continuum theory described by the supereigenvalue
model we will have to discuss some results obtained in super--Liouville
theory. This subsection will by no means give a review on this subject,
we simply highlight some formulas relevant for the interpretation of
our results. For more details see refs.\ \cite{DHK,HP,Book}.
\par
In the following we consider \Math{N=1} superconformal field theories
with \Math{\widehat{c}=d} coupled to 2d supergravity. For a base
manifold of fixed genus $g$ the partition function is given by
\beq
Z_g= \sum_s\, \int [{\schnorkel D} E^A_M ] \, [{\schnorkel D} {\schnorkel
X}^I] \, e^{ -S[{\schnorkel X}, E]},
\eeq[Zg]
where the sum runs over the spin structures. Here
\Math{E^A_M} denotes the super--zweibein, \Math{{\schnorkel X}^I} the
free, conformal superfield and the matter action reads
\beq
S[{\schnorkel X}, E] = \frac{1}{2} \, \int_{M_s} d^2z\, d^2\q D_\a 
{\schnorkel X}^I \, D^\a{\schnorkel X}^I,
\eeq[SXE]
using the superspace notation of ref.\ \cite{DHK}. Here $D_\a$ are 
superdifferentials
and \Math{I=1,\ldots, d}. The action and the measures 
\Math{{\schnorkel D} E^A_M} and \Math{{\schnorkel D} {\schnorkel X}^I} are
invariant under superdiffeomorphisms and local Lorentz transformations.
These have to be divided out of the path integral \gl{Zg}
which is symbolized by square brackets around the volume elements.
The action \gl{SXE} is also invariant under the group of local superconformal
transformations while the measure is not. Thus a 
superconformal gauge fixing to the reference frame 
\Math{E=e^{a\,{\eckig \f}}\, \widehat{E}} will yield a Jacobian which 
is assumed to be \cite{DHK}
 \beq
 J({\eckig \f}, \widehat{E}) = \exp (-S_{\mbox{\scriptsize SL}}[{\eckig \f}, 
 \widehat{E}]\, ).
\eeq[JfE]
The action functional in the exponent is the super--Liouville action 
\beq
S_{\mbox{\scriptsize SL}}[{\eckig \f}, \widehat{E}] = \frac{1}{4\p}\, \int
d^2z\, d^2\q \, \widehat{E}\, ( \frac{1}{2} \, \widehat{D}_\a{\eckig \f}\,
\widehat{D}^\a {\eckig \f} - \sqrt{\frac{9-\widehat{c}}{2}}\, \widehat{Y}\,
{\eckig \f} + \k\, {\cal O}_{\mbox{\scriptsize min}}\, 
e^{a\, {\eckig \f} }\, )
\eeq[sLiouville]
where \Math{{\eckig \f}} is the Liouville superfield and \Math{\widehat{Y}}
the curvature superfield. The cosmological constant $\k$ couples to
the operator of lowest--dimension in the Neveu--Schwarz sector of
the matter theory denoted by \Math{{\cal O}_{\mbox{\scriptsize min}}}
\cite{BerKleb}. For unitary theories 
\Math{{\cal O}_{\mbox{\scriptsize min}}} is unity.
Moreover \Math{a} 
is a \Math{\widehat{c}} dependent constant.
The precise form of the matter action \gl{SXE} is not of importance, all
that is needed is that one has some superconformal field theory with 
central charge $\widehat{c}$. One can now study the partition
function \gl{Zg} in the form
\beq
Z_g = \int_0^\infty dL  \, Z_g(L),
\eeq[ZgL]
where $L$ is a characteristic length scale of the super--Riemann 
surface \footnote{Note that $L$ has the dimension of a length. In
contrast to the bosonic random surface which has an intrinsic area
defined by \Math{\int d^2 z\, \sqrt{g}}, the analogous object for a
super--random surface is given by \Math{\int d^2 z\, d^2 \q\, E = L}.
Because of the Grassmann integrations the dimension of L is lowered
by 1 compared to the bosonic case.}. The partition function for the 
super--random surface of fixed length is defined to be
\beq
Z_g(L)= \sum_s \int {\schnorkel D} {\schnorkel X} \, {\schnorkel D}{\eckig
\f}\, e^{- S_{\mbox{\scriptsize SL}}-S_{\mbox{\scriptsize M}}}\,
\d\Bigl ( \int d^2z\, d^2\q\, \widehat{E} -L \Bigr ),
\eeq[Fgdef]
where the symbol \Math{{\schnorkel D} {\schnorkel X}} denotes the 
integration over all the super--matter, super--ghost and super--moduli
contributions, contained in the matter action 
\Math{S_{\mbox{\scriptsize M}}}. Similarly to the bosonic case the
scaling behaviour of \Math{Z_g(L)} may be computed \cite{DHK}
\beq
Z_g(L)= e^{(\k_c-\k)\, L}\, L^{(1-g)\, (\g_{\mbox{\scriptsize str.}}
 -2)-1}\, f_g ,
\eeq[FgLscal]
where $\k_c$ and $f_g$ are undetermined and $L$ independent. Here the 
string  susceptibility \Math{\g_{\mbox{\scriptsize str.}}} of a unitary 
superconformal field theory takes the form
\beq
\g_{\mbox{\scriptsize str.}}=2 + \frac{1}{4}\, ( \,\widehat{c} -9 - \sqrt{(
9- \widehat{c}\, )\, (1-\widehat{c}\, )} \, ).
\eeq[gstring]
The minimal \Math{N=1} superconformal field theories with \Math{
\widehat{c}\leq 1} are classified by a pair of integers \Math{(p,q)}
and have central charge
\beq
\widehat{c}= 1- \frac{2\, (p-q)^2}{p\, q}.
\eeq[cpq]
The unitary theories (with \Math{\widehat{c}\geq 0}) are given by 
the pair \Math{(p,q)=(m+2,m)} with \Math{m\geq 2}. For
a non--unitary theory formula \gl{gstring} is no longer valid and
instead one has \cite{BerKleb,Book}
\beq
\g_{\mbox{\scriptsize str.}}=2 - \frac{2\, (p+q)}{p+q-2}.
\eeq[gstring2]
In any case the integral over $L$ in eq.\ \gl{ZgL} with the \Math{Z_g(L)}
of eq.\ \gl{FgLscal} may be performed to give
\beq
Z_g = (\k -\k_c)^{(2-\g_{\mbox{\scriptsize str.}})\, (1-g)} \, \G[ (g-1)(2-
\g_{\mbox{\scriptsize str.}})]\, f_g , 
\eeq[Zgres]
note the analogy to the bosonic result of eq.\ \gl{ZQGres}. Guided by the
approach in the Hermitian matrix model, one now identifies the
free energy of the supereigenvalue model with the partition function
of 2d supergravity coupled to superconformal matter, i.e.
\beq
N^2 \, F_g [g_k, \x_{k+1/2}] \, \widehat{=} \, Z_g,
\eeq[superid]
and sets \Math{(g_k - g_k^c)\sim (\k-\k_c)}. Using our result of eq.\ 
\gl{Fscaling} and the scaling behaviour of the bosonic couplings in eq.\
\gl{ygscale} one finds
\beq
N^2\, F_g\sim (g_k -g_k^c)^{(2+1/m)\, (1-g)},
\eeq[Fgsuperscale]
hence we see that \Math{\g_{\mbox{\scriptsize str.}}=-1/m}. The computation
of a string susceptibility exponent is of course not enough to identify
the continuum model described by the double scaled supereigenvalue
model. One has to establish a connection between Liouville operators
to supereigenvalue correlators.
The Neveu--Schwarz and Ramond sectors of the super--Liouville theory are 
reflected in the even and odd correlators of the supereigenvalue model
\Math{\langle \sum_i {\l_i}^k \rangle} and 
\Math{\langle \sum_i \q\,{\l_i}^k \rangle} respectively. Adequate linear
combinations of these correlators in the scaling limit can be identified
with super--Liouville amplitudes, as was shown by Abdalla and
Zadra in ref.\ \cite{Zad} by using the planar supereigenvalue results of
ref.\ \cite{Alv}. The outcome is that the double scaled supereigenvalue
model at its $m$'th critical point describes the coupling of a minimal
superconformal field theory of type \Math{(p,q)=(4m,2)} to 2d supergravity
having the central charge
\beq
\widehat{c}= 1 - \frac{(2m-1)^2}{m}= 0, -\frac{7}{2}, -\frac{22}{3},\ldots
\qquad m\geq 1.
\eeq[chatmrel]
The only unitary model in this series is pure supergravity with \Math{m=1}. 
Note, however,
that this case must be treated separately, as in our approach \Math{m=1}
corresponds to no restrictions on the moments (cf.\ eq.\ \gl{MJcritical}),
and thus leads to no critical behaviour.
This is completely analogous to the bosonic case, where a string
susceptibility of \Math{\g_{\mbox{\scriptsize str.}}=-1} corresponding to 
\Math{c=-2} may not be reached by the Hermitian one matrix
model, but rather through the model studied in chapter II. For the
higher non--unitary superconformal theories one easily convinces oneself 
that \Math{(p,q)=(4m,2)} yields \Math{\g_{\mbox{\scriptsize str.}}=-1/m} by 
using eq.\ \gl{gstring2}.
\Subkapitel{Conclusions}
We have constructed the supereigenvalue model by imposing the 
super--Virasoro constraints on its partition function.
By integrating out its fermionic degrees of freedom we saw that the bosonic
part of the free energy of the supereigenvalue model is simply twice the
free energy of the corresponding hermitian matrix model.
First the model was solved away from the double scaling limit.
The superloop correlators obey a set of integral equations, the
superloop equations. These two equations could be split up into a set of four
equations, sorted by their order in fermionic coupling constants. By a change
of variables from coupling constants to moments we were
able to present the planar solution of the superloop equations
for general potentials in a very compact form.
The remarkable structure of the superloop equations enabled us to develop an
iterative procedure for the calculation of higher--genera contributions to the
free energy and the superloop correlators. Here it proved sufficient to solve
the two lowest--order equations at genus $g$ for 
the purely bosonic $u_g(p)$ and
the fermionic $v_g(p)$ (up to a zero mode contribution). The zero mode
as well as the doubly fermionic part of the free energy could then be
found by rewriting $v_g(p)$ as a total derivative in the fermionic potential.
The purely bosonic part of the free energy can be calculated
with the methods of ref.\ \cite{Amb93}.
In principle the application of loop insertion operators to the free
energy then yields
arbitrary multi--superloop correlators. As we demonstrated for genus one,
in practice these expressions become quite lengthy. 
\par
We then turned to the double scaling limit of the model in the moment
description. The $m$'th multi--critical point was identified and the
scaling properties of the moments and basis functions were derived. The 
iterative scheme for the calculation of higher genus  contributions 
could be optimized to produce only terms relevant in the double scaling 
limit. The general form of the free energy and the one--superloop 
correlators at genus $g$ were stated. We presented explicit
result up to genus two. Finally we commented on
the identification of the supereigenvalue model at its $m$'th 
multi--critical point with 2d supergravity coupled to \Math{(4m,2)}
minimal superconformal matter.
\par
The analogy of structures in the Hermitian one matrix model and in the
supereigenvalue model is rather impressive. This lends hope to find
an answer to further interesting questions one should address, such as the
supersymmetric generalization of two-- and multi--matrix models or 
the generalization of matrix models in external fields, which we shall
be discussing in the next chapter.

%% file: external
%
%
So far the partition functions of the Hermitian matrix and the 
supereigenvalue model acted as
generating functionals at the same time due to the presence of 
arbitrary polynomial potentials. This is why solving for the free energy
of these models really meant completely determining all possible
correlators as well. As we know from field theory
an alternative way to build generating functionals
is to include external fields as sources coupled linearly to the fields
of the theory. In the language of Hermitian matrix models this leads to 
partition functions depending on the eigenvalues of an external matrix.
It is thus not surprising that interrelations between ordinary and 
external matrix models may be established. The key point here is to
find a translation scheme from the coupling constants of the ordinary
theory to the eigenvalues of the external matrices known as Miwa
transformations.
The study of matrix models in external fields has received much attention
following the work of Kontsevich \cite{Kont} whose external field
model was shown to be equivalent to the Hermitian matrix model in the
double scaling limit. No limiting procedure is necessary as the model
directly produces only the scaling relevant terms. 
\par
Moreover external field models are closely related to Hermitian
two matrix models, which in their scaling limit are capable of describing
{\it all} the \Math{(p,q)} minimal conformal field theories coupled to 2d 
gravity. Although the form of the two matrix model is obvious on the
matrix level, it is not on the eigenvalue level as two arbitrary Hermitian
matrices may not be simultaneously diagonalized by one \Math{U(n)}
transformation.
\par
In regard of these results it would be quite desirable to find the correct
supersymmetric generalizations of external Hermitian matrix models.
However, in that respect this chapter has the form of an outlook, as we 
have not been able to perform this program completely. Nevertheless some 
encouraging results may be reported.
\par
Due to simplicity we study the discrete case, i.e.\ away from the 
double scaling limit. After a review of the bosonic case we turn to the
supersymmetrization of the observed structures.
\Subkapitel{The Bosonic Case}{BosCas}
Let us consider a \Math{N \times N} Hermitian matrix model with a 
potential \Math{V_0(\MX)} and an external Hermitian matrix field \Math{\ML}
as source term 
\beq
Z_N[ \ML] = \int_{N\times N} {\cal D}\MX\, \exp [ \, \tr \ML\, \MX 
- \tr V_0(\MX)\, ].
\eeq[EFModel]
By taking matrix derivatives of \Math{Z_N[\ML]} with 
respect to the source and defined by
\beq
(\DL)_{ab} \equiv  \frac{\del}{\del \ML_{ba}} \qquad \Rightarrow \qquad
(\DL)_{ab}\, Z_N[\ML] = \langle\, \MX_{ab}\,\rangle\, Z_N[\ML]
\eeq[MatDelDef]
one may generate arbitrary correlators. 
In this sense the source \Math{\ML} replaces
the infinite set of coupling constants $g_k$ of the Hermitian 
matrix model of eq.\ \gl{hermmmodel}.
\par
External field models may be solved in the planar limit
and in their genus expansions through the method of Schwinger--Dyson
equations \cite{Gro,MakSem1}. These equations are easily derived.
Consider the shift of the matrix $\MX$ in eq.\ \gl{EFModel}
\beq
\MX \ra \MX + \e_n\, \MX^{n+1} \qquad n\geq -1,
\eeq[Eshift]
with $\e_n$ infinitesimal. Then the invariance of the integral under
a renaming of integration variables yields 
\beqx
\int_{N\times N} {\cal D}\MX \, \tr \Bigl [ \ML\, \MX^{n+1} - \MX^{n+1}\,
V_0^\prime(\MX ) + \sum_{k=0}^n \tr [ \,\MX^k\,] \, \MX^{n-k}\, \Bigl ]
\eeqx
\beq
\quad \cdot\: \exp [ \, \tr \ML\, \MX - \tr V_0(\MX)\, ] = 0.
\eeq[ESD1]
in first order $\e_n$.
\Comment{
See this by calculating the Jacobian of the transformation \gl{Eshift}
to first order $\e_n$:
\beqx
\Det \frac{\del \,{\fett X}^\prime}{\del\MX^T}= e^{\mbox{\scriptsize Tr}
 \ln (1 + \e_n\, \frac{\del \,\MX^{n+1}}{\del \MX^T}\, ) }\zeile
= e^{\mbox{\scriptsize Tr}\e_n\,\frac{\del\MX^{n+1}}{\del \MX^T} }= 
1+ \e_n\,\tr \frac{\del\MX^{n+1}}{\del \MX^T} 
\eeqx
Now 
\beqx
\tr \frac{\del\,\MX^{n+1}}{\del\MX^T}= \frac{\del}{\del\MX_{ba}}\, 
\MX_{b\, i_1}\,\MX_{i_1 i_2} \ldots \MX_{i_n a} = \sum_{k=1}^n 
\tr \MX^k\, \tr \MX^{n-k},
\eeqx
and combining this with the variation of the potential and the source 
term yields eq.\ \gl{ESD1}
}
Using derivatives with respect to the source eq.\ \gl{ESD1} may be 
rewritten as
\beq
\tr \Bigl [ \ML\, (\DL)^{n+1} - (\DL)^{n+1}\, V_0^\prime(\DL) +
\sum_{k=0}^n (\DL)^k \, \tr (\DL)^{n-k}\, \Bigr ] \, Z_N [\ML] =0,
\eeq[ESD2]
and by pulling out all matrix derivatives we have
\beq
\tr (\DL)^{n+1}\, \Bigl[ \ML - V^\prime_0(\DL)\, \Bigr ] \, Z_N [\ML] =0.
\eeq[ESD3]
This system is in fact equivalent to a single matrix valued equation
\beq
\Bigl[ \, V^\prime_0(\DL) -\ML \, \Bigr ] \, Z_N[\ML] =0,
\eeq[GrossNewman]
known as the Gross--Newman equation \cite{Gro}.
\par
Interestingly enough the external field model of eq.\ \gl{EFModel} with 
the potential
\beq
V_0(\MX)= \frac{1}{2} \, \MX^2 - n\, \ln \MX
\eeq[KostMeh]
is closely related to the \Math{n\times n} Hermitian one matrix model 
with general polynomial potential
\beq
{\cal Z}_n[\, g_k]=\int_{n\times n} {\cal D}\MX \, \exp [- \tr 
\sum_{k=0}^\infty g_k\, \MX^k\, ].
\eeq[HerM]
The exact relation between the partition functions \gl{EFModel} and
\gl{HerM} was noted by Chekhov and Makeenko \cite{ChekMak} and reads 
\beq
{\cal Z}_{n} [\, g_k] = \exp [- \frac{1}{2}\, \tr \ML^2\, ]\, Z_N [\ML],
\eeq[Relation]
provided \Math{\ML} and $g_k$ are related by the Miwa--transformation
\beql
g_k &=& \frac{1}{k}\, \tr \ML^{-k} - \frac{1}{2}\, \d_{k,2}
\qquad \mbox{for} \, k\geq 1  \zeile
g_0 &=& -\tr \ln \ML
\eeql[MiwaTrafo]
in the limit \Math{N\ra\infty} which turns the $g_k$'s into independent 
variables.
Note that  in eq.\ \gl{Relation} the \Math{N\times N} external field model 
is related to a \Math{n\times n} Hermitian one matrix model.
The proof of eq.\ \gl{Relation} is based on the fact that the 
Gross--Newman equation \gl{GrossNewman} for the potential
\gl{KostMeh} is equivalent to the Virasoro constraints of eq.\ 
\gl{Virasoroconstraint} for the Hermitian one matrix model \gl{HerM}.
\par
Let us prove the relation \gl{Relation}. Taking the Gross--Newman
equation \gl{GrossNewman} and applying one more matrix derivative
\Math{\DL} to get rid of the singular \Math{\DL^{-1}} term coming from
the derivative of the logarithm we find
\beq
\Bigl (\, \DL^2 - (n +1) - \ML\, \DL \, \Bigr )\, Z_N[\ML] =0,
\eeq[Pr1]
which may be restated as
\beq
\Bigl ( \, \DL^2 + L\, \DL - n \, \Bigr ) \,\exp [- \frac{1}{2}\, \tr \ML^2\, ]
\, Z_N[\ML]=0.
\eeq[Pr2]
Now insert the Miwa transformation into the matrix differential 
operator of this equation. By using the chain rule we have
\beq
\DL = \sum_{k=0}^\infty \frac{\del g_k}{\del \ML^T}\, \del_{g_k}=
-\sum_{k=0}^\infty \ML^{-k-1}\, \del_{g_k}
\eeq[DLgk]
and 
\beql
(\DL)^2 &= &- \sum_{k=0}^\infty ( \, \DL\, \ML^{-k-1}\, ) \, \del_{g_k}
+ \sum_{k,l} \ML^{-k-l-2}\, \del_{g_k}\,\del_{g_l} \label{Pr3} \\
&=& \sum_{k=0}^\infty \,\Bigl \{ \sum_{a=1}^{k+1} \frac{1}{\ML^{k+2-a}}\,
\tr \Bigl (\frac{1}{\ML^a}\Bigr )\, \del_{g_k} + 
\frac{1}{\ML^{k+2}}\, \sum_{n=0}^k \del_{g_n}\, \del_{g_{k-n}}\, \Bigl \}.
\nonumber
\eeql
Rewriting the double sum in the first term of eq.\ \gl{Pr3} leads us to
\beq
(\DL)^2 = \sum_{m=-1}^\infty \frac{1}{\ML^{m+2}} \sum_{a=1}^\infty
\tr \Bigl ( \frac{1}{\ML^a} \Bigr )\, \del_{g_{m+a}} +
\sum_{m=0}^\infty \frac{1}{\ML^{m+2}} \sum_{k=0}^m \del_{g_k}\, 
\del_{g_{m-k}},
\eeq[Pr31]
which is already strongly reminiscent to the Virasoro constraints if one
would ignore the delta function addition in eq.\ \gl{MiwaTrafo}. Putting
all this together we see that
\beql
(\DL^2 + \ML\, \DL -n) &= &\sum_{m=-1}^\infty \frac{1}{\ML^{m+2}}
\Bigl \{ \sum_{k=1}^\infty \tr (\ML^{-k} )\, \del_{g_{m+k}} -
\del_{g_{m+2}} + \sum_{k=0}^m \del_{g_k}\, \del_{g_{m-k}}\, \Bigr \}\zeile
&& \quad - (\, n + \del_{g_0}\, ),
\eeql[Pr4]
so by identifying \Math{k\, g_k= \tr( \ML^{-k}) - \d_{k,2}} in the above
we recover the Virasoro generator \Math{\VL{m}}
\beql
(\DL^2 + \ML\, \DL -n) &= &\sum_{m=-1}^\infty \frac{1}{\ML^{m+2}}
\Bigl \{ \underbrace{\sum_{k=0}^\infty k\, g_k\, \del_{g_{m+k}}
 + \sum_{k=0}^m \del_{g_k}\, \del_{g_{m-k}}\,}_{=\VL{m}} \Bigr \}\zeile
&& \quad - (\, n + \del_{g_0}\, ),
\eeql[Virid]
of eq.\ \gl{LVir}. Therefore the Virasoro constraints obeyed by \Math{{\cal
Z}_n[\, g_k]} are equivalent to the Gross--Newman equation for 
\Math{Z_N[\ML]}, provided that 
\beq
\del_{g_0}\,{\cal Z}_{n} [g_k] = - n \, {\cal Z}_{n} [\, g_k]
\eeq[alpharel]
is valid. But this is obviously true for an \Math{n\times n} Hermitian
matrix model, as \Math{\del_{g_0}\,{\cal Z}_{n} [g_k]=-\langle \tr 1 
\rangle\, {\cal Z}_{n} [g_k]} seen from the definition in eq.\ \gl{HerM}.
Hence we have proven the relation \gl{Relation}.
Amazingly enough this identity 
does not depend on the dimension $N$ of the external matrix field.
\par
Of course these identities may also be stated in the eigenvalue
language. As the partition function of the external field model
\gl{EFModel} only depends on the eigenvalues of the matrix source
\Math{\ML}, the integral over the angular variables may be performed
with the help of the Itzykson--Zuber formula \cite{ItzZub} to give the 
external eigenvalue model
\beq
Z_N[ l_i ] = \Bigl ( \int \prod_{i=1}^N dx_i\Bigr ) \, 
\frac{\prod_{i<j}(x_i-x_j)}
{\prod_{i<j}(l_i-l_j)}\, \exp \Bigl [ \sum_{i=1}^N ( - V_0(x_i) + l_i\, x_i \, )
\Bigr ],
\eeq[EEVM2] 
modulo an irrelevant multiplicative constant. The $x_i$ denote the 
eigenvalues of $\MX$ and $l_i$ those of $\ML$. Integrating \Math{Z_N[\, l_i\,]}
with the measure \Math{\prod_{i<j}(l_i-l_j)^2\, \exp[ -\sum_i V_1(l_i)\, ]}
over $l_i$ yields the eigenvalue formulation of the Hermitian two
matrix model.
\par
In terms of eigenvalues the 
Schwinger--Dyson equation \gl{Pr2} now takes the form
\beqx
\sum_{i=1}^N \Bigl [ \, \frac{\del^2}{\del{l_i}^2} + \sum_{i\neq j}
\frac{1}{l_i-l_j}\, \Bigl ( \,\frac{\del}{\del l_i} -\frac{\del}{\del l_j}
\, \Bigr )
+ l_i\, \frac{\del}{\del l_i} - n\, \Bigr ]
\eeqx
\beq\times \: \exp [ -\frac{1}{2}\,
 \sum_{k=1}^\infty {l_k}^2\, ]\, Z_N[ l_i ] =0,
\eeq[SDEV]
which may be derived by considering the variation \Math{\d l_i= \e\, l_i}
in eq.\ \gl{EEVM2} or by studying the eigenvalue dependence of the
matrix derivatives \Math{\DL} and \Math{\DL^2}.
\par
The identity \gl{Relation} may also be proven directly on the level of
the eigenvalue models, by inserting the Miwa--transformations into
eq.\ \gl{HerM} and using integral identities for Hermite polynomials
\cite{KMMM}.
\par
The equivalence of external field and Hermitian matrix models continues
to hold true in the continuum limit. Here the Kontsevich model \cite{Kont},
which is built from a cubic \Math{V_0(\ML )}, is the matrix model 
describing the Hermitian matrix model in the double scaling limit, as
shown in \cite{MMM}. Let me just mention, that in the framework of
generalized Kontsevich models a unification of all \Math{(p,q)}  minimal
models coupled to 2d gravity could be achieved \cite{GKM}. 
 \Subkapitel{The Super--Miwa Transformations}{SMiwa}
 Our analysis of the supereigenvalue model in chapter III revealed
 a striking number of analogies to the Hermitian matrix model. It is thus
 rather tempting to ask whether there exists an external field formulation
 of the supereigenvalue model as well, generalizing the results of the previous
 section. The partition function of such an external supereigenvalue
 model should depend on a set of even and odd external ``supereigenvalues''
 $l_i$ and $\q_i$ respectively. Similar to the Gross--Newman equation
 the Schwinger--Dyson equation of this model would then be represented
 as a differential operator in the external fields $l_i$ and $\q_i$
 annihilating the partition function. After performing a super--Miwa 
 transformation of this differential operator from the $l_i$ and $\q_i$ 
 to the coupling constants $g_k$ and $\x_{k+1/2}$ one should recover
 the super--Virasoro generators, thus proving the equivalence of these
 two models.
 \par
 Unfortunately we have not succeeded in performing this program so far.
 However, we can report on some preliminary results. 
 The correct differential operator in external fields could
 be identified, leading to the super--Virasoro constraints through a
 set of super--Miwa transformations. This operator may even be 
 formulated on the matrix level, thus giving hope to find a true 
 supersymmetric external {\it matrix} model. 
 \par
Let us consider a pair of external \Math{N\times N} matrices, the 
Grassmann even matrix \Math{\ML} accompanied by the matrix
\Math{\MQ} with Grassmann odd entries. These shall be related to
the coupling constants $g_k$ and $\x_{k+1/2}$ of the supereigenvalue
model via the super--Miwa transformations
\beql
g_k&=& \frac{1}{k}\, \tr \ML^{-k} \qquad k\geq 1 \qquad\qquad
\x_{k+1/2}= - \tr \MQ\, \ML^{-k-1} \qquad k\geq 0 \zeile
g_0 &=& -\tr \ln \ML,
\eeql[SuperMiwaTrafo]
generalizing eq.\ \gl{MiwaTrafo}. Note that at this stage we have not
included delta function additions as in the bosonic case of eq.\ 
\gl{MiwaTrafo}. The matrix derivatives 
\beq
\DL = \frac{\del}{\del \ML^T} \qquad \mbox{and} \qquad
\DQ = \frac{\del}{\del \MQ^T}
\eeq[SMDef]
may then be reexpressed as differential operators in the coupling
constants by using the chain rule as
\beq
\DQ= -\sum_{k=0}^\infty \frac{\del\, \tr\MQ\, \ML^{-k-1} }{\del\MQ^T}
\, \del_{\x_{k+1/2}} = -\sum_{k=0}^\infty \ML^{-k-1}\, \del_{\x_{k+1/2}}
\eeq[SCR1]
and
\beql
\DL&=& \sum_{k=0}^\infty \Bigl ( \frac{1}{k}\, \frac{\del \, \tr \ML^{-k}}
{\del \ML^T} \, \del_{g_k}- \frac{\del \, \tr \MQ\, \ML^{-k-1}}{\del \ML^T}\,
\del_{\x_{k+1/2}}\, \Bigr ) \zeile
&=& \sum_{k=0}^\infty \Bigl ( - \ML^{-k-1}\, \del_{g_k} + \sum_{a=1}^{k+1}
\ML^{-a}\, \MQ\, \ML^{-k-2+a}\, \del_{\x_{k+1/2}}\, \Bigr ) .
\eeql[SCR2]
As the super--Virasoro generators \Math{\SG{n+1/2}} and \Math{\SL{n}}
are maximally quadratic in the \Math{\del_{g_k}} and 
\Math{\del_{\x_{k+1/2}}}
it is obvious that we should study quadratic matrix differentials in
\Math{\ML} and \Math{\MQ}. The first candidate is \Math{\DQ\, \DL}
which after some calculation may be found to be
\beql
 \tr ( \DQ\, \DL ) &=& \sum_{m=-1}^\infty \tr ( \,\ML^{-m-2}\, ) \, 
 \Bigl [ \sum_{k=0}^\infty k\, g_k \, \del_{\x_{k+m+1/2}} + 
 \sum_{k=0}^m \del_{g_{m-k}}\, \del_{\x_{k+1/2}} \, \Bigl ] \zeile
 && + \sum_{m=-1}^\infty \tr ( \,\MQ \,\ML^{-m-2}\, ) \,  \Bigl [
 \sum_{k=0}^{m-1} k \, \del_{\x_{m-k-1/2}}\, \del_{\x_{k+1/2}} \, 
 \Bigr ].
\eeql[DQDL]
Note that already a number of factors present in the super--Virasoro
generators of eq.\ \gl{SuperVirGen} appear. The next candidate to consider
is the 
expression \Math{\MQ\, \DL^2}. A rather tedious computation reveals
that
\beql
\tr( \MQ\, \DL^2 )&=& \sum_{m=-1}^\infty \tr (\, \MQ\, \ML^{-m-2}\, )
\, \Bigl [ \sum_{k=0}^\infty k\, g_k\, \del_{g_{k+m}} + \sum_{k=0}^m
\del_{g_k}\, \del_{g_{m-k}} \zeile
&& \phantom{\sum_{m=-1}^\infty \tr (\, \MQ\, \ML^{-m-2}\, )
\, \Bigl [ }  + \sum_{k=0}^\infty k\, \x_{k+1/2}\, \del_{\x_{m+k+1/2}}\, 
\Bigr ],
\eeql[MQDLDL]
where all terms of order 3 in \Math{\MQ} cancel due to their fermionic
structure. Additionally one proves the following identities
\beql
0 &=& \sum_{m=-1}^\infty \tr ( \, \MQ\, \ML^{-m-2}\, ) \, \Bigl [
\sum_{k=0}^\infty (k+m+1)\, \x_{k+1/2}\, \del_{\x_{k+m+1/2}}\, \Bigr ]
\zeile && \zeile
0 &=& \sum_{m=-1}^\infty \tr ( \, \ML^{-m-2}\, ) \, \Bigl [ \sum_{k=0}^\infty
\x_{k+1/2}\, \del_{g_{m+k+1}}\, \Bigr ] \zeile
&& + \sum_{m=-1}^\infty \tr (\, \MQ\, \ML^{-m-2}\, ) \, \Bigl [ 
\sum_{k=0}^\infty k\, g_k\, \del_{g_{k+m}}\, \Bigr ].
\eeql[SMId]
So by putting eq.\ \gl{DQDL}, \gl{MQDLDL} and \gl{SMId} together we
see that the correct differential operator yielding the super--Virasoro
constraints is as simple as
\beqy
\tr [ \, \DQ\,\DL + \MQ\, \DL^2\, ] =
\eeqy
\beql
\sum_{m=-1}^\infty \tr (\, \ML^{-m-2}\, ) && \Bigl \{ \sum_{k=0}^\infty
k\, g_k\, \del_{\x_{k+m+1/2}} + \sum_{k=0}^\infty \x_{k+1/2}\, 
\del_{g_{m+k+1}} \zeile &&
+ \sum_{k=0}^m \del_{g_{m-k}}\, \del_{\x_{k+1/2}}\, 
\Bigr \} \zeile
+ \sum_{m=-1}^\infty 2 \tr (\, \MQ\,\ML^{-m-2}\, ) && \Bigl \{ 
\sum_{k=0}^\infty k\, g_k\, \del_{g_{k+m}} + \frac{1}{2}\sum_{k=0}^m
\del_{g_k}\, \del_{g_{m-k}} \zeile
&& + \sum_{k=0}^\infty (k + \frac{m+1}{2})\, 
\x_{k+1/2}\, 
\del_{\x_{k+m+1/2}}\zeile &&
 + \frac{1}{2}\sum_{k=0}^{m-1} k \, \del_{m-k-1/2}
\, \del_{k+1/2}\, \Bigr \} \nonumber
\eeql
\beq
= \sum_{m=-1}^\infty \tr( \, \ML^{-m-2}\, )\, \SG{m+1/2} + 2
\sum_{m=-1}^\infty \tr( \, \MQ\,\ML^{-m-2}\, )\, \SL{m},
\eeq[SVirSMiwa]
with the super--Virasoro generators \Math{\SG{m+1/2}}  and
\Math{\SL{m}} defined in eq.\ \gl{SuperVirGen} \footnote{
Note that the factors of $N^2$ are recovered through the rescalings
\Math{g_k\ra N\, g_k} and \Math{\x_{k+1/2}\ra N\, \x_{k+1/2}}.}.
\par
The relation \gl{SVirSMiwa} may also be stated on the level of 
``supereigenvalues'' $l_i$ and $\q_i$, despite the fact that we do not
know how the $\q_i$ should be related to the matrix $\MQ$.
With the obvious supereigenvalue version of the super--Miwa 
transformations
\beql 
g_k&=& \frac{1}{k}\, \sum_i {l_i}^{-k} \qquad k\geq 1 \qquad\qquad
\x_{k+1/2}= - \sum_i \q_i\, {l_i}^{-k-1} \qquad k\geq 0 \zeile
g_0 &=& -\sum_i \ln l_i,
\eeql[SuperMiwaTrafo2]
one has the following identity
\beqy
\dq{i}\, \dl{i} + \q_i \frac{\del^2}{\del {l_i}^2} + 
\sum_{j\neq i} \frac{1}{l_i-l_j-\q_i\,\q_j}
\,\Bigl [ \dq{i}-\dq{j} + (\,2\, \q_i -\q_j\, )\, (\, \dl{i}-\dl{j}\, ) \Bigr ]=
\eeqy
\beq
\sum_{m=-1}^\infty {l_i}^{-m-2}\, [\, \SG{m+1/2} + 2 \, \q_i\, \SL{m}\, ],
\eeq[EvSMSV]
as already noted by Huang and Zhang \cite{HuaZha} in a different 
context.
So we find ourselves in the situation where we know the matrix and
eigenvalue based formulation of the Schwinger--Dyson equations of
a model, but do not know the model they belong to. Including delta function
additions in the super--Miwa transformations of eqs.\ \gl{SuperMiwaTrafo}
and \gl{SuperMiwaTrafo2} as they appear in the bosonic transformations
of eq.\ \gl{MiwaTrafo} leads to terms linear in $\DQ$ and $\DL$.
Unfortunately we have not succeeded in formulating the correct
external supersymmetric model so far.
\Subkapitel{Properties of External Eigenvalue Models}
In section \Sub{BosCas} we saw how the Virasoro constraints emerged 
out of the Schwinger--Dyson equation associated with the shift
\Math{\d\MX=\e\, \MX} for the external field model
of eq.\ \gl{EFModel} with the special potential \gl{KostMeh}. There is yet
another way in which the Virasoro structure is hidden in the 
Schwinger--Dyson equations \gl{ESD3} of the external field models
with arbitrary potentials as noted by Makeenko and Semenoff in 
ref.\ \cite{MakSem1}. 
It turns out that this structure may be generalized to
an external supereigenvalue model. 
\par
Let us first study the bosonic case in the eigenvalue description. 
The external field model of eq.\ \gl{EEVM2} reads
\beq
Z_N[\, l \, ] = \Bigl ( \int \prod_{i=1}^N dx_i\Bigr ) \, 
\frac{\D[\,x\, ]}
{\D[\,l\, ]}\, \exp \Bigl [\, \sum_{i=1}^N ( - V_0(x_i) + l_i\, x_i \, )
\, \Bigr ],
\eeq[EEVM]
where \Math{\D[\,l\, ]= \prod_{i<j}(l_i-l_j)} is the van der Monde determinant.
The Schwinger--Dyson equations associated to the shift \Math{x_i\ra
x_i + \e_n\, {x_i}^{n+1}} for \Math{n\geq -1} may be written in the
form
\beq
L_n\, \D[\,l\, ]\, Z_N[\,l\, ] = 0 \qquad \mbox{for} \quad n\geq -1
\eeq[ViraltCon]
with
\beql
L_n&=& -\sum_i V^\prime_0 \Bigl ( \dl{i} \Bigr )\, \Bigl ( \dl{i}\Bigr )^{n+1}
+ \sum_i \Bigl (\dl{i}\Bigr )^{n+1}\, l_i  \zeile
&&\qquad  + \frac{1}{2} \sum_{k=0}^n \, \sum_{j\neq i}\, \Bigl (\dl{i}\Bigr )^k
\, \Bigl (\dl{j}\Bigr )^{n-k}.
\eeql[VirGenalt1]
The differential operators $L_n$ obey the Virasoro algebra without 
central extension, as one verifies by direct computation
\beq
[\,\, L_n , L_m\, \, ]= (n-m) \, L_{n+m}.
\eeq[VirAlgalt]
Note that the generators \Math{L_n} annihilate \Math{\D[\,l\, ]\, Z_N[\,l\, ]}.
One may easily construct generators \Math{{\cal L}_n} which annihilate 
the partition function \Math{Z_N[\,l\, ]} itself. For this purpose introduce
the ``long derivatives''
\beq
\nabla_{l_i}\equiv \D^{-1}[\,l\, ]\, \dl{i}\, \D[\,l\, ] = \dl{i} + \sum_{i\neq j}
\frac{1}{l_i-l_j},
\eeq[Boslongder]
which commute with each other. The Virasoro constraints of eqs.\ 
\gl{ViraltCon} and \gl{VirGenalt1} now read
\beq
{\cal L}_n\, Z_N[\, l \, ]=0 \qquad \mbox{for}\quad  n\geq -1
\eeq[ViraltCon2]
and
\beql
{\cal L}_n &=& \sum_i \Bigl ( - V^\prime_0(\nabla_{l_i})\, 
(\nabla_{l_i} )^{n+1}
+ (\nabla_{l_i} )^{n+1}\, l_i \zeile
&&\qquad + \frac{1}{2}\, \sum_{k=0}^n \, \sum_{i\neq j}
(\nabla_{l_i})^k\, (\nabla_{l_j})^{n-k}\, \Bigr ).
\eeql[VirGenalt2]
Due to the commutativity of the $\nabla_{l_i}$'s the generators of eq.
\gl{VirGenalt2} obey the Virasoro algebra \gl{VirAlgalt} as well.
\par
For the supersymmetric case let us take the following ansatz for an
external supereigenvalue model
\beq
Z_N[\, l,\m \, ] = \int \Bigl( \prod_i d\l_i\, d\q_i \Bigr ) \, \frac{\D^\a
[\,\l,\q\, ]}
{\D^\a[\,l,\m\, ]}\, \exp \Bigl [ \sum_{i=1}^N ( -V(\l_i) + \l_i\, l_i + \q_i 
\, \m_i
\, )\, \Bigr ],
\eeq[extsupereigen]
with the measure
 \Math{\D^\a[\,\l,\q\, ]= \prod_{i<j} (\l_i-\l_j-\q_i\,\q_j)^\a}
and $\a$ undetermined. Note that the fermionic variables $\q_i$ enter
only through the source term and the measure.
One can now study the Schwinger--Dyson equations of this model
associated to the shift in integration variables
\beq
\d\l_i = -\e_n\, \q_i\, {\l_i}^{n+1} \qquad \mbox{and} \qquad
 \d\q_i= \e_n\, {\l_i}^{n+1}
\eeq[SDshift]
with \Math{n\geq -1} and $\e_n$ an Grassmann odd parameter. The resulting
correlator may be rewritten in the form
\beq
G_{n+1/2}\, \D^\a[\,l,\m \, ]\, Z_N[\, l, \m \, ] = 0 \qquad \mbox{for}
\quad n\geq -1
\eeq[sVirConalt]
where
\beql
G_{n+1/2}&=& - \sum_i \dm{i}\, V^\prime\Bigl (\dl{i}\Bigr )\, 
\Bigl (\dl{i}\Bigr )^{n+1} +
\sum_i \dm{i}\,\Bigl ( \dl{i}\Bigr )^{n+1}\, l_i \zeile
&& + \sum_i \m_i\, \Bigl (\dl{i}\Bigr )^{n+1} + \a \sum_{k=0}^n\, 
\sum_{i\neq j}
\dm{i}\, \Bigl (\dl{i}\Bigr )^k\, \Bigl (\dl{i}\Bigr )^{n-k}.
\eeql[Galt]
And in fact these generators obey the super--Virasoro algebra which one
may directly check
\beql
\{\, G_{n+1/2}, G_{m+1/2}\, \} &=& 2\, L_{n+m+1} \zeile
[\,\, L_n, L_m\, \, ] &=&\Bigl (n-m\Bigr )\, L_{n+m} \zeile 
[\,\, L_n, G_{m+1/2}\, \, ] &=& \Bigl ( \frac{n-1}{2} - m \Bigr )
\, G_{n+m+1/2},
\eeql[SViralgalt]
and where the generators $L_n$ take the more complicated form
\beql
L_n &=& -\sum_i V^\prime\Bigl (\dl{i}\Bigr )\, \Bigl (\dl{i}\Bigr )^{n+1} + 
\sum_i l_i\, \Bigl (\dl{i}\Bigr )^{n+1}
+ \frac{1}{2}\, \sum_i (n+1)\, \m_i\, \dm{i}\,\Bigl (\dl{i}\Bigr )^n \zeile
&& + \frac{1}{2}\, \sum_i (n+1)\, \Bigl (\dl{i}\Bigr )^n + 
\frac{\a}{2}\, \sum_{k=0}^n
\, \sum_{i\neq j}\Bigl (\dl{i}\Bigr )^k\, \Bigl (\dl{j}\Bigr )^{n-k} \zeile
&& + \frac{\a}{2}\, \sum_{k=0}^{n-1}\, \sum_{i\neq j} k\, \dm{i}\, 
\Bigl (\dl{i}\Bigr )^{n-k-1}\, \dm{j}\, \Bigl (\dl{j}\Bigr )^k.
\eeql[sLalt]
In fact no constraints on the constant $\a$ in eq.\ \gl{extsupereigen}
arise.
In order to obtain super--Virasoro generators directly annihilating the
partition function \Math{Z_N[\, l,\m\, ]} introduce the ``long
derivatives'' 
\beql
\nabla_{l_i}&\equiv& \D^{-\a}[\, l,\m\, ] \, \dl{i}\, \D^\a [\, l,\m\, ] =
\dl{i} +  \sum_{i\neq j} \frac{\a}{l_i-l_j-\m_i\,\m_j} \zeile
\nabla_{\m_i}&\equiv& \D^{-\a}[\, l,\m\, ] \, \dm{i}\, \D^\a [\, l,\m\, ] =
\dm{i} - \sum_{i\neq j} \frac{\a\, \m_j}{l_i-l_j}.
\eeql[superlong]
Note that the \Math{\nabla_{l_i}} commute and the \Math{\nabla_{\m_i}} 
anticommute with each other. 
Replacing the derivatives \Math{\del_{l_i}} by \Math{\nabla_{l_i}} 
and the \Math{\del_{\m_i}} by \Math{\nabla_{\m_i}} 
in eqs.\ \gl{Galt} and \gl{sLalt} then yields the super--Virasoro
generators \Math{{\cal G}_{n+1/2}} and \Math{{\cal L}_n} which
annihilate the partition function \Math{Z_N[\, l,\m\, ]}. 
\par
So we have seen that the ``naive'' ansatz of eq.\ \gl{extsupereigen} is
capable of producing a generalization of the properties found in the
external Hermitian matrix model. However all efforts to use this ansatz
as the model generating the Schwinger--Dyson equation \gl{EvSMSV} 
associated to the super--Miwa transformations have failed.
\par
This concludes our short outlook on external supereigenvalue and supermatrix
models.